\documentclass[useAMS,usenatbib]{mn2e}
\usepackage{graphicx}
\usepackage{placeins}
\usepackage{float}
\usepackage{subfigure}
\usepackage{amsmath}
\usepackage{mathrsfs}
\usepackage{amssymb}
\usepackage{color,epstopdf}
\usepackage{pdflscape}

\newcommand{\CII}{[C{\scriptsize II}]}
\newcommand{\CIItwelve}{[$^{12}$C{\scriptsize II}]}
\newcommand{\CIIthirteen}{[$^{13}$C{\scriptsize II}]}

\newcommand{\cmcube}{cm$^{-3}$}

\newcommand{\Tkin}{$T_{\text{kin}}$}
\newcommand{\Tex}{$T_{\text{ex}}$}

\newcommand{\HII}{H{\scriptsize II}}
\newcommand{\HI}{H{\scriptsize I}}
\newcommand{\LFIR}{$L_{\text{FIR}}$}
\newcommand{\CIIFIR}{$L_{\text{\CII}}/L_{\text{FIR}}$}
\newcommand{\COFIR} {$L_{\text{CO(1--0)}}/L_{\text{FIR}}$}
\newcommand{\CIICO}{$L_{\text{\CII}}/L_{\text{CO(1--0)}}$}
\newcommand{\LCII}{$L_{\text{\CII}}$}
\newcommand{\LCO}{$L_{\text{CO(1--0)}}$}
\newcommand{\Td}{$T_{\text{d}}$}
\newcommand{\LMHtwo}{$L_{\text{FIR}}/M_{\text{H}_2}$}
\newcommand{\TexCII}{$T_{\text{ex,\CII}}$}
\newcommand{\TexCO}{$T_{\text{ex,CO(1--0)}}$}
\newcommand{\tauCII}{$\tau_{\text{\CII}}$}
\newcommand{\tauCO}{$\tau_{\text{CO(1--0)}}$}

\usepackage{color}

\def\ESOGarching{1}
\def\Illinois{2}
\def\MPIfR{3}
\def\UPenn{4}
\def\ESOChile{5}
\def\Diego{6}
\def\JPL{7}
\def\Cambridge{8}
\def\KICPChicago{9}
\def\EFIChicago{10}
\def\PhysicsUChicago{11}
\def\AAUChicago{12}
\def\Dal{13}
\def\Davis{14}
\def\UFlorida{15}
\def\UCL{16}
\def\Stanford{17}
\def\Bristol{18}
\def\UCLA{19}
\def\Arizona{20}
\def\IPAC{21}

\def\Berkeley{23}
\def\CfA{24}
\def\Oxford{25}

\title[The nature of the \CII \ emission in dusty star-forming galaxies from the SPT-survey]{The nature of the [CII] emission in dusty star-forming galaxies from the SPT survey}
\author[B. Gullberg et al.]
{\parbox{\textwidth}{B.~Gullberg$^{\ESOGarching}$\thanks{E-mail: bgullber@eso.org}, 
C.~De~Breuck$^{\ESOGarching}$,
J.~D.~Vieira$^{\Illinois}$,
A.~Wei\ss$^{\MPIfR}$,
J.~E.~Aguirre$^{\UPenn}$,
M.~Aravena$^{\ESOChile,\Diego}$,
M.~B\'ethermin$^{\ESOGarching}$,
C.~M.~Bradford$^{\JPL}$,
M.~S.~Bothwell$^{\Cambridge}$,
J.~E.~Carlstrom$^{\KICPChicago,\EFIChicago,\PhysicsUChicago,\AAUChicago}$, 
S.~C.~Chapman$^{\Dal}$,
C.~D.~Fassnacht$^{\Davis}$,
A.~H.~Gonzalez$^{\UFlorida}$, 
T.~R.~Greve$^{\UCL}$,	
Y.~Hezaveh$^{\Stanford}$,
W.~L.~Holzapfel$^{\Berkeley}$, 
K.~Husband$^{\Bristol}$,
J.~Ma$^{\UFlorida}$, 
M.~Malkan$^{\UCLA}$,
D.~P.~Marrone$^{\Arizona}$,   
K.~Menten$^{\MPIfR}$,
E.~J.~Murphy$^{\IPAC}$,
C.~L.~Reichardt$^{\Berkeley}$, 
J.~S.~Spilker$^{\Arizona}$,
A.~A.~Stark$^{\CfA}$, 
M.~Strandet$^{\MPIfR}$, 
N.~Welikala$^{\Oxford}$
}\vspace{0.5cm}\\
\parbox{\textwidth}{
$^{\ESOGarching}${European Southern Observatory, Karl Schwarzschild Stra\ss e 2, 85748 Garching, Germany}\\
$^{\Illinois}${Department of Astronomy and Department of Physics, University of Illinois, 1002 West Green Street, Urbana, IL 61801, USA}\\
$^{\MPIfR}${Max-Planck-Institut f\"{u}r Radioastronomie, Auf dem H\"{u}gel 69 D-53121 Bonn, Germany}\\
$^{\UPenn}${University of Pennsylvania, 209 South 33rd Street, Philadelphia, PA 19104, USA}\\
$^{\ESOChile}${European Southern Observatory, , Alonso de Cordova 3107, Casilla 19001 Vitacura Santiago, Chile.}\\
$^{\Diego}${N\'ucleo de Astronom\'{\i}a, Facultad de Ingenier\'{\i}a, Universidad Diego Portales, Av. Ej\'ercito 441, Santiago, Chile}\\
$^{\JPL}${Jet Propulsion Laboratory, 4800 Oak Grove Drive, Pasadena, CA 91109, USA}\\
$^{\Cambridge}${Cavendish Laboratory, University of Cambridge, JJ Thompson Ave, Cambridge CB3 0HA, UK}\\
$^{\KICPChicago}${Kavli Institute for Cosmological Physics, University of Chicago, 5640 South Ellis Avenue, Chicago, IL 60637, USA}\\
$^{\EFIChicago}${Enrico Fermi Institute, University of Chicago, 5640 South Ellis Avenue, Chicago, IL 60637, USA}\\
$^{\PhysicsUChicago}${Department of Physics, University of Chicago, 5640 South Ellis Avenue, Chicago, IL 60637, USA}\\
$^{\AAUChicago}${Department of Astronomy and Astrophysics, University of Chicago, 5640 South Ellis Avenue, Chicago, IL 60637, USA}\\
$^{\Dal}${Dalhousie University, Halifax, Nova Scotia, Canada}\\
$^{\Davis}${Department of Physics,  University of California, One Shields Avenue, Davis, CA 95616, USA}\\
$^{\UFlorida}${Department of Astronomy, University of Florida, Gainesville, FL 32611, USA}\\
$^{\UCL}${Department of Physics and Astronomy, University College London, Gower Street, London WC1E 6BT, UK}\\
$^{\Stanford}${Kavli Institute for Particle Astrophysics and Cosmology, Stanford University, Stanford, CA 94305, USA}\\
$^{\Bristol}${H.H. Wills Physics Laboratory, University of Bristol, Tyndall Avenue, Bristol BS8 1TL, UK}\\
$^{\UCLA}${Department of Physics and Astronomy, University of California, Los Angeles, CA 90095-1547, USA}\\
$^{\Arizona}${Steward Observatory, University of Arizona, 933 North Cherry Avenue, Tucson, AZ 85721, USA}\\
$^{\IPAC}${Infrared Processing and Analysis Center, California Institute of Technology, MC 220-6, Pasadena, CA 91125, USA}\\
$^{\CfA}${Harvard-Smithsonian Center for Astrophysics, 60 Garden Street, Cambridge, MA 02138, USA}\\
$^{\Berkeley}${Department of Physics, University of California, Berkeley, CA 94720, USA}\\
$^{\Oxford}${Department of Physiscs, Oxford University, Denis Wilkinson Building, Keble Road, Oxford, OX1 3RH, UK}\\
}}
\begin{document}

\date{\today}

\pagerange{\pageref{firstpage}--\pageref{lastpage}} \pubyear{2014}

\maketitle


\label{firstpage}

\begin{abstract}
We present \CII \ observations of 20 strongly lensed dusty star forming galaxies at $2.1<z<5.7$ using APEX and \textit{Herschel}.  
The sources were selected on their 1.4\,mm flux ($S_{\text{1.4\,mm}}>20\,$mJy) from the South Pole Telescope survey, with far-infrared (FIR) luminosities determined from extensive photometric data.
The \CII \ line is robustly detected in 17 sources, all but one being spectrally resolved.  
Eleven out of 20 sources observed in \CII \ also have low-$J$ CO detections from ATCA.  A comparison with mid- and high-$J$ CO lines from ALMA reveals consistent \CII \ and CO velocity profiles, suggesting that there is little differential lensing between these species.
The \CII, low-$J$ CO and FIR data allow us to constrain the properties of the interstellar medium.
We find \CII \ to CO(1--0) luminosity ratios in the SPT sample of
$5200\pm1800$, with significantly less scatter than in other samples.  
This line ratio can be best described by a medium of \CII\ and CO emitting gas with a higher \CII\ than CO excitation temperature, high CO optical depth \tauCO~$\gg$~1, and low to moderate \CII\ optical depth \tauCII~$\lesssim$~1.
The geometric structure of photodissociation regions allows for such conditions.
\end{abstract}



\section{Introduction}\label{sec:intro}
The discovery of (sub)millimeter-selected dusty star forming galaxies (DSFGs) at high redshifts 
(e.g. \citealt{smail97,hughes98,barger98}) fundamentally changed our view of galaxy formation and evolution. DSFGs are massive ($M_*\sim10^{11}$\,M$_{\odot}$; e.g. \citealt{hainline11,michalowski12}) and gas-rich ($M_{\text{gas}}\sim3-5\times10^{10}$\,M$_{\odot}$; e.g. \citealt{greve05,tacconi10,bothwell13}), and have star formation rates $\gtrsim1000$\,M$_{\odot}$yr$^{-1}$ (e.g. \citealt{chapman05}). 
The properties of these galaxies remain a challenge for conventional galaxy formation models \cite[e.g.][]{baugh05,lacey10, benson12b, hayward13}. 

Thanks to the availability of new space- and ground-based sub-millimetre facilities, our knowledge of the interstellar medium (ISM) in massive gas-rich galaxies at high redshift has dramatically improved in the past decade (\citealt{solomon05,carilli13,casey14}). 
The most commonly used lines for studying the ISM in DSFGs at high redshift are the rotational transitions of carbon-monoxide (CO) and the \CII \ $\lambda158\,\mu$m fine structure line.
The latter is the most important cooling line in the ISM \citep{dalgarno72}, and traces neutral gas exposed to ultraviolet photons from young stars. The \CII \ line can therefore be used to probe the stellar radiation field and how it affects the physical conditions of the gas. 
The bulk of the \CII \ emission line (70\% in \citealt{stacey91,stacey10}) is believed to originate from photodissociation regions (PDRs), and the remainder from X-ray dominated regions (XDRs), cosmic ray dominated regions (CRDRs), ionised regions (H{\scriptsize II} regions) \citep{meijerink07}, low density warm gas and/or diffuse HI clouds \citep{madden97}. 

The \CII\ fine-structure transition ($\nu_{\text{rest}}^{\text{\CII}}=1900.54\,$GHz) is nearly unobservable from the ground at $z<1$ due to strong atmospheric absorption. The only low-$z$ \CII \ samples have been observed with airborne \citep{crawford86,stacey91} or space based observatories \citep{malhotra01, brauher08, diazsantos13, farrah13, 	
sargsyan14, delooze14}. Submillimetre atmospheric windows provide some access to the line from the ground at $z>1$, with atmospheric transparency continuing to improve towards higher redshifts (longer wavelengths).
As a consequence, the first high-$z$ \CII \ detection was reported for the $z=6.42$ quasar host galaxy SDSSJ1148+5251 a decade ago \citep{maiolino05}. 
The number of \CII \ detections has been steadily increasing since then, thanks to facilities such as the Caltech Submillimeter Observatory (CSO), the Submillimeter Array, 
the IRAM\footnote{Institute for Radio Astronomy in the Millimeter Range} Plateau de Bure Interferometer, the Atacama Pathfinder EXperiment (APEX), the \textit{Herschel Space Observatory}, the Combined Array for Research in Millimeter-wave Astronomy, and the Atacama Large Millimeter/submillimeter Array (ALMA). 
At $1<z<2$, 21 \CII \ detections were made using the redshift ($z$) and Early Universe Spectrometer (ZEUS) on CSO \citep{hailey-dunsheath10,stacey10,brisbin14}, while \textit{Herchel} detected three \CII \ lines at $1.5<z<3$ \citep{ivison10,valtchanov11,george13}.
Seventeen have been added to the number of $z>4$ \CII \ detections in the past decade \citep{maiolino05,maiolino09,iono06,wagg10,ivison10,debreuck11,cox11,swinbank12,venemans12,walter12,riechers13,wang13,rawle14,debreuck14,neri14}.
Many of these objects have been selected based on the presence of a luminous active galactic nuclei  \citep[AGN; e.g.][]{maiolino05, stacey10, wang13}, while others have been selected as starburst galaxies \citep[e.g.][]{ivison10,hailey-dunsheath10,stacey10, cox11,swinbank12,walter12,riechers13,brisbin14}. 
This leads to a heterogeneous sample of high-$z$ \CII \ detections, containing a mixture of AGN and starburst-dominated systems. The heterogeneity of the sample complicates the interpretation of trends within it.
\cite{stacey10} suggest, based on the \CII/FIR and \CII \ to CO(1--0) luminosity ratios, that \CII \ emission originates mainly from PDRs and that the ISM and stellar radiation field in these $z\sim1-2$ galaxies resemble that observed for local starburst systems. These studies have followed the conclusion of \cite{crawford85} that the \CII \ emission is optically thin or reaching unity opacity ($
\tau \lesssim 1$); however, this was recently challenged by \cite{neri14}, who argued for optically thick \CII.

Here we present \CII \ observations of 20 
gravitationally lensed DSFGs in the redshift range $z\sim2.1 - 5.7$ discovered by the South Pole Telescope \citep[SPT;][]{carlstrom11, vieira10}. These 20 sources are a subset of those selected from the first 1300~deg$^2$ of the 2500~deg$^2$ SPT-SZ survey. Followup observations with ALMA have provided spectroscopic redshifts for all of these objects \citep{weiss13,vieira13}. We also include two new sources observed in Cycle~1 (see Appendix~\ref{app:new_sources} for details).
The bright ($S_{1.4\text{mm}}>20\,$mJy) flux selection of the SPT sample ensures that virtually all sources will be gravitationally magnified, with a bias towards $z>2$ \citep{hezaveh11,weiss13}. 
The magnified emission allows us to study the ISM in these DSFGs in greater detail, using fine structure and molecular lines such as \CII \ and CO. 
By including low-$J$ CO observations for 11 sources in our analysis, we determine the physical state of the ISM by studying the \CII \ and CO(1--0) line intensity ratios.

This paper is organised as follows: in $\mathsection$\ref{sec:obs} we describe  the \CII \ and CO observations, and the results for these observations are given in $\mathsection$\ref{sec:results}. In $\mathsection$\ref{sec:analysis} we present our analysis, and discuss the implications in $\mathsection$\ref{sec:discussion}. Our conclusions and summary are given in $\mathsection$\ref{sec:conclusion}. Throughout this paper we adopt the cosmology: $H_0=71$\,km s$^{-1}$  Mpc$^{-1}$, $\Omega_{\Lambda}=0.73$ and $\Omega_m=0.27$ \citep{komatsu11}.

\section{Observations}\label{sec:obs}
\subsection{Supporting ALMA and ATCA observations}\label{sec:supporting_obs}
The 20 DSFGs presented here are a subset of  100 strongly lensed DSFGs selected over the 2500 deg$^2$ SPT-SZ survey. See Table 1 in \citet{weiss13} and Table~\ref{table:newSPTz} in Appendix~\ref{app:new_sources} for the full names and positions. 
Mid- and high-$J$ CO rotational lines (i.e. CO(3--2), CO(4--3), CO(5--4) and/or CO(6--5)) were detected for 23 \footnote{One source SPT0538-50 \citep{greve12,bothwell13}, was not observed with ALMA.} DSFGs with the Atacama Large Millimeter/submillimeter Array (ALMA). The redshifts of all our \CII\ targets are determined by one or more CO lines plus the \CII\ line itself and are therefore robust.

In addition, low-$J$ CO emission lines (CO(1--0) or CO(2--1)) were observed with the Australia Telescope Compact Array (ATCA) for 
11 of the SPT DSFGs for which \CII\ observations are presented in this paper \citealt{aravena13}; Aravena \textit{et al. in prep}). 
Absolute flux calibration of the ATCA data is estimated to be accurate to within 15\%. Details of the observations, fluxes and associated uncertainties will be presented in a forthcoming paper by Aravena \textit{et al.}

\subsection{APEX/FLASH}\label{sec:flash}
We targeted all galaxies in the SPT DSFG sample with known redshifts that place the \CII\ line at frequencies which are observable with good atmospheric transparency using the First Light APEX Submillimetre Heterodyne receiver (FLASH, \citealt{heyminck06}). 
Eleven sources at $4.2<z<5.7$ were observed in the 345\,GHz channel between 2012 August and 2014 June, and six sources at $3.1<z<3.8$ were observed with the 460\,GHz channel between 2013 March and August during Max Planck time. 
All observations were done in good weather conditions with an
average precipitable water vapour $<1.0$\,mm, yielding typical
system temperatures of 230 and 170\,K for the 345 and 460\,GHz
observations, respectively. 
The beam sizes/antenna gains are $22.0''/40$\,Jy\,K$^{-1}$ and $13.5''/48$\,Jy\,K$^{-1}$ for the lowest and highest observed frequencies of the \CII \ line, respectively. The beam size is much larger than the observed Einstein radii of these sources and thus they are unresolved \citep{vieira13, hezaveh13}.
The 82 hours of observations were done in wobbler switching mode, with switching frequency of 1.5\,Hz and a wobbler throw of $50''$ in
azimuth. Pointing was checked frequently and was found to be stable to within $2.5''$. Calibration was done every $\sim10$\,min using the standard
hot/cold-load absorber measurements.  The data were recorded with the
MPIfR Fast Fourier Transform spectrometers (FFTS; \citealt{klein06}) providing $4\times2.5\,$GHz of bandwidth 
to cover the full 4\,GHz bandwidth in each of the upper and lower sidebands of the sideband-separating FLASH receiver. 

The data were processed with the Continuum and Line Analysis Single-dish Software (CLASS\footnote{\tt http://www.iram.fr/IRAMFR/GILDAS/}). 
We visually inspected the individual scans and omitted scans with
unstable baselines, resulting in $<10\%$ data loss.
We subtracted linear baselines from the individual spectra in each of the two FFTS units, 
and regridded to a velocity resolution of $\sim90$\,km\,s$^{-1}$ in the averaged spectra. 
On-source integration times were between 1.5 and 5 hours. Table~\ref{table:CII_data} summarises the line intensities, and Figure~\ref{fig:compare_spec} shows the spectra.

We detect \CII\ emission in 16 out of 17 sources observed with FLASH. The only non-detection is in the highest redshift source, SPT0243-49 at $z=5.699$. This source has an unambiguous redshift confirmed by two high-$J$ CO lines with ALMA \citep{weiss13} and a CO(2--1) line with ATCA (Aravena \textit{et al. in prep.}). The observed \CIIFIR \ ratio ($<1.3\times10^{-3}$) is close to the median of our sample (see Table~\ref{table:CII_data}), suggesting that a \CII\ detection for this source is feasible with a moderately deeper integration.

\subsection{\textit{Herschel}/SPIRE}\label{sec:SPIRE}
For three $z\sim2$ sources in the SPT sample (see Table~\ref{table:CII_data}), the \CII\ line falls in a frequency range (500-610 GHz) where the atmosphere is opaque. We thus observed these sources with the SPIRE Fourier Transform Spectrometer (FTS; \citealt{griffin10}) onboard \textit{Herschel}\footnote{\textit{Herschel} is an ESA space observatory with science instruments provided by European-led Principal Investigator consortia and with important participation from NASA.}. \citep{pilbratt10}. For more detail about the observations and results for SPT0538-50, see \cite{bothwell13b}. 
The observations of SPT0551-50 and SPT0512-59 were carried out on 2013 March 2 in single-pointing mode using both the short wavelength (SSW) and long wavelength (SLW) bands covering 194-313\,$\mu$m and 303-671\,$\mu$m. The observations were done in high spectral resolution mode equivalent to 0.04\,cm$^{-1}$ (1.2\,GHz) with 100 repetitions, resulting in an on-source integration time of 13752\,s (3.8\,h) per source. 

The data were reduced and calibrated using an updated SPIRE FTS pipeline in the \textit{Herschel} reduction tool HIPE v11, which includes all detectors in the observation and uses new calibration files. 
The detection of lines fainter than 1\,Jy is very challenging because thermal emission from the warm optics contributes as 
much as 1000\,Jy at 1000\,GHz. 
We subtract the average of the off-target pixels in order to remove this excess emission from the telescope. Another possibility is to subtract a `dark sky' observation
made on the same day with the same exposure time, but in our case the noise level was lowest by subtracting the average of the off-target pixels. 
As the model of the telescope has an uncertainty of 0.1\%, there remains a residual continuum uncertainty of $\sim1\,$Jy in the continuum. 
We therefore subtract an additional second order polynomial from the SSW and the SLW spectral part separately, and look for the \CII \ line at the expected frequency \citep{valtchanov11}. Figure~\ref{fig:herschelCII} shows the resulting spectra.  


\section{Results}\label{sec:results}
Figures~\ref{fig:compare_spec} and 2 show the \CII \ emission lines and Table~\ref{table:CII_data} lists the \CII \ luminosities obtained from APEX/FLASH and \textit{Herschel} SPIRE FTS observations, along with far-IR luminosities ($L_{\text{FIR}}$).
We have 17 \CII \ emission line detections (16 with FLASH and one with \textit{Herschel} SPIRE FTS) and three non-detections (one with FLASH and two with \textit{Herschel} SPIRE FTS). 

\subsection{Velocity profiles and line fluxes}

Despite an increasing number of high-$z$ \CII \ detections in the literature, only a few of these have sufficient spectral resolution and S/N and supporting data to compare line profiles with other bright lines like CO (e.g. \citealt{rawle14}). 
This is the first sample of sources with spectrally resolved data with S/N$\geq3$ in \textit{both} \CII \ and CO making it possible to compare the shapes of velocity profiles. In the following, we consider only the 17 sources observed with FLASH, as the 3 sources observed with SPIRE are not spectrally resolved.

\begin{figure*}
\includegraphics[scale=0.915]{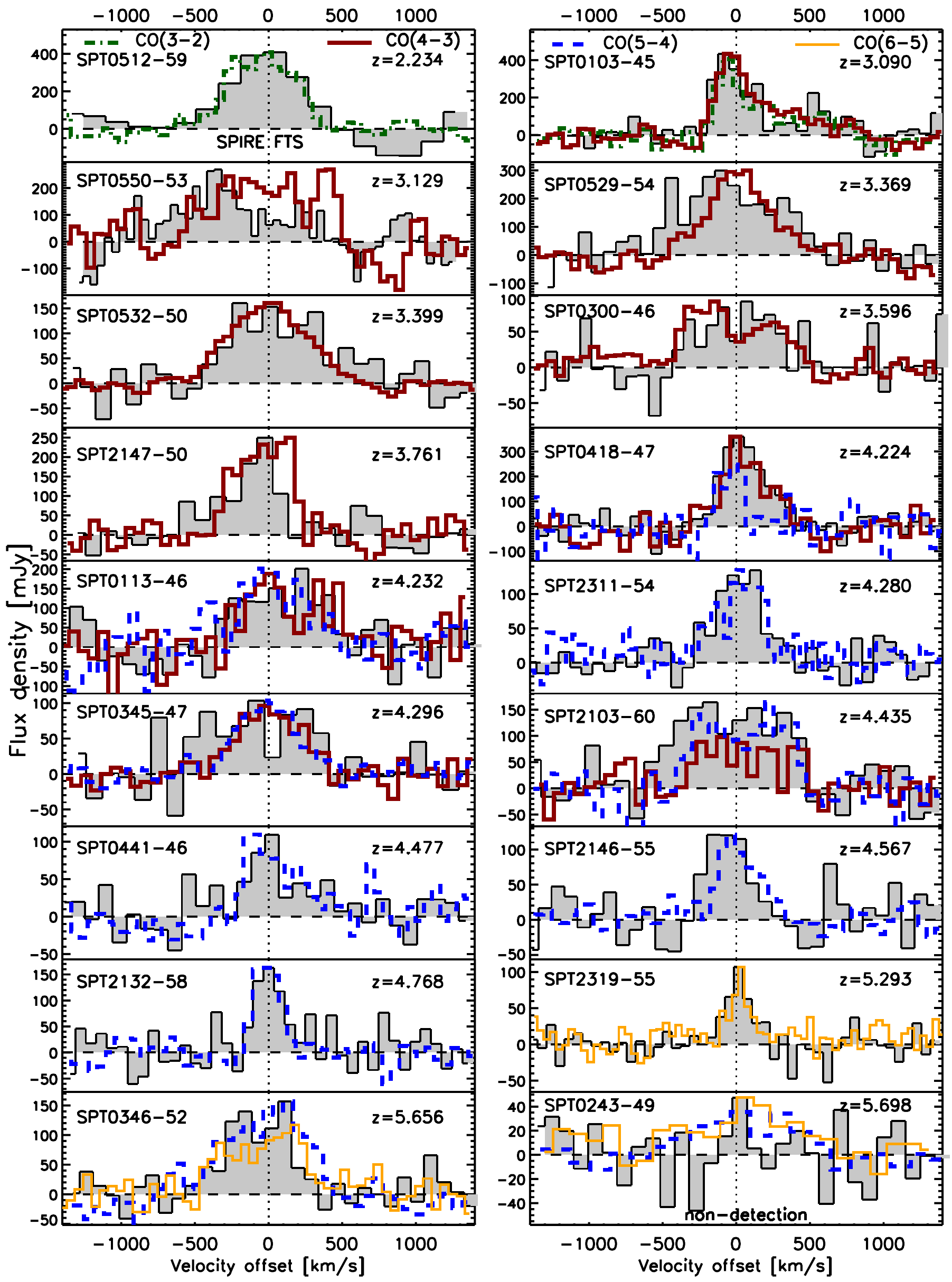}
\caption{Comparison of the velocity profiles of the \CII \ lines detected with APEX/FLASH and SPIRE FTS (grey filled profiles) and  mid-$J$ CO line observed with ALMA in Cycle 0 \citep[coloured lines;][]{weiss13}.The CO lines have been scaled to match the \CII peak flux. The 
similarities between the CO and \CII lines for individual sources suggest that the spatial distributions are similar and differential lensing is not significant.}
\label{fig:compare_spec}
\end{figure*}

Figure~\ref{fig:compare_spec} shows the velocity profiles of the \CII \ lines compared with the mid- and high-$J$ CO lines observed with ALMA. The CO lines have been scaled to match the peak flux of the \CII \  line, in order to facilitate the comparison of the velocity profiles. We first fit the CO and the \CII \ lines independently with single Gaussian functions. We accept the single Gaussian fit if the reduced $\chi^2$ does not exceed 1+5$\times \sqrt{2 n_\textrm{dof}}$, where $n_\textrm{dof}$ is the number of degrees of freedom. This quantity corresponds to 5 standard deviations of the $\chi^2$ distribution. If the single Gaussian fit does not match the above criterion, we use a double Gaussian function, i.e. two Gaussian functions with displaced central positions. The double Gaussian function is sufficient to describe the line profiles that do not match a single Gaussian. In practice, this happens in SPT0103-45 and SPT0418-47, which have lines that display a slight asymmetry on the red side of the lines; the velocity difference between the CO and \CII \ peaks is $<$150\,km\,s$^{-1}$. 

We then simultaneously fit the single or double Gaussian profile (depending on what is necessary to fit the profiles individually) to the CO and \CII \ velocity profiles assuming the profiles are similar with just one free scaling parameter. This allows us to test if the two profiles are consistent or not. Only the \CII \ and CO lines for SPT0532-50 have different velocity profiles. 
The remaining 16 sources have consistent line profiles with $\chi^2 < 1+5\times \sqrt{2 n_\textrm{dof}}$. 

We obtain the line widths (FWHM) listed in Table~\ref{table:CII_data} by fitting a single Gaussian. To test the reliability of our method we also fit the spectra by taking velocity-weighted moments, and find fully consistent results. 
The resulting FWHMs are in the range $\sim210-820\,$km/s. This is comparable to the typical CO line width of $\sim460$\,km\,s$^{-1}$ found for SPT DSFGs by stacking 22 spectra \citep{spilker14}. 
Nine out of 20 sources have FWHM $> 500$\,km/s, which is large compared to other high-$z$ \CII \ detections (e.g. 360\,km/s on average in the sample of \citealt{wang13}). 

We obtain the velocity-integrated fluxes for the FLASH/APEX detections by summing the observed line profiles over the $3\sigma$ limits obtained from the single Gaussian fits. 
The \CII \ apparent luminosities range from 1.4 to $9.2\times10^{10}\,$L$_{\odot}$ which is 1-2 orders of magnitude higher than their mid- and high-$J$ CO luminosities ($\sim3-30\times10^8\,$L$_{\odot}$), where both are uncorrected for lensing.

\begin{figure*}
\includegraphics[scale=0.385]{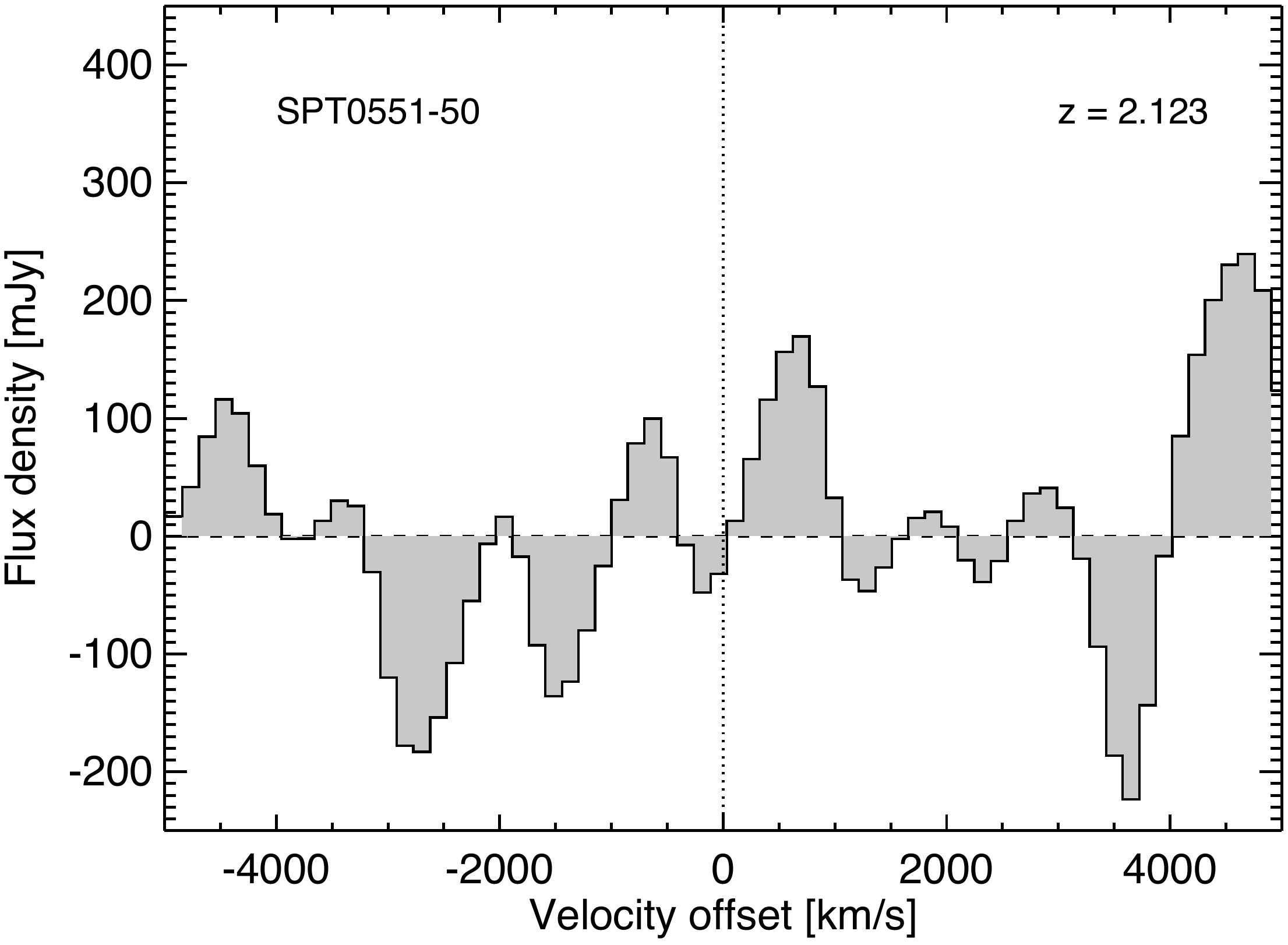}
\includegraphics[scale=0.385]{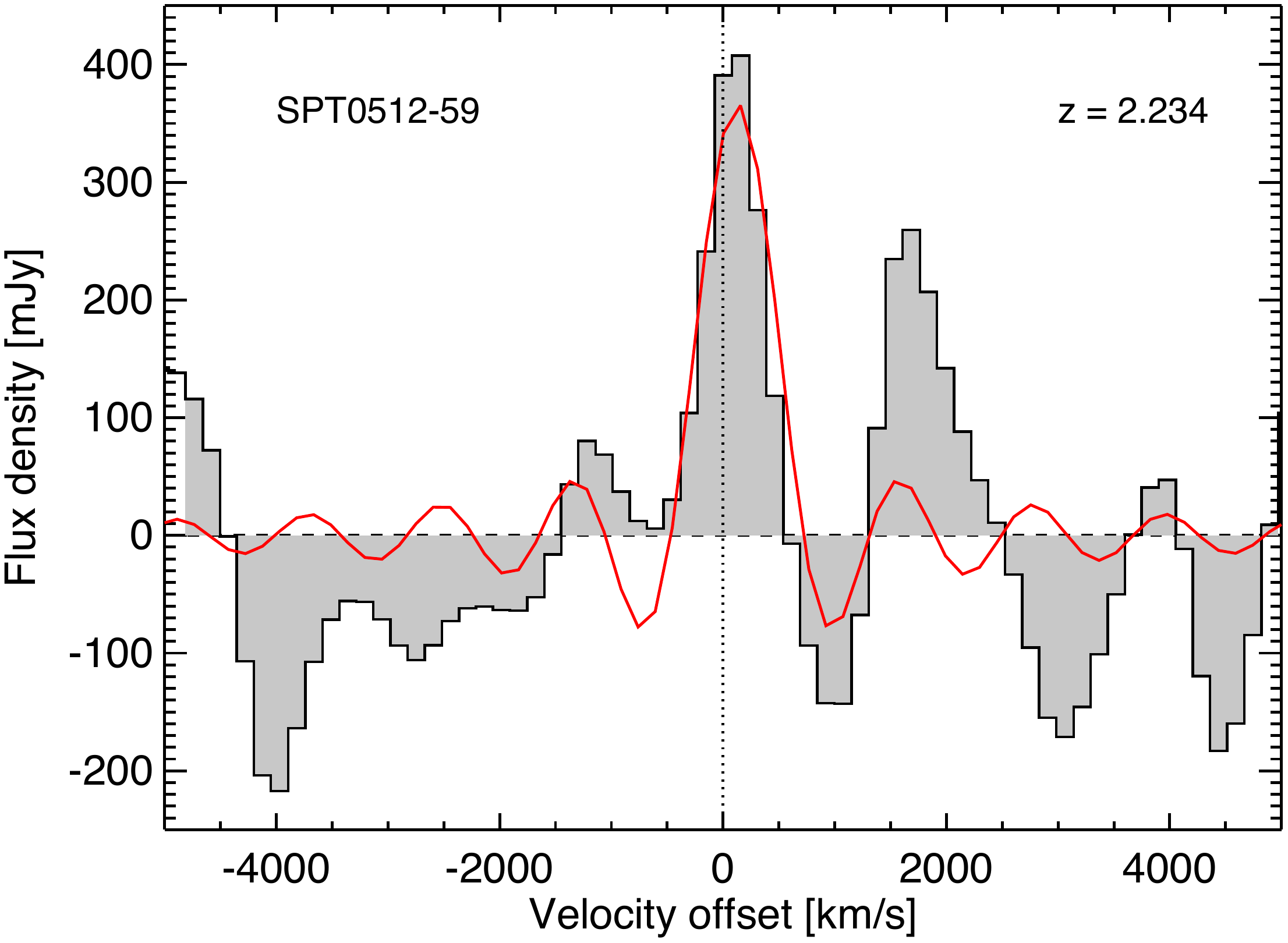}
\caption {\textit{Herschel} SPIRE FTS spectra for SPT0551-50 and SPT0512-59. \textit{Left:} Non-detection of the \CII \ emission line for SPT0551-50. \textit{Right:} Detection of \CII \ emission line for SPT0512-59. The continuous red curve is the sinc-function used to fit the \CII \ line (see \citealt{valtchanov11}). \label{fig:herschelCII}}
\end{figure*}

The \CII \ detection in SPT0512-59 with SPIRE FTS (see Figure~\ref{fig:herschelCII} \textit{right}) confirms the redshift at $z=2.234$.
We determine the integrated line flux in the same manner as \cite{valtchanov11} by fitting the emission lines with a sinc-function, and calculate the RMS using the standard deviation of each channel within $\pm 5000$\,km\,s$^{-1}$ of the expected line centroid (see \citealt{valtchanov11} for more details).

Our SPIRE FTS observation for SPT0551-50\footnote{Note that is is one of the rare cluster lenses in the SPT sample \citep{vieira13}. This, however, should not have any effect on the detectability of \CII\ in this source.} does not detect any \CII \ emission (see Figure~\ref{fig:herschelCII} \textit{left}). 
The redshift was reconfirmed by weak CO(1--0) emission observed with ATCA (Aravena \textit{et al. in prep.}).
We therefore take the 3$\sigma$ upper limit of the expected line peak to be 3 times the RMS noise.

\begin{table*}
\begin{tabular}{@{}llcccccccccl}
 \hline
 Source$\dagger$ & $z$ &  $SdV_{\text{\CII}}$ & $dV$ &  $L'_{\text{\CII}}/10^{10}$ &  $L_{\text{\CII}}$ & $L_{\text{FIR}}$ & \Td & $\frac{L_{\text{[CII]}}}{L_{\text{FIR}}}$ & Instrument & Time \\
& & & (FWHM) & & & & & &\\
& &  [Jy\,km/s] &  [km/s] &  [K\,km/s/pc$^2$] & $[10^{10}$\,L$_{\odot}]$ & $[10^{13}$\,L$_{\odot}]$ & [K] &  $[10^{-4}]$ & & [h] \\
 \hline
SPT0551-50$^1$ & 2.123 & $ <180 \ (3\sigma) $ &         --- & $ <13.4                $ & $ <3.0               $ & $1.1\pm 0.1$ & $27.2\pm1.0$ &$ < 26.6              $ & SPIRE FTS &3.8\\  
SPT0512-59     & 2.234 & $ 227 \pm     43$    &         --- & $     18.4\pm      3.5$ & $      4.0 \pm      0.8$ & $2.8\pm0.2$ & $33.2\pm1.2$& $      14.4\pm       2.9$ & SPIRE FTS &3.8\\
SPT0538-50$^1$ & 2.782 & $ <465 \ (3\sigma)$  &          --- & $ < 81.9         $ &     $ < 18.0         $ & $ 5.8\pm0.3$ & $36.9\pm1.4$ & $ < 31.0           $ & SPIRE FTS$^{\bigstar}$ &3.8\\ 
SPT0103-45     & 3.090 & $ 125 \pm     17$    & $    304\pm     47$ & $     17.5\pm      2.3$ & $      3.8 \pm      0.5$ & $3.4\pm 0.2$ & $33.5\pm1.1$ &$      11.3\pm       1.6$ & FLASH & 2.7\\
SPT0550-53     & 3.129 & $ 129 \pm     25$    & $    719\pm    124$ & $     18.5\pm      3.6$ & $      4.1 \pm      0.8$ &   $1.6\pm0.1$ &  $34.5\pm1.8$ &          $25.4\pm5.2$ &  FLASH & 17.3\\
SPT0529-54     & 3.369 & $ 217 \pm     18$    & $    823\pm     92$ & $     35.1\pm      3.0$ & $      7.7 \pm      0.7$ & $3.0\pm 0.2$ & $33.1\pm1.2$ &$      25.7\pm       2.8$ & FLASH &3.8\\
SPT0532-50     & 3.399 & $ 113 \pm     18$    & $    767\pm    124$ & $     18.6\pm      3.0$ & $      4.1 \pm      0.7$ & $6.5\pm 0.4$ & $37.9\pm1.4$ & $       6.3\pm       1.1$ & FLASH &4.6\\
SPT0300-46     & 3.596 & $  41.5 \pm   10.4$  & $    583\pm    138$ & $      7.5\pm      1.9$ & $       1.6 \pm      0.4$ &  $3.3\pm0.2$ & $39.2\pm1.5$ &          $5.0\pm1.3$ & FLASH &12.8\\
SPT2147-50$^2$ & 3.761 & $  80.5 \pm   11.7 $ & $    329\pm     56$ & $     15.5\pm      2.3$ & $      3.4 \pm      0.5$ &$3.2\pm 0.2$& $41.4\pm1.7$ &  $      10.7\pm       1.7$ & FLASH & 5.0\\ 
SPT0418-47$^2$ & 4.224 & $ 127 \pm   10$      & $    347\pm     29$ & $     29.5\pm      2.3$ & $      6.5 \pm      0.5$ &$5.9\pm 0.4$& $47.3\pm2.5$ & $      11.0\pm       1.1$ & FLASH &1.5\\ 
SPT0113-46$^2$ & 4.232 & $  91 \pm   19$      & $    619\pm    132$ & $     21.1\pm      4.4$ & $      4.6 \pm      1.0$ &$2.1\pm 0.1$& $32.9\pm1.4$ & $      22.1\pm       4.8$ &FLASH & 4.5\\ 
SPT2311-54     & 4.281 & $  46.4 \pm    5.3$  & $    360\pm     44$ & $     11.0\pm      1.2$ & $      2.4 \pm      0.3$&  $3.3\pm0.3$ & $43.3\pm3.3$ & $        7.3\pm1.1       $ & FLASH &3.0\\
SPT0345-47$^2$ & 4.296 & $  63.7 \pm    8.3$  & $    810\pm    200$ & $     15.2\pm      2.0$ & $      3.3 \pm      0.4$ & $9.2\pm 0.8$ & $51.8\pm3.2$ & $3.6\pm0.6$ &FLASH & 2.3\\ 
SPT2103-60$^2$ & 4.435 & $ 129 \pm     18$    & $    780\pm    125$ & $     32.2\pm      4.4$ & $      7.1 \pm      1.0$ & $3.4\pm 0.2$& $39.2\pm1.5$ & $      20.8\pm       3.1$ & FLASH &1.5\\ 
SPT0441-46$^2$ & 4.477 & $  42.5 \pm   10.6$  & $    581\pm    162$ & $     10.8\pm      2.7$ & $      2.4 \pm      0.6$ &$3.7\pm 0.2$& $39.9\pm1.9$ & $       6.6\pm       1.7$ & FLASH & 3.2\\ 
SPT2146-55$^2$ & 4.567 & $  39.0 \pm    9.0$  & $    302\pm     62$ & $     10.2\pm      2.4$ & $      2.2 \pm      0.5$ &$2.7\pm 0.3$& $39.2\pm2.0$ & $       8.3\pm       2.1$ & FLASH &3.1\\ 
SPT2132-58     & 4.768 & $  34.9 \pm    6.9$  & $    212\pm     43$ & $      9.7\pm      1.9$ & $      2.1 \pm      0.4$ &$3.1\pm0.3$& $39.5\pm1.9$ & $       6.9\pm   1.5$ & FLASH &2.1\\
SPT2319-55     & 5.293 & $  19.1 \pm    3.2$  & $    198\pm     34$ & $      6.3\pm      1.1$ & $      1.4 \pm      0.2$ &   $2.5\pm0.2$ & $42.0\pm3.1$ & $  5.4\pm    1.1$ & FLASH & 9.5\\
SPT0346-52$^2$ & 5.656 & $  63.3 \pm    8.7$  & $    502\pm     72$ & $     22.8\pm      3.2$ & $      5.0 \pm      0.7$ & $12.3\pm 0.5$& $52.4\pm2.2$ & $       4.1\pm       0.6$ &FLASH &1.4\\ 
SPT0243-49$^2$ & 5.699 & $ < 51 \ (3\sigma) $ &         --- & $ < 21.0                $ & $ < 4.5               $ & $ 3.3\pm 0.3$ & $35.3\pm1.6$ & $ < 13.6             $ & FLASH & 2.2\\ 
\hline
\end{tabular}
\caption{Observed \CII \ and FIR properties. All luminosities are uncorrected for the lensing amplification. The upper limits for the velocity integrated fluxes given for the non-detections are obtained by assuming the FWHM observed for the CO lines. The total integration time with SPIRE FTS and FLASH is 92 hours. The integration time per sources is given in the last column.\newline
$\dagger$ Full source names are listed in Table~1 of \citet{weiss13} or in Table~\ref{table:newSPTz}.\newline
$^1$ Has CO(1-0) observations (Aravena \textit{et al., in prep.}).\newline
$^2$ Has CO(2-1) observations (Aravena \textit{et al., in prep.}).\newline
$\bigstar$ See \citet{bothwell13b} for more details.}
\label{table:CII_data}
\end{table*}

\subsection{Lensing}\label{sec:lensing}
The lensing magnification of the SPT DSFGs allows us to study the ISM in galaxies at high redshifts, but also introduces the possibility of differential lensing. 
The compactness and location of a region relative to the lensing caustic determines the magnification of the emission. 
Differential lensing amplification may thus occur between compact and extended emitting regions, or components occupying different regions \citep[e.g.][]{hezaveh12a}. 
From observations of the Milky Way and nearby galaxies, we might expect the \CII \ emission to originate from more extended and diffuse media than the more optically thick low-$J$ CO emission in giant molecular clouds (GMCs) \citep{fixsen99}. If the \CII \ was dominated by emission from such diffuse regions, it could be subject to differential lensing compared to the more compact CO emission. However, the similar \CII\ and CO velocity profiles suggest that such differential lensing is not significant. 

Modelling the lensing magnification factor is of great importance, and has been performed for some of the DSFGs in this sample using ALMA data. 
The lens modelling is performed in the $(u,v)-$plane to properly represent the parameter uncertainties in the interferometric data (see \citealt{hezaveh13} for more details). The four sources with lens models have lensing magnification factors $\mu$ between $\sim5.4-21.0$ \citep{hezaveh13} and a mean of $\langle \mu \rangle=14.1$. In cases where the lensing magnification factor is unknown, we use the mean magnification factor to estimate the intrinsic luminosity. 
We then conservatively choose to span the uncertainty on the lensing magnification factor from the smallest ($\mu=5.4$) to the largest ($\mu=21.0$) giving the mean magnification an uncertainty of 7.8.
We assume that the magnification factors derived from the dust continuum are also appropriate for the line emitting gas. A lens modelling analysis of the remainder of the ALMA Cycle 0  imaging data is currently under way.


\section{Analysis}\label{sec:analysis}

\subsection{Comparison sample of nearby and distant galaxies}\label{sec:compare_sample}
We compare the SPT sample with that of Gracia-Carpio et al. (in prep), which is comprised of 333 sources. These sources include LIRGs and ULIRGs from the Great Observatories All-sky Survey (GOALS, \citealt{diazsantos13}) and normal and Seyfert galaxies from \cite{brauher08} with spatially unresolved \CII \ detections integrated over the entire individual galaxies. The sample contains 308 sources at $z<0.4$ and 25 at $z>1$. In addition, we have searched the literature for additional $z>1$ DSFGs with \CII\ observations, which we list in Appendix~\ref{app:high_z_sample}.
In constructing this low and high-$z$ comparison sample, we have paid particular attention to ensure that the photometric data are integrated over the full galaxies, as all line and continuum data for the SPT sample are also integrated values.

\subsubsection{Conversion to CO(1--0) luminosities}\label{subsec:CO10_conversion}
Like the SPT sample (\S~\ref{sec:supporting_obs}), many of the sources in the comparison sample also have published CO observations. 
In cases where the CO(1--0) emission lines have not been observed, Gracia-Carpio et al. (in prep) derive the $L_{\text{CO(1--0)}}$ by converting the observed mid-$J$ CO luminosities to $L_{\text{CO(1--0)}}$ using scaling factors from \cite{stacey10}. When multiple $J>1$ CO lines were observed, we take the average of the scaled $L_{\text{CO(1--0)}}$. 
These scaling factors are based on previous studies involving several rotational lines from both nearby galaxies, ULIRGs and high-$z$ galaxies \citep{stacey10}, which allow for the assumption of fixed CO line ratios up to CO(4--3)/CO(1--0). 
The conversion factors used by \cite{stacey10} assume an integrated line flux (W\,m$^{-2}$) ratio of CO(2--1)/CO(1--0) = 7.2, equivalent to 90\% of the thermalised optically thick emission (i.e. a brightness temperature ratio of 0.9),  CO(3--2)/CO(2--1) = 3.0 (90\% of the thermalised optically thick emission) and CO(4--3)/CO(2--1) = 6.4 (80\% of the thermalised optically thick emission). 

The assumed CO(2--1) to CO(1--0) brightness temperature ratio of 0.9 for DSFGs is in agreement with observations of \cite{bothwell13}, who find a ratio of 0.85 for 32 luminous submm galaxies at $z\sim1.2-4.1$. 
The [CO(2--1)/CO(1--0)] ratio has also been observed for four normal star forming galaxies by \cite{aravena14}, resulting in a slightly lower [CO(2--1)/CO(1--0)] brightness temperature ratio of $0.7\pm0.16$ for four BzK galaxies at $z\sim1.5-2.2$. 
Even though the ratio is consistent with \cite{bothwell13} within the uncertainties, a lower [CO(2--1)/CO(1--0)] ratio is expected for normal star forming galaxies than for starburst galaxies, because the molecular gas in starburst galaxies is expected to be more highly excited. 
\cite{spilker14} find an average CO(2--1)/CO(1--0) brightness temperature ratio of $1.1\pm0.1$ for 22 of the SPT DSFGs by stacking ATCA spectra, after scaling them by their 1.4\,mm continuum flux density. 
Given the range of values of these three methods, and for consistency with previous literature, we adopt the scaling factor of 0.9 from \cite{stacey10}.

\subsubsection{FIR luminosities}\label{subsec:FIR_luminosities}
\begin{figure}
\includegraphics[trim=0cm 0cm 0cm 0cm, clip=true, scale=0.66]{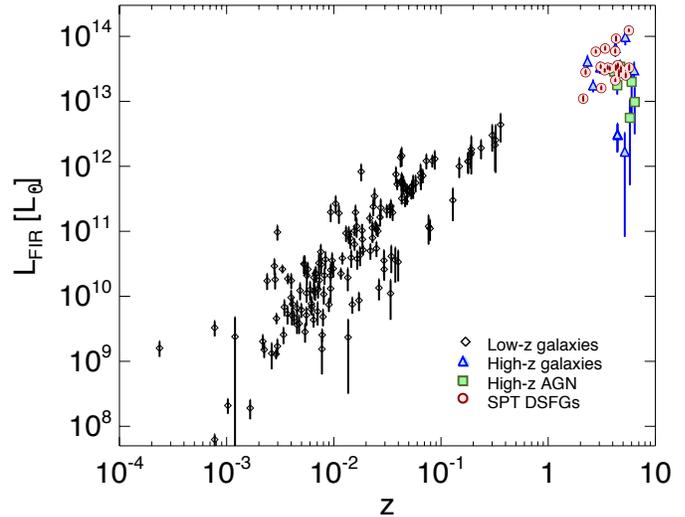}
 \caption{Observed \LFIR vs redshift for the 20 SPT sources and the comparison sample. No \LFIR \ have been corrected for lensing magnification factors. The distribution shows the Malmquist bias where high-$z$ galaxies require either lensing magnification or very high intrinsic FIR luminosity of \LFIR \ $\gtrsim 10^{12}$\,L$_{\odot}$ to be included in the parent sample. The evolution of the luminosity function and the smaller comoving volume at low redshifts imply that objects with similar high intrinsic \LFIR \ are missing from the low-$z$ sample. However, the most highly lensed DSFGs may have similar intrinsic \LFIR than the most luminous sources in the local sample.}
  \label{fig:DL_JGC}
\end{figure}

We obtain the FIR luminosities \LFIR\ and the dust temperature \Td\ for each source in the SPT DSFG sample by fitting the well-sampled spectral energy distributions (SEDs) with a greybody law fixing the emissivity index ($\beta$) at 2.0, $\mu_0=100$ and fitting $\lambda_{\text{rest}}<50\,\mu$m, following \cite{greve12} and Strandet et al. (in prep). We integrate the SED between $\lambda_{\text{rest}}=42-500\,\mu m$ in the rest frame to obtain \LFIR.
Our SPT DSFGs all have 7 photometric data points covering observed wavelengths from 250 to 3000\,$\mu$m \citep[e.g.][]{weiss13}. This allows for a uniform determination of the FIR luminosity using a parametrised SED fitting\footnote{Other studies \citep[e.g.][]{stacey10,helou88} use the two band definition for \LFIR: $F_{\text{FIR}}=1.26\times10^{-14} \times (2.58 f_{60} +f_{100})[\text{W\,m$^{-2}$}]$ which is equivalent to the 42.5-122.5\,$\mu$m luminosity \citep{helou88}.}.  The SPT sources, which have well-sampled SEDs, thus have a smaller uncertainty in \LFIR\ than sources with poor photometric coverage (see Figure~\ref{fig:DL_JGC}).

To compare the low-$z$ and high-$z$ sample with the SPT sources in a consistent way, we compiled published FIR photometry (Gracia-Carpio et al. in prep; Appendix~\ref{app:high_z_sample}), and derived \LFIR\ and \Td\ using the same fitting code we used for the SPT sources \citep{greve12}.  
IRAS and ISO data are available for sources published by \cite{brauher08}, 
and FIR data are also available for a large number of the GOALS sources at NASA/IPAC Extragalactic Database (NED\footnote{http://ned.ipac.caltech.edu}). The comparison sample contains 
165 sources with sufficiently good photometry to derive \LFIR\ and \Td\ using the method in \citet{greve12}; 14 of these are $z>1$ sources  (see Figure~\ref{fig:DL_JGC}). Sources with insufficient available photometry to derive \LFIR\ using our procedure are not included in analyses requiring \LFIR\, but are still included in the \CIICO\ ratios.

Figure~\ref{fig:DL_JGC} presents the observed \LFIR \ distribution of our combined sample. There is a clear gap between the low and high-$z$ samples due to the limited sensitivity of the nearby samples, and the atmospheric transparency prohibiting \CII \ observations at $z<1$ from the ground. The selection function of the high-$z$ sample is quite complex because many sources have been pre-selected to have a good chance of detection in \CII\ (e.g. by having strong dust continuum and/or CO emission). In addition, the improved atmospheric transparency at lower frequencies mostly compensates for the distance dimming in the more distant sources. 
The evolution of the luminosity function and the smaller co-moving volume at low redshifts imply that objects with similar high intrinsic \LFIR \ are missing from the low-$z$ sample.

The SPT sources in this sample have FIR luminosities in the range \LFIR$=(1.2-11.8)\times10^{13}$L$_{\odot}$. After de-magnification the range is \LFIR$/\mu = (1.1-21.9)\times10^{12}$\,L$_{\odot}$, with a mean de-magnified \LFIR \ of $5.2\times10^{12}$\,L$_{\odot}$, which is similar to the most luminous sources in the low-$z$ sample. 

The uniform sensitivity of the ALMA mm spectroscopy, continuum, low-$J$ CO and \CII \ observations (all with S/N$>$3) lead to a high completeness of the SPT sample. This, combined with the lensing amplification factor which is about one order of magnitude \citep{hezaveh13}, make the SPT DSFG sample one of the most representative samples to date for massive, IR-luminous starburst galaxies at high redshifts.

\subsubsection{AGN content}\label{subsec:AGN_content}
The comparison sample contains both starburst galaxies and luminous AGN. In the high-redshift sample, we distinguish the AGN-dominated sources from those without any known AGN. These high-redshift AGN are quite rare sources, mostly selected based on their bright optical emission lines (e.g. \citealt{wang13}).
The SPT DSFGs are selected solely on their lensed 1.4\,mm continuum flux, and direct mm imaging and spectroscopy avoids any radio/optical identification steps that may introduce biases towards AGN-dominated systems. Optical spectroscopy of the SPT DSFG sample to derive the redshifts of the foreground lensing galaxies has not shown any indications of type-1 or type-2 AGN. Strongly obscured type-2 AGN may still be present in some SPT DSFGs. Supported by the discussion in \S\ref{sec:obs_ratio}, we will assume in the following that the AGN contributions in the SPT DSFGs are negligible.

\subsection{Observed \CII \ to FIR  ratios}\label{sec:deficit}
\begin{figure}
\begin{center}
\includegraphics[trim=0cm 0cm 0cm 0cm, clip=true, scale=0.66]{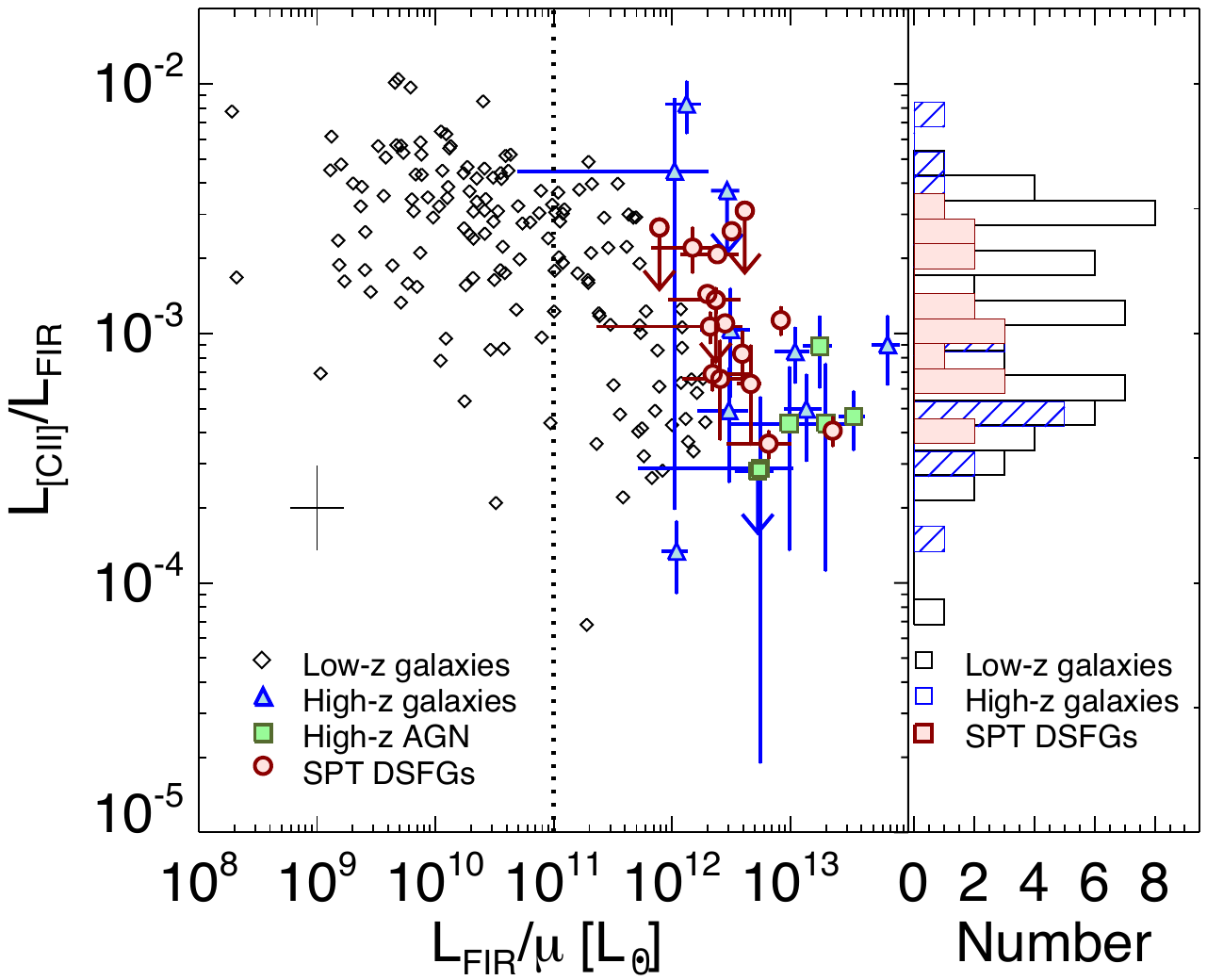}
 \caption{\CIIFIR \ vs \LFIR \ for SPT sources and the comparison sample. As reported in previous versions of this plot, the \CIIFIR \ is anti-correlated with \LFIR. In particular at $L\gtrsim10^{11}$\,L$_{\odot}$,  the \CIIFIR ratio drops and has a larger spread. For the SPT sources without known lensing models, we assume a lensing magnification factor of 14.1 (and an uncertainty which encompasses the range of 5 to 22 from the known models). The typical error bar for the literature sources is represented by the black cross in the lower left. The histogram on the right shows the distribution of galaxies with \LFIR \ $\gtrsim10^{11}$\,L$_{\odot}$. }
  \label{fig:CII_deficit}
  \end{center}
\end{figure}

Figure~\ref{fig:CII_deficit} presents the \CIIFIR\ ratio against \LFIR\ for the SPT sources and the comparison sample. 
The typical error bar for the low-$z$ sources (in this and the following plots) is illustrated by the cross in the bottom left. This typical error bar includes the quoted uncertainties of the lines \citep[e.g.][]{young95,negishi01}, the absolute and statistical uncertainty of the FIR photometry, and of the FIR luminosity determined by our own SED fitting. The SPT DSFGs have been corrected for lensing amplification either using the known lens model \citep{hezaveh13}, or assuming a mean magnification factor $\langle\mu\rangle=14.1\pm7.8$ for the sources without a lens model (see section \ref{sec:lensing}).

Sources with \LFIR \ $\lesssim10^{11}\,$L$_{\odot}$ appear to have a roughly constant \CIIFIR \ ratio ($\sim4\times10^{-3}$). At \LFIR \ $\gtrsim10^{11}\,$L$_{\odot}$, the ratio drops to $6\times10^{-4}$ for the $z<1$ sources in the comparison sample, similar to what has been reported in previous studies \citep[e.g.][]{maiolino09,stacey10}. 
Whether this is an intrinsic or observational effect due to limited sensitivity of the \CII \ and/or FIR photometry in the low-$z$ samples is not clear, and investigating this is beyond the scope of this paper.
The $z>1$ sources from the comparison sample are scattered over two orders of magnitude, which may be due to the heterogeneous mix of the parent samples. The highly complete (82\% detections) and uniformly observed SPT sources have a  smaller scatter and an average \CIIFIR \ ratio of $\sim10^{-3}$. 
A Kolmogorov-Smirnov test results in a probability of 0.7 that the \CIIFIR \ values for \LFIR \ $\gtrsim10^{11}\,$L$_{\odot}$ from the SPT and the low-$z$ samples are drawn from the same distribution.
\noindent

\begin{figure}
\includegraphics[trim=0cm 0cm 0cm 0cm, clip=true, scale=0.66]{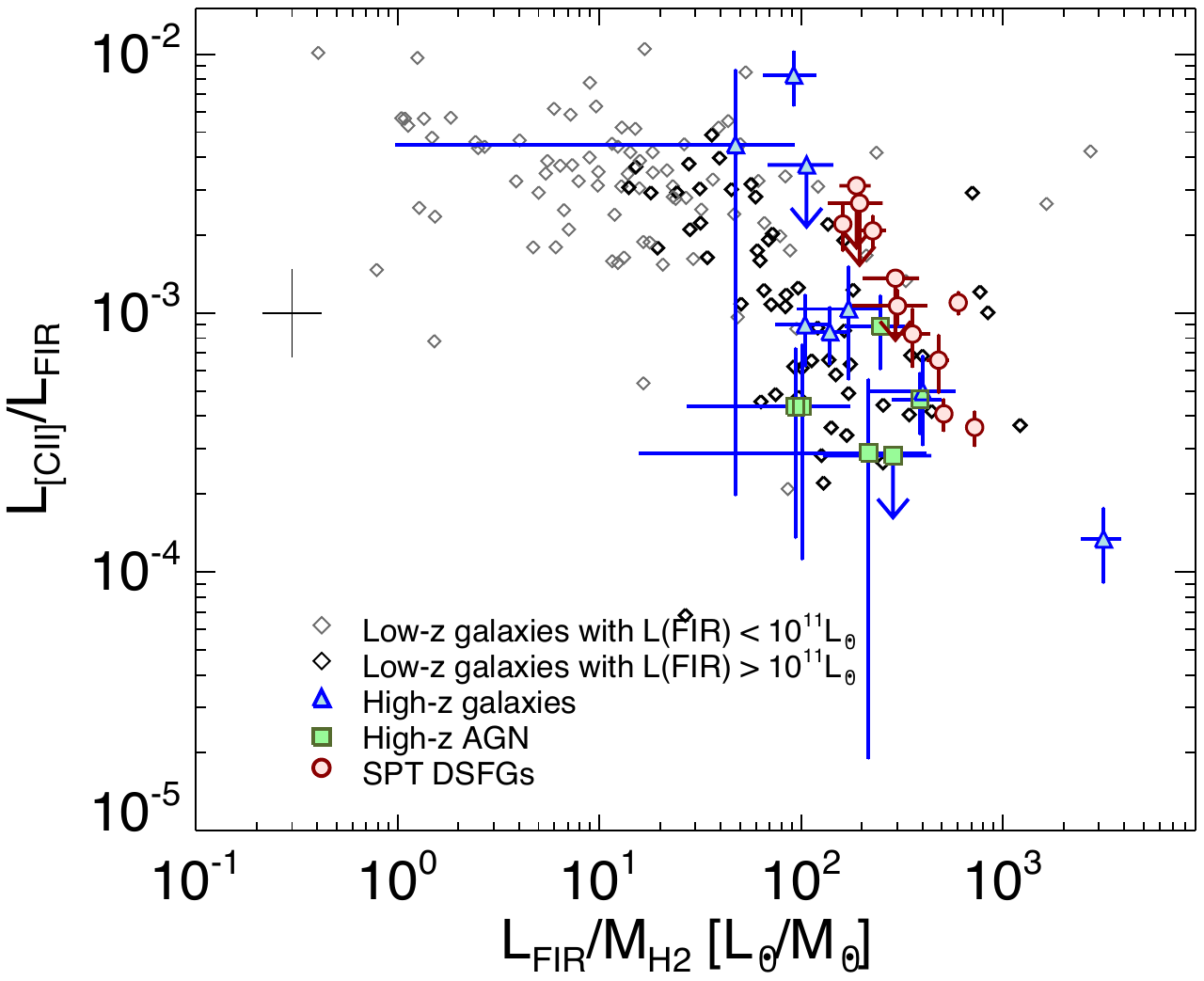}
 \caption{\CIIFIR vs the FIR luminosity normalised by the molecular gas mass. The molecular gas mass is derived assuming a conversion factor of $\alpha_{\text{CO}}=0.8$M$_{\odot}$\,(K\,km\,s$^{-1}$\,pc$^{-2}$)$^{-1}$. 
The $L_{\text{FIR}}/M_{\text{H}_2}$ ratio is expected to be proportional to the number of stars formed in a galaxy per unit molecular gas mass \citep{graciacarpio11}. This $M_{\text{H}_2}$ normalisation removes the uncertainty due to the unknown lensing magnification factors, and reduces the scatter seen in Figure~\ref{fig:CII_deficit}, but the deficit in the \CIIFIR\ ratio still persists. The typical error bar is represented by the black cross.}
  \label{fig:CII_deficitMH2}
\end{figure}

Normalising the FIR luminosity by the molecular gas mass reduces the scatter seen in Figure~\ref{fig:CII_deficit} for the \LFIR \ $\gtrsim10^{11}$\,L$_{\odot}$ sources \citep{graciacarpio11}. The $L_{\text{FIR}}/M_{\text{H}_2}$ ratio is expected to be proportional to the star formation efficiency, specifically to the number of stars formed in a galaxy per unit molecular gas mass.  This ratio has the additional advantage that the lensing magnification factor cancels out. We estimate the molecular gas masses for the 11 sources for which we also have low-$J$ CO line detections (Aravena \textit{et al, in prep.}). We determine the molecular gas mass by: 
\begin{align}\label{equ:mol_mass}
M_{\text{H}_2}=\alpha_{\text{CO}}L'_{\text{CO(1-0)}},
\end{align}
where $\alpha_{\text{CO}}$ is the CO-to-H$_2$ conversion factor. To be consistent with \cite{graciacarpio11}, we assume $\alpha_{\text{CO}}=0.8\,$M$_{\odot}\,($K\,km/s\,pc$^{2})^{-1}$ determined by \cite{downes98} for the SPT DSFGs and the low- and high-$z$ comparison samples. Figure~\ref{fig:CII_deficitMH2} plots \CIIFIR \ as a function of \LMHtwo \ and shows that the DSFGs lie among the local LIRGs. 
Low-$z$ sources with \LFIR \ $\gtrsim10^{11}\,$L$_{\odot}$ and the high-$z$ sources have similar \LCII/\LFIR \ ratios.

\subsection{Dust temperatures}\label{sec:Td}
\begin{figure}
\includegraphics[trim=0cm 0cm 0cm 0cm, clip=true, scale=0.66]{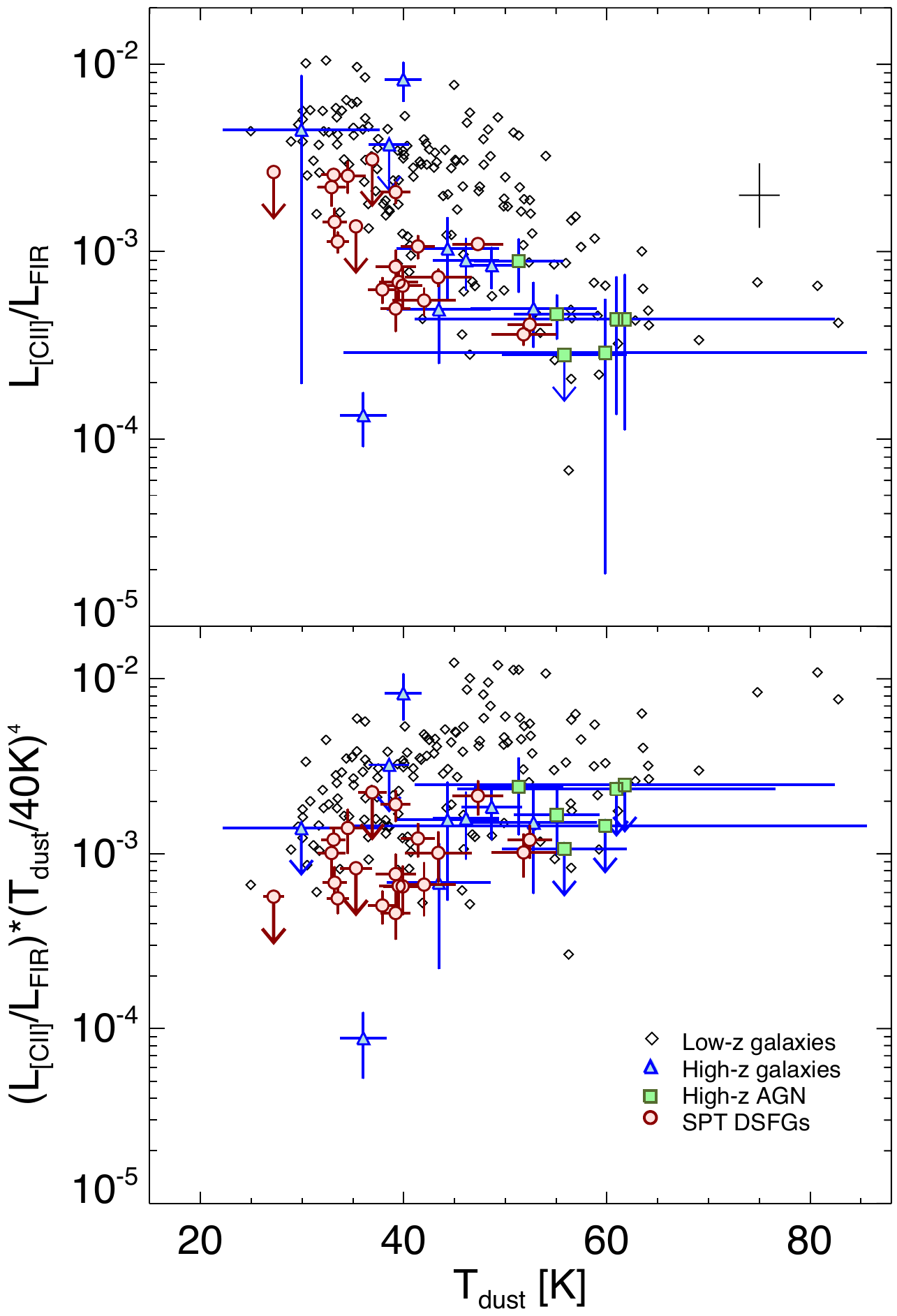}
 \caption{\textit{Top panel:} \CIIFIR \ vs \Td \ for the SPT DSFGs and the low and high-$z$ comparison sample. The anti-correlation between the \CIIFIR \ ratios and the dust temperatures is seen for both low- and high-$z$ sources, and is expected because the Stefan-Boltzmann law predicts  \LFIR~$\propto$~\Td$^4$.
 \textit{Bottom panel:} \CII$\times$\Td$^4$/\LFIR vs \Td\ for the SPT DSFGs and the low and high-$z$ comparison sample. Multiplying the \CIIFIR \ ratio with \Td$^4$ cancels out the temperature dependence of the Stefan-Boltzmann law.  All lensing magnification factors and beam filling factors cancel in both panels, and the typical error bar is represented by the black cross.}
  \label{fig:Td_vs_LCIILFIR}
\end{figure}

 As first shown by \citet{malhotra97}, the \LCII/\LFIR\ ratio shows a strong anti-correlation with \Td. They attributed this trend to an increase in the $G_0/n$ ratio (far-UV ionising field over density) in the hotter, more active galaxies, and hence a lower efficiency of gas heating reducing the \CII\ flux while increasing the dust temperature.
We here revisit this trend using the uniformly-derived set of \LFIR\ and \Td\ values for the SPT sources and comparison samples, derived from our SED modelling (\S~\ref{sec:compare_sample}).
The top panel of Figure~\ref{fig:Td_vs_LCIILFIR} compares the \LCII/\LFIR\ ratio to \Td, with both quantities being independent of the lensing magnification. 
The strong anti-correlation of these quantities is visible in all three samples. However, if the dust and \CII\ emission are coming from the same regions, and \LCII \ is less dependent on \Td, a significant part of this anti-correlation can be explained with the Stefan-Boltzmann law which predicts \LFIR~$\propto$~\Td$^4$.

\noindent

In order to look for a residual correlation, we cancel out the \LFIR~$\propto$~\Td$^4$ dependence by plotting \LCII\ $\times$ \Td$^4$/\LFIR \ against \Td\ (see bottom panel of Figure~\ref{fig:Td_vs_LCIILFIR}).
We note that a small systematic offset in \Td\ would get propagated as \Td$^4$.
To test the presence of a correlation taking into account the \CII\ upper limits, we use the generalised Kendall's tau method \citep{lavalley92}. We find a probability that both variables are not correlated of 0.036, 0.602 and 0.151 for the low-$z$, high-$z$ and SPT samples, respectively.
The small displacement of the SPT sources relative to the comparison sample in Figure~\ref{fig:Td_vs_LCIILFIR} is likely due to Malmquist bias and evolution effects resulting in the most luminous sources being absent in the low-$z$ comparison sample (see \S~\ref{subsec:FIR_luminosities}). While there is marginal evidence for a small positive residual correlation (especially in the low-$z$ sample), most of the correlation seen in the top panel has canceled out. This confirms our assumptions that the dust and \CII \ emission are originating from the same regions and that \LCII \ is less dependent on \Td. We therefore conclude that the observed correlation is mostly dominated by the Stefan-Boltzmann law.

\begin{figure}
\includegraphics[trim=0cm 0cm 0cm 0cm, clip=true, scale=0.66]{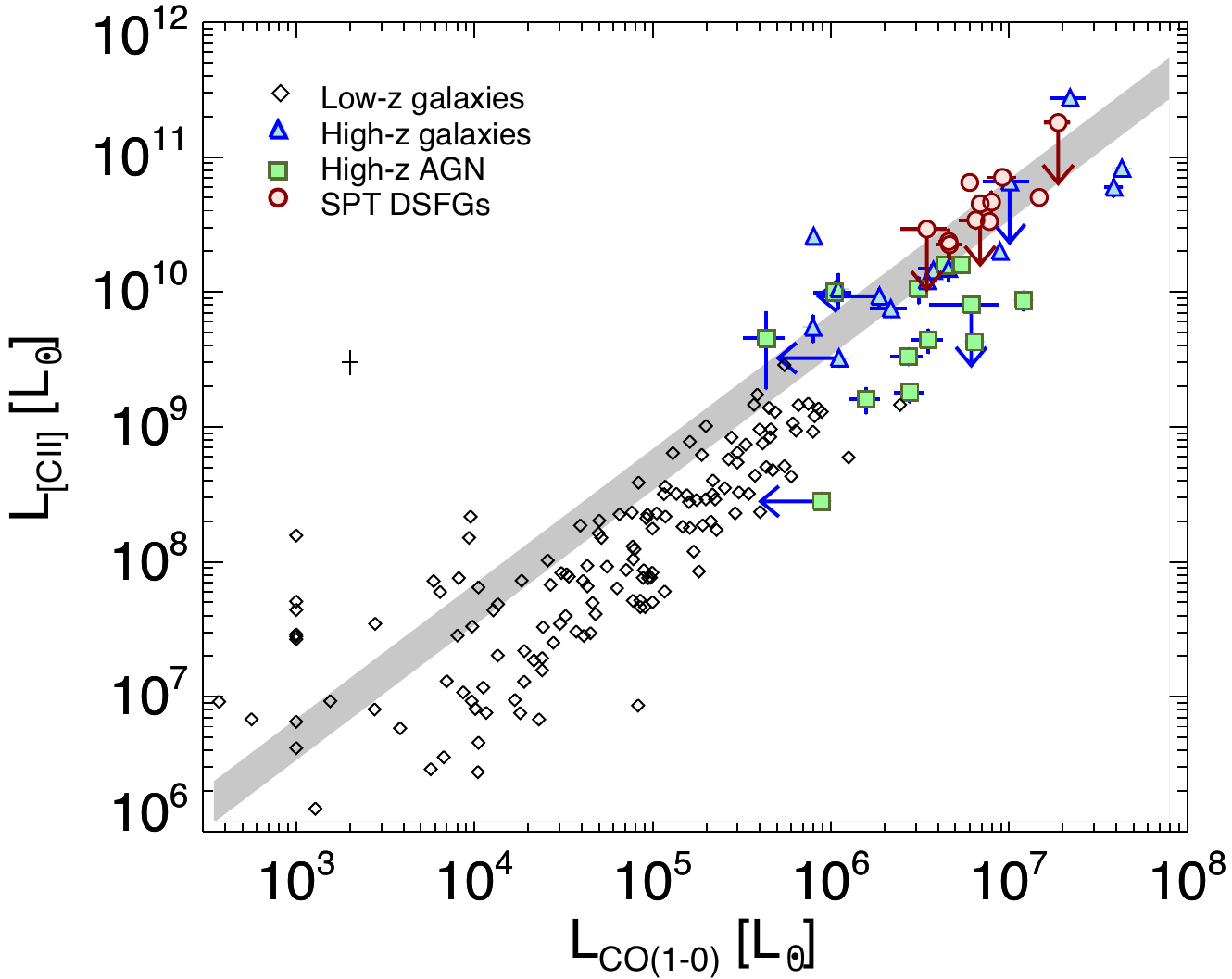}
 \caption{The \CII \ luminosity versus the CO luminosity for the SPT DSFGs and the comparison low and high-$z$ samples. These star forming systems show a correlation between the \CII \ and CO(1-0) luminosities. Fitting a ratio to 11 SPT sources with \CII \ detections and CO(1--0) data yields a slope of $\sim5200$. The width of the grey shaded area represents a 1$\sigma$ spread, $\sim5200\pm1800$. Fitting a ratio to the low-$z$ sample yields a slope of $1300 \pm 440$. The typical error bar for the low-$z$ sources is represented by the black cross.}
  \label{fig:CII_CO}
\end{figure}

\begin{figure}
\includegraphics[trim=0cm 0cm 0cm 0cm, clip=true, scale=0.66]{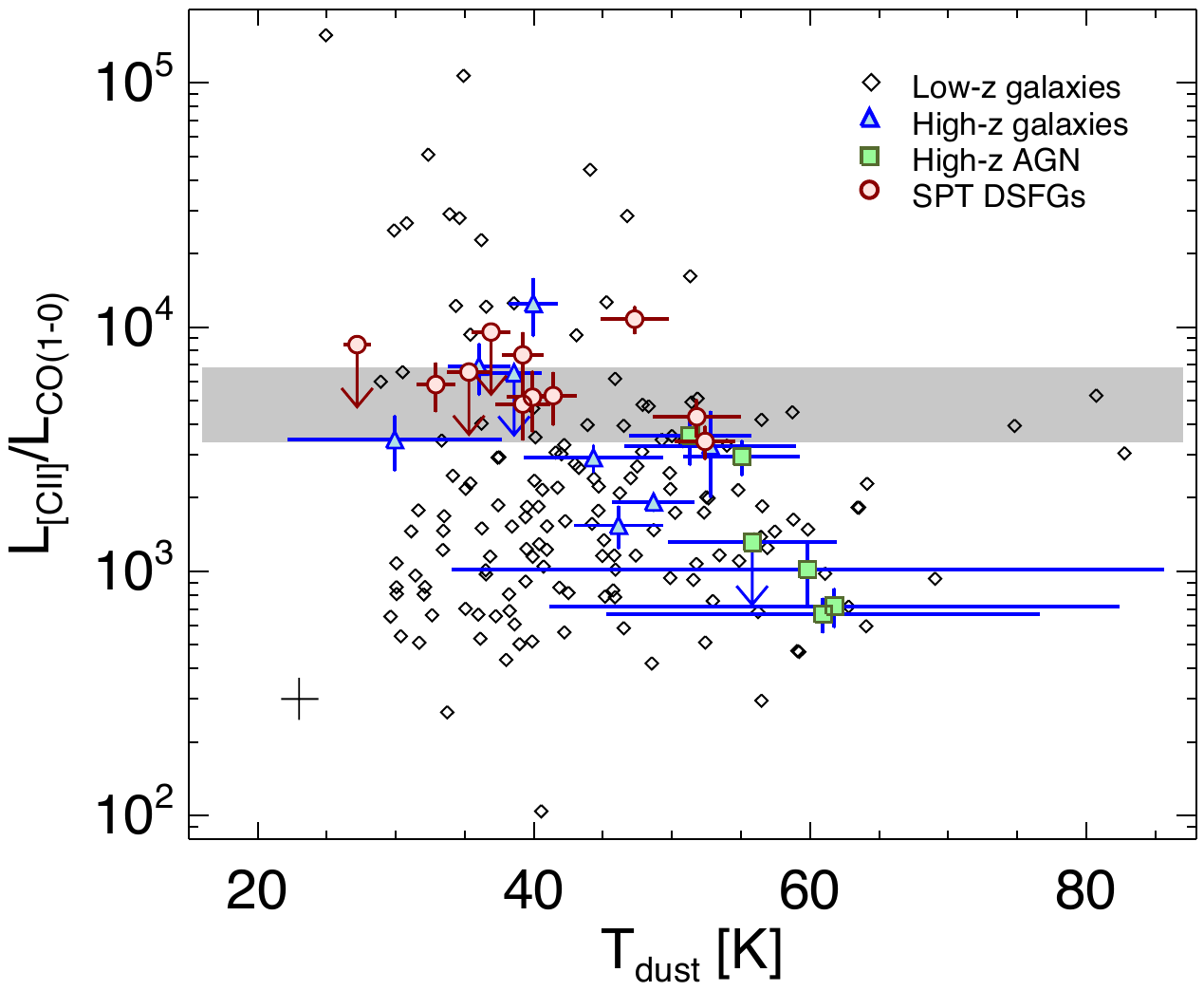}
 \caption{\CIICO \ vs \Td \ for the SPT sources and the low and high-$z$ comparison sample. The typical error bar is represented by the black cross. The grey shaded area represents the 1$\sigma$ spread of the \CIICO\ ratio in the SPT sample. Both the SPT sources and the low-$z$ sample do not show any dependence on \Td. The high-$z$ AGN-dominated  sources are both warmer and have fainter \CII\ relative to CO. }
  \label{fig:CIICO_Tdust}
\end{figure}

\subsection{Observed \CII \ to CO ratios}\label{sec:obs_ratio}
One of the strengths of the SPT DSFG sample is that more than half of the sources have both \CII \ and low-$J$ CO detections.
Figure~\ref{fig:CII_CO} plots the \CII \ vs CO(1--0) luminosities for these 11 sources, along with high-$z$ sources in the comparison sample\footnote{Note that Figure~\ref{fig:CII_CO} presents a larger number of high-$z$ 
sources than Figures~\ref{fig:CII_deficitMH2}, \ref{fig:Td_vs_LCIILFIR}, and \ref{fig:CIICO_Tdust}, as it does not involve quantities derived from continuum photometry. The calculation of \LFIR\ and \Td\ require photometric data that are unavailable for several of the high-$z$ comparison objects (see section~\ref{sec:compare_sample} and Appendix~\ref{app:high_z_sample}).}.
All high-$z$ sources fall close to the \CIICO\ $\sim4400$ relation for local galaxies reported by \cite{crawford85,wolfire89,stacey91,stacey10,swinbank12,neri14}. This relation has been explained in terms of PDR models \citep{wolfire89,wolfire93,stacey91}.
Using the eight SPT DSFGs with \CII \ and CO(1--0) detections, we determine the \CIICO \ ratio to be $5200\pm1800$. 
In section \ref{sec:origin}, we derive the physical conditions that can be derived from this ratio from first principles and eventually compare it with the PDR models.

Finally, in Figure~\ref{fig:CIICO_Tdust}, we test if the \CIICO \ ratio\footnote{Note that both parameters are  independent of the lensing magnification.} depends on \Td. We find no correlation in either the low-$z$ or the SPT DFSG sample. However, the four high-$z$ sources with the lowest \CIICO\ ratio are all AGN dominated. 
This is consistent with the observations of \citet{stacey10} and \citet{sargsyan14} that AGN dominated sources have lower \LCII/\LFIR \ ratios. Although our \Td\ determinations are rather crude, these AGN-dominated sources are also those with the warmest \Td, as expected (see, e.g., Figure 3 in \citealt[]{greve12}). 
Interestingly, only 2 out of 11 SPT DSFGs fall near the AGN-dominated sources in Fig.~\ref{fig:CIICO_Tdust}, supporting our conclusion (\S~\ref{subsec:AGN_content}) that the SPT sample does not contain many strongly AGN-dominated sources.


\section{Discussion}\label{sec:discussion}
\subsection{Possible origins of the \CII \ to CO correlation}\label{sec:origin}
\subsubsection{Determining the optical depth of the \CII\ line}\label{sec:optDepth}
The origin of the correlation between the \CII \ and CO(1--0) luminosities and the impact of their relative optical depths was first discussed by \cite{crawford85}, and later by others in both low and high-$z$ objects (e.g.,\citealt{wolfire89,stacey91,stacey10,swinbank12,neri14}).
\cite{crawford85} assumed \CII\ excitation temperatures \TexCII\ $\gg 92$\,K\footnote{The \CII\ ground state energy level is 92\,K.}, and could therefore apply the Rayleigh-Jeans approximation. 
This assumption was supported by independent estimates of the gas temperature of order $\sim 300$\,K using other fine-structure lines and assuming optically thin \CII \ emission \citep{ellis84}.
They remarked that their observed \LCII/\LCO \ ratio $\sim4400$ is close to the \CII \ to CO(1--0) frequency ratio cubed, suggesting optically thick \CII \ emission. 
However, using supporting data in Orion, the Galactic centre and M82, they independetly derived \CII\ optical depths $\tau_{\text{\CII}}=0.03-1$. This conclusion that \CII\ is mostly optically thin has been assumed several times since \citep[e.g.][]{stacey91,hailey-dunsheath10, stacey10, debreuck11, rawle14}.

The most accurate way to determine the optical depth of the \CIItwelve \ emission line is through observations of the isotopic line ratios. The two \CII \ emission lines for $^{12}$C and $^{13}$C have been observed in local star forming regions, e.g. in M42 \citep{stacey91b,boreiko96}, NGC2024 \citep[in the Orion nebula;][]{graf12}, the Orion Bar, Mon R2, NGC\,3603, the Carina Nebula and NGC\,7023 \citep{ossenkopf13}. These observations find optical depths ranging from $\tau \sim 1$ to $\tau \sim 3$, with on average a moderate optical dept $\tau \sim 1.4$. However, one should keep in mind that Orion and Carina are very bright \CII\ emitters (to allow for a detection of the faint $^{13}$C), with very strong FUV fields ($G_0 \sim 10^4 - 10^5$), implying large \CII\ columns and hence rather high opacities. The galaxy-scale average may result in lower optical depths near unity.
Unfortunately, this isotope ratio technique cannot be applied to galaxy-integrated \CII \ observations as the \CIItwelve \ fine structure line and the brightest \CIIthirteen \ hyperfine structure line are only separated by $\sim110$\,km/s, which is smaller than the typical widths of these lines.
The much weaker $^{13}$C line is then indistinguishable from features in the profile of the $^{12}$C line.

The optical depth of \CII\ in very distant galaxies must therefore be estimated by other methods. 
The idea of optically thick \CII \ emission at high redshifts was recently proposed by \cite{neri14} for the high-$z$ sub-millimetre source HDF\,850.1. Assuming the \CII \ line excitation temperature is the same as the dust and gas kinetic temperatures, they argue for high line optical depth of the \CII \ line (i.e. $\tau_{\text{\CII}}\gtrsim1$).
The observed \LCII \ and predicted \LCO \ by \cite{walter12}, yield a \LCII/\LCO \ ratio of $\sim5200$, in agreement with the ratio derived from the SPT sample of $5200\pm1800$  (\S~\ref{sec:obs_ratio}). We now generalise this line ratio method by comparing the source functions. This does not require  the CO and \CII \ excitation temperatures  to be the same, nor that the Rayleigh-Jeans approximation applies (i.e. we also consider cases where \TexCII\ is close to or below 91\,K). 

From the source functions of both lines, the luminosity ratio depends on 
\begin{align}
\frac{L_{\text{[CII]}}}{L_{\text{CO(1--0)}}}= & \left(\frac{\nu_{\text{[CII]}}}{\nu_{\text{CO(1--0)}}}\right)^3 \times \left(\frac{\Delta \nu_{\rm [CII]}}{\Delta \nu_{\rm CO(1-0)}}\right) \times \\
\nonumber
&\frac{e^{h\nu_{\text{CO(1--0)}}/kT_{\text{ex,CO(1--0)}}}-1}{e^{h\nu_{\text{[CII]}}/kT_{\text{ex,[CII]}}}-1}  \cdot  \frac{1-e^{-\tau_{\text{[CII]}}}}{1-e^{-\tau_{\text{CO(1--0)}}}},
\end{align}
where we have assumed that the \CII\ and CO emitting gas have the same filling factors (see \S\ref{sec:Td}). This assumption is consistent with the very similar \CII \ and CO velocity profiles (Figure~\ref{fig:compare_spec}). In this case, $\frac{\Delta \nu_{\rm [CII]}}{\Delta \nu_{\rm CO(1-0)}}=\frac{\nu_{\text{[CII]}}}{\nu_{\text{CO(1--0)}}}$. 
Also in nearby galaxies, \CII \ and CO have been found to trace each other both morphologically and kinematically (e.g. \citealt{braine12} and \citealt{mittal11}). Equation (2) has four free parameters: two excitation temperatures and two opacities. In order to obtain constraints, we therefore have to fix some of these.

\subsubsection{Same \CII \ and CO excitation temperatures}\label{sec:sameTex}
\begin{figure}
\includegraphics[trim=0cm 0cm 0cm 0cm, clip=true, scale=0.66]{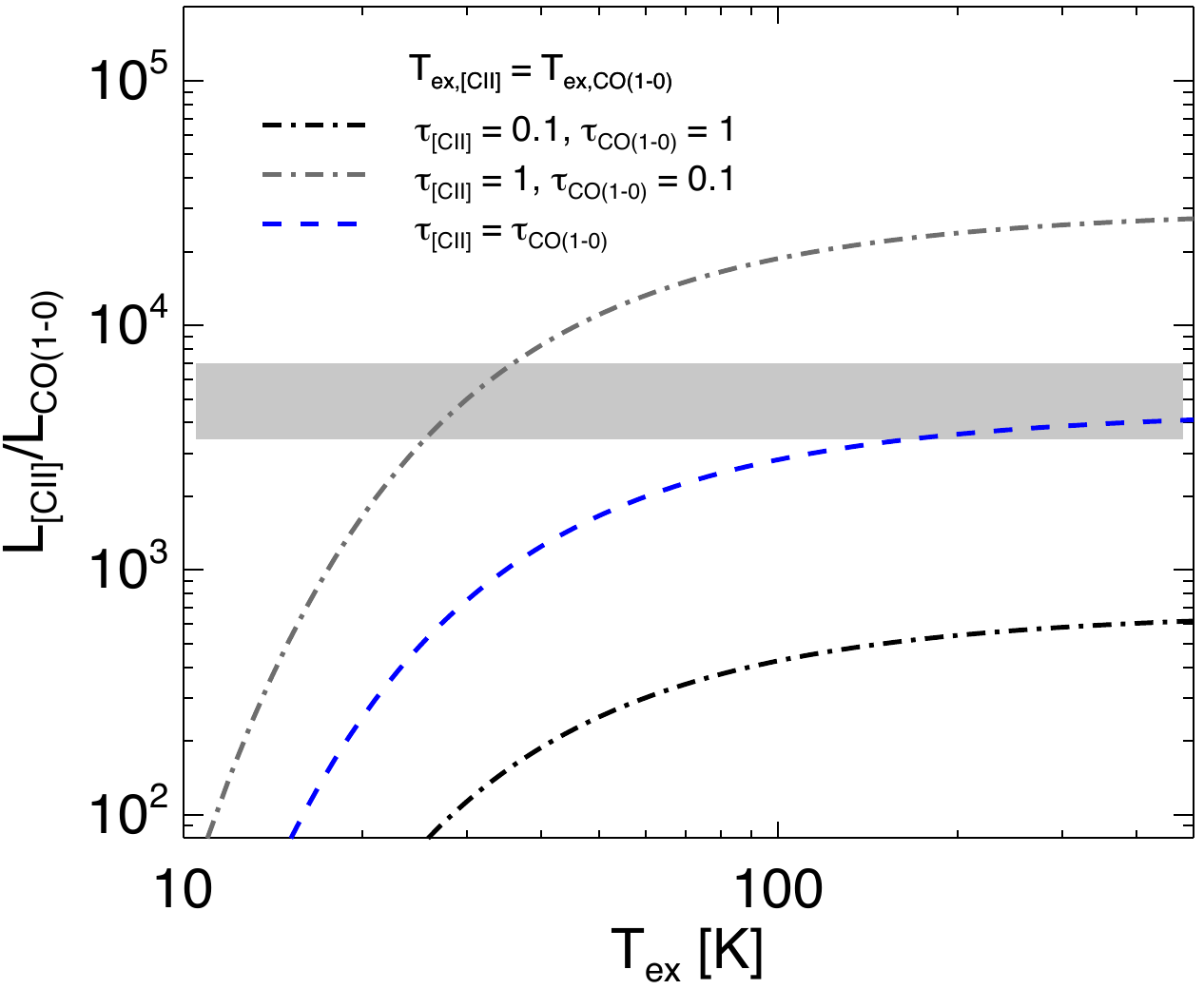}
 \caption{The \LCII/\LCO \ ratio predicted from the source functions versus equal \CII \ and CO(1--0) excitation temperatures (\TexCII \ = \TexCO), for three different cases of the optical depth: 
 \textit{i}) optically thin \CII \ and optically thick CO(1--0) (the black dot-dashed curve),
 \textit{ii}) optically thick \CII \ and optically thin CO(1--0) - (the grey dot-dashed curve),
 \textit{iii}) same optical depth of \CII \ and CO(1--0) (blue dashed curve).
The grey shaded area represents the 1$\sigma$ spread of the \CIICO\ ratio in the SPT sample.
Case \textit{i} underpredicts the ratio by an order of magnitude. Cases \textit{ii} and \textit{iii} can both reproduce the observed ratio. However, we know from $^{12}$CO to $^{13}$CO ratios that CO is optically thick \citep[e.g.][]{spilker14}, which rules out case \textit{ii}. Only case \textit{iii}, implying optically thick CO and \CII, is consistent with all observational data.}
  \label{fig:Lratio_eq_Tex}
\end{figure}

We first consider the case suggested by \cite{neri14} of equal excitation temperatures (i.e. $T_{\text{ex,\CII}}=T_{\text{ex,CO(1--0)}}$), Figure~\ref{fig:Lratio_eq_Tex} plots \LCII/\LCO \ vs $T_{\text{ex}}$ for the three scenarios:
\textit{i}) optically thin \CII \ and (nearly) optically thick CO(1--0) emission, 
\textit{ii}) (nearly) optically thick \CII \ and optically thin CO(1--0) emission, 
\textit{iii}) same optical depth for \CII \ and CO(1--0).

Scenario \textit{i} shown by the black dot-dashed curve underpredicts the observed ratio (grey shaded area) by an order of magnitude. In the opposite case, scenario \textit{ii} (grey dot-dashed curve), the observed ratio is reached only for very low excitation temperatures. 
However, this optically thin CO scenario can be ruled out because both the $^{12}$CO to $^{13}$CO ratios and the low-$J$ $^{12}$CO line ratios in the SPT sample imply that CO is moderately optically thick with \tauCO$=1-10$ \citep{spilker14}, similar to what is seen in our own galaxy \citep{penzias72, goldreich74}.
Scenario \textit{iii}, where the \CII \ and CO(1--0) optical depths are in the same regime is thus the only one that can fit the observed ratios, but only for excitation temperatures $\gtrsim180$\,K. 
While we cannot distinguish mathematically between low and high optical depth, the known \tauCO$=1-10$ and the equality of both the excitation temperatures and opacities would imply that also \CII\ would need to be (nearly or fully) optically thick.

The only way the above scenario \textit{iii} can fit the observed \LCII/\LCO\ ratio is in a thermalised region where both \CII\ and CO(1--0) have excitation temperatures $\gtrsim180$\,K (see Fig.~\ref{fig:Lratio_eq_Tex}). However, the average CO excitation temperatures in the SPT sample are $\lesssim50$\,K \citep{spilker14}, well below the values required to fit the observed ratio (Fig.~\ref{fig:Lratio_eq_Tex}). The only way to reconcile the model in equation (2) with the known physical parameters of the CO (\tauCO$>1$ and $T_{\rm ex, CO} \lesssim 50$\,K) is to allow for different excitation temperatures of \CII\ and CO. Such different excitation temperatures also imply different \CII\ and CO emitting regions within the GMCs.

\begin{figure}
\includegraphics[trim=0cm 0cm 0cm 0cm, clip=true, scale=0.66]{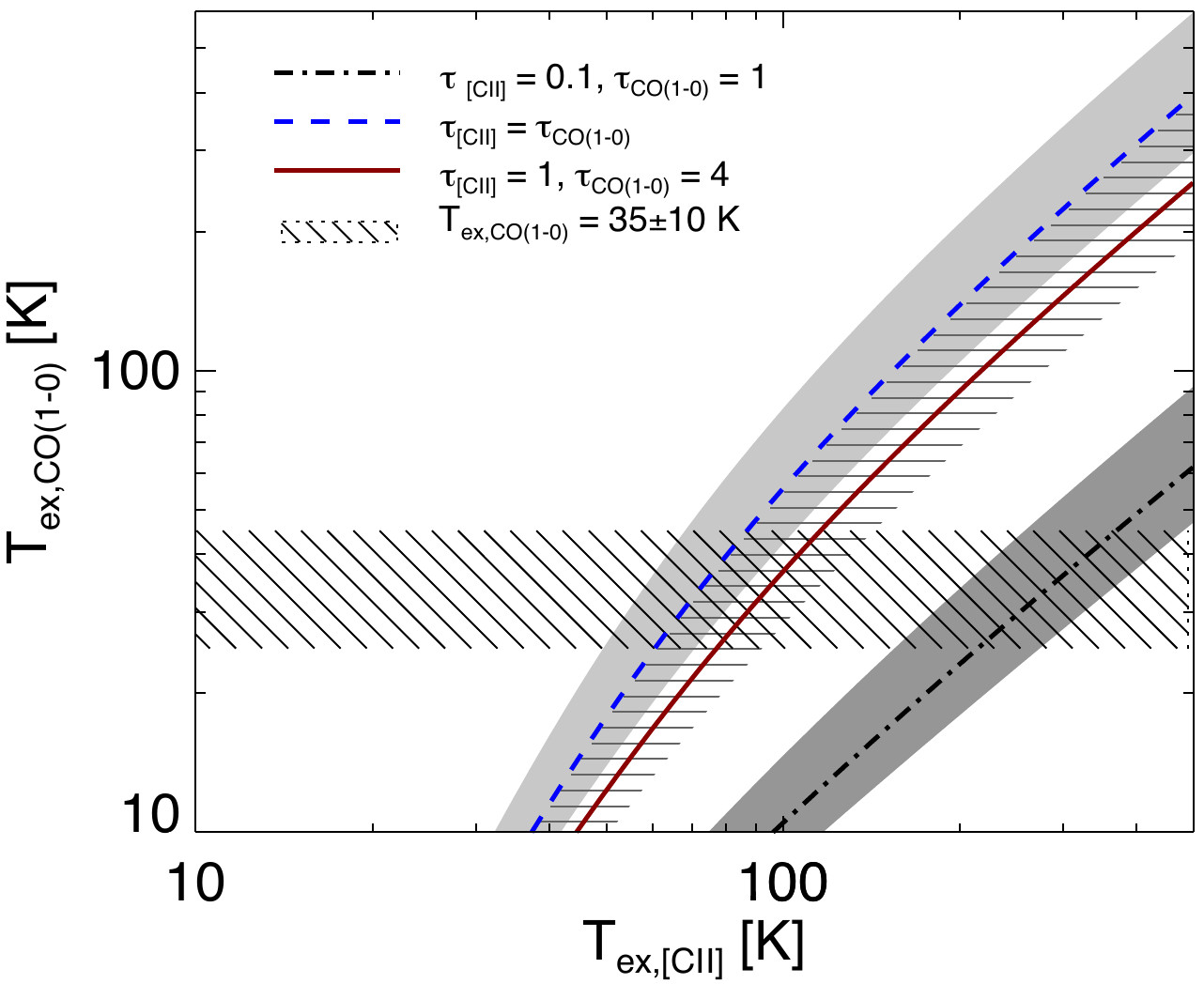}
 \caption{\TexCO \ as a function of \TexCII. 
In all cases, the excitation temperature of \CII\ is higher than for CO(1--0). 
The \textit{blue dashed} curve with the \textit{light grey} shaded area shows the observed \CIICO \ $=5200\pm1800$ range of SPT DSFGs in the case of equal \CII\ and CO optical depths. The red continuous curve and hashed area illustrates that the difference between the excitation temperatures becomes even more significant for \tauCII=1 and \tauCO=4. The \textit{dot-dashed} curve and the \textit{dark grey} shaded area illustrate the case of \tauCII$=0.1$ and \tauCO$=1$. The hatched horizontal area marks \TexCO$=35\pm10$\,K. }
  \label{fig:Tex}
\end{figure}

\subsubsection{Different \CII \ and CO excitation temperatures}\label{sec:differentTex}
To examine the model from equation (2) with different excitation temperatures, we have to fix the observed \LCII/\LCO \ ratio and at least two parameters (two optical depths or an optical depths and a temperature). We first consider the case when \tauCII\ = \tauCO, and plot \TexCO\ as a function of \TexCII\ in Figure~\ref{fig:Tex} (blue dashed line with grey shaded area illustrating the observed \LCII/\LCO\ range).  We conclude that \TexCII \ $>$ \TexCO\ throughout. 
In the optically thick case, we can also allow for  $1 \le$ \tauCII \ $<$ \tauCO. The red curve in Figure~\ref{fig:Tex} illustrates this for \tauCO=4. Any further increase of the difference  between \tauCII\ and \tauCO, will also increase the difference in \Tex.  We warn that once both the \CII\ and CO become strongly optically thick (i.e. $1\ll\tau_{\text{\CII}} < \tau_{\text{CO(1--0)}}$), one can no longer determine any differences between the optical depths, and hence no longer determine \TexCII.

\begin{figure}
\includegraphics[trim=0cm 0cm 0cm 0cm, clip=true, scale=0.66]{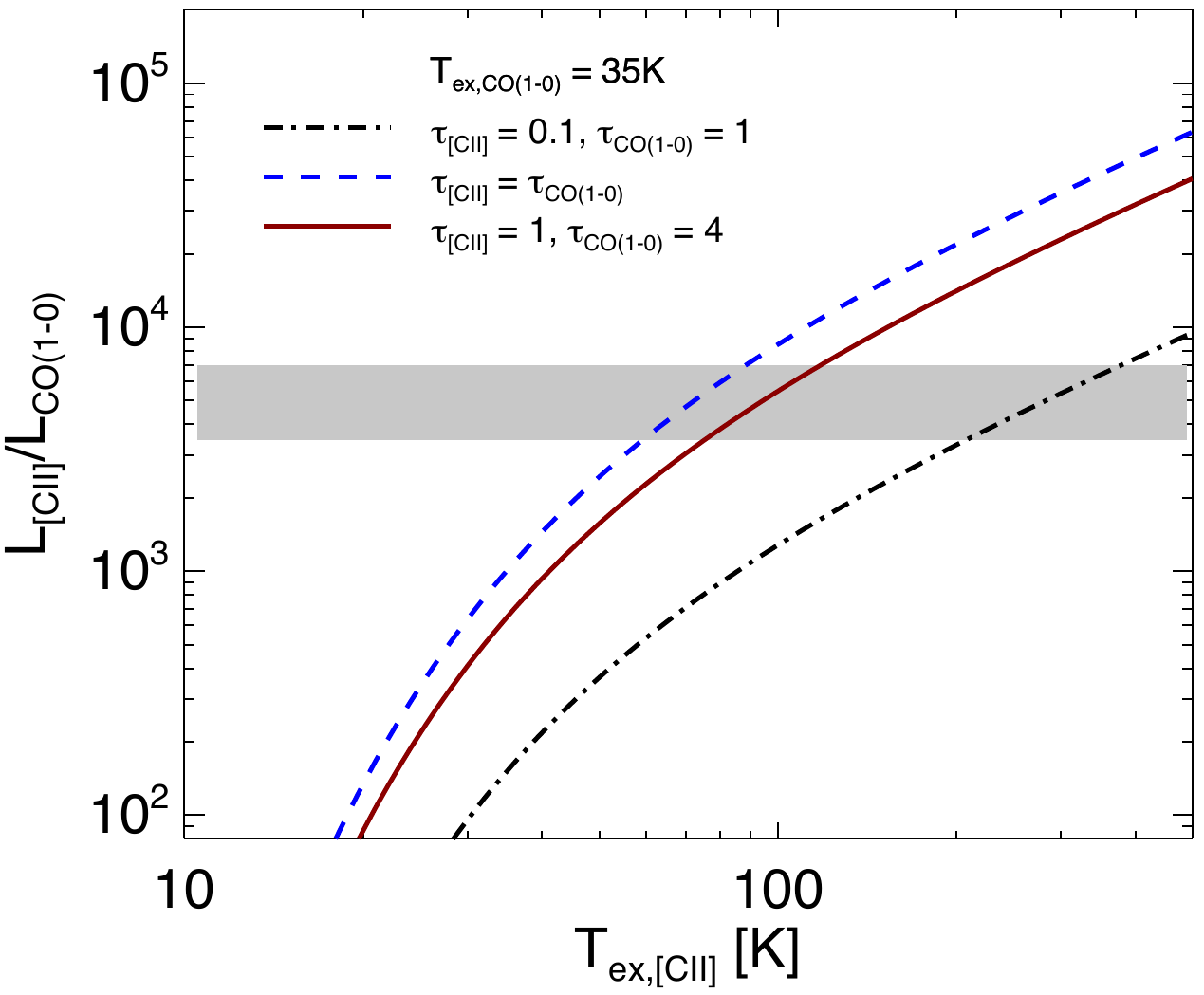}
 \caption{The \LCII/\LCO \ ratio as a function of the \CII \ excitation temperature for a fixed \TexCO=35\,K. The observed \LCII/\LCO \ ratio in the SPT DSFG sources (grey shaded area) is achieved within \TexCII \ $\sim60-90$\,K for equal \CII\ and CO optical depth (blue curve). In the case of \tauCII \ $=1$ and \tauCO \ $=4$, the \LCII/\LCO \ ratio is achieved for \TexCII \ $=85-110$\,K. The optically thin \tauCII \ $=0.1$ and nearly optically thick \tauCO \ $=1$ case is reached by \CII \ excitation temperatures in the range $\sim240-330$\,K (black dashed curve).}
  \label{fig:Lratio_diff_Tex}
\end{figure}

Alternatively, we can also fix the CO opacity and excitation temperature based on existing observations of the SPT DSFG sample, and determine which \TexCII\ values are predicted for a given \tauCII. Assuming the molecular gas is traced by CO(1--0) and the dust is thermalised allows us to fix the \TexCO \ = \Td \ $\simeq35\,$K \citep{weiss13}. This value is consistent with the \Tkin\  determined from the stacked ALMA spectrum of the SPT sample \citep{spilker14}. This \TexCO \ =35\,K case is illustrated in Figure~\ref{fig:Lratio_diff_Tex}. If \tauCII \ = \tauCO, this would imply \TexCII \ $\sim60-90$\,K. Raising \tauCO \ = 4 while keeping \tauCII \ = 1 would imply \TexCII \ $\sim80-110$\,K.
The observed \LCII/\LCO \ ratios can also be reproduced with optically thin \CII\ and (nearly) optically thick CO (\tauCII \  = 0.1 and \tauCO \ = 1) when the \CII\ excitation temperatures are $\sim240-330$\,K. A determination of \TexCII\ is therefore needed to determine \tauCII. If the gas densities are higher than the critical density for \CII\ \citep[$2400-6100$\,cm$^{-3}$ for the above range of \TexCII;][]{goldsmith12}, \TexCII$ \sim T(\rm gas)$. The gas temperatures can be obtained from observed \CII/[OI] ratios \citep[e.g.][]{stacey83,lord96,brauher08}, inferred through the peak \CII\ antenna temperatures \citep[e.g.][]{graf12,ossenkopf13}, or by theoretical modelling \citep[e.g.][]{kaufman99}. These studies obtain $T(\rm gas) = 100-500$\,K, which from Fig.~\ref{fig:Lratio_diff_Tex} would imply \CII\ optical depths ranging from 0.1 to 1. However, the densities we derive from a comparison with PDR models (see \S\ref{sec:PDR}) are in the range $100 - 10^5$ (Table~\ref{table:PDRsize}), with half of our sources below the \CII\ critical density, so the \TexCII$ \sim T(\rm gas)$ may not be valid for a significant part of our sample. Hence, optical depths of $0.1<\tau<1$ are consistent with the observed line ratios in the SPT DSFG sample.

We also compare the model predictions in Figure~\ref{fig:Lratio_diff_Tex} with the low- and high-$z$ comparison samples. The lower average \LCII/\LCO \ ratios in these samples (though with a much larger spread than for the SPT DSFGs) imply \TexCII\ between $\sim30$ and $\sim200$\,K. In particular for the low-$z$ sample, optically thick \CII\ and CO emission would imply very low \TexCII \ $\sim40$\,K, well below the ground state energy. The difference between the SPT DSFG and the low-$z$ comparison sample could therefore be ascribed to a lower optical depths in the nearby sources. In the context of PDR models, this can also be understood as a decrease of the $G_0/n$ ratio. Reducing this parameter implies smaller \CII\ emitting columns and a smaller effective optical depth. The lower $G_0/n$ ratio in the low-$z$ sample could be an effect of the lower far-UV fields found in galaxies forming stars at a more modest rate \citep{stacey91,stacey93,kaufman99}.

Finally, we note that Figure~\ref{fig:CIICO_Tdust} also contains some low-$z$ sources with \CIICO \ ratios $>$15,000. Such values are difficult to explain with standard PDR models. Low metallicity has been invoked to explain these sources \cite[e.g.][]{maloney88,stacey91,madden97}. Considering this effect is beyond the scope of this paper. 

In summary, the observed \LCII/\LCO \ ratios in the SPT DSFGs are best described by a non-uniform medium of \CII \ and CO(1--0) emitting gas with \TexCII $>$ \TexCO, \tauCO~$\gg$~1 and \tauCII~$\lesssim$~1. 

\subsection{Implications of different \CII \ and CO(1--0) excitation temperatures}\label{sec:implications}
In section~\ref{sec:sameTex} we concluded that a homogeneous region with thermalised CO and \CII\ gas is incompatible with the observed \CIICO \ ratios.
Even in the optically thick case, the only way to reproduce the observed ratio is for higher \CII \ than CO(1--0) excitation temperatures (see Figure~\ref{fig:Tex}).
The cases of uniform and separated \CII\ and CO gas was recently studied by \cite{mashian13}.
They explore four different models to explain the observations in the high-$z$ submillimetre source HDF\,850.1:
1) separate CO - \CII\ virialised gas, 
2) separate CO - \CII\ unvirialised gas, 
3) uniformly mixed CO - \CII\ virialised gas and 
4) uniformly mixed CO - \CII\ unvirialised gas. 
Based on cosmological constraints due to the dark matter halo abundance in the standard $\Lambda$ cold dark matter cosmology, they rule out three of the models and conclude that the preferred model is an unvirialised molecular cloud model with independent CO and \CII\ emitting gas with a average kinetic temperature of 100\,K and density of 10$^3$\,cm$^{-3}$ for the molecular gas. 
Both our conclusions and those of \citet{mashian13} are completely consistent with the structure described by PDR models.

PDRs are clouds of molecular gas associated with star-forming regions, as they are often found near young massive O and B stars, acting as the source of the FUV photons that determine the temperature and chemical composition of the gas \citep{meijerink07}.
Schematically, in a PDR the increasing extinction ($A_V$) with depth into the cloud creates a layer-structure, where the surface of the cloud with $A_V \sim 1$ is dominated by H$^+$, C$^+$ and O{\scriptsize I}. As the gas becomes more self-shielded against the dissociating FUV photons deeper in ($A_V \sim 2-4$), layers of \HI \ and H$_2$ form and a transition region of C$^+$, C and CO is present. 
At the centre of the cloud, the molecular gas is so opaque that the chemistry and heating are dominated by cosmic rays. Hence, the \CII \ fine structure line probes the surface of a PDR where $A_V \lesssim 1$  and $T\gtrsim 100$\,K, while CO traces the core of the cloud. To derive detailed physical parameters from the PDR models therefore requires spatially resolved observations of different species such as C, C$^+$, CO, O, H$_2$, polycyclic aromatic hydrocarbons and dust continuum emission \citep[e.g.][]{hollenbach99,orr14}.

\subsubsection{Implications of the photodissociation region structure}\label{sec:PDR}	
\begin{figure*}
\includegraphics[trim=0cm 0cm 0cm 0cm, clip=true, scale=0.96]{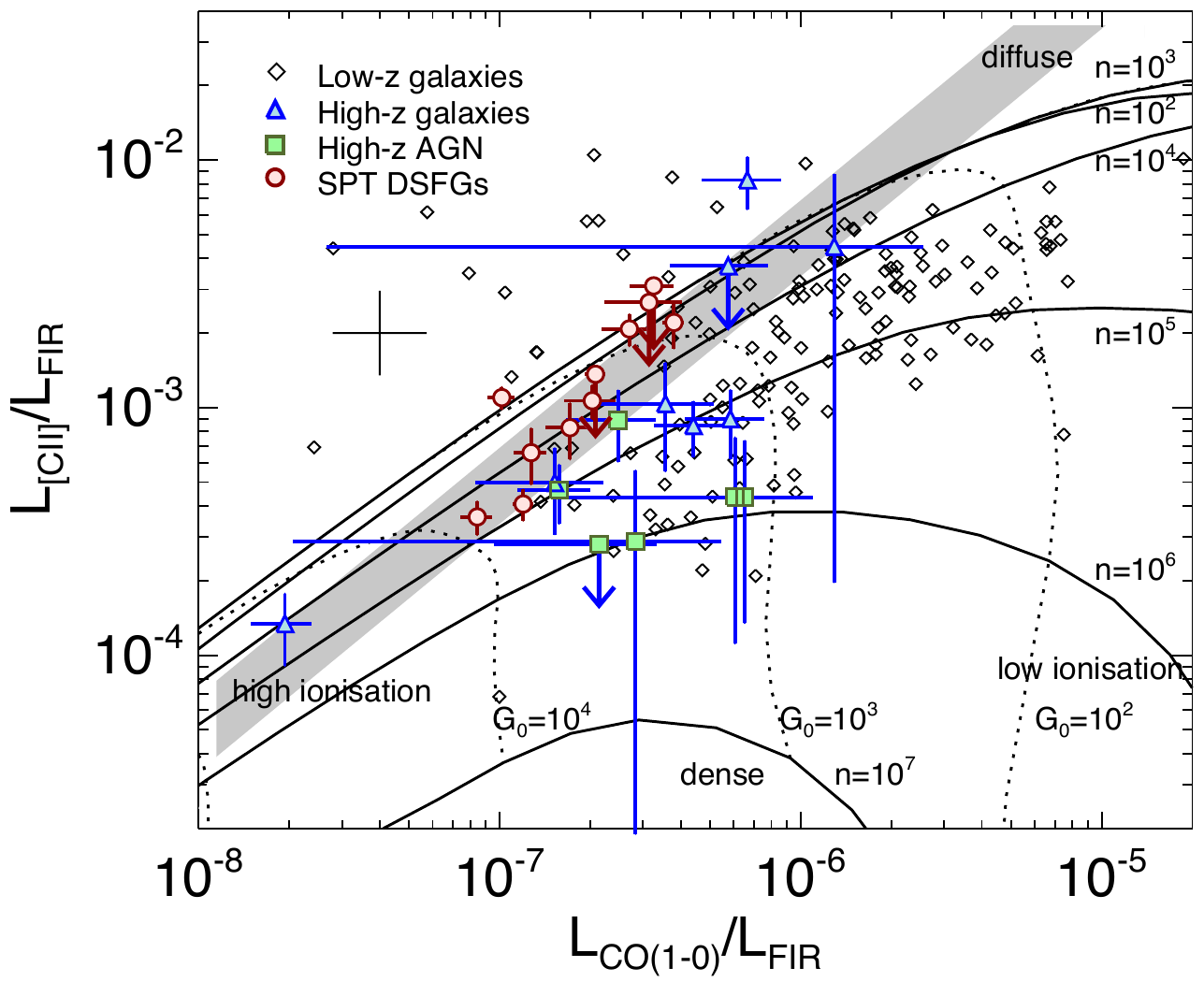}
\caption{\CIIFIR \ vs \COFIR \ for the SPT sources and the low and high-$z$ comparison sample. The figure \citep[e.g.][]{wolfire89,stacey10} compares the values of the strength of the radiation field $G_0$ and the density $n$ for low and high-$z$ sources. The diagram is independent of lensing magnification factors for high-$z$ source and beam filling factors for low-$z$ sources as both the \CII \ and CO(1--0) emission is normalised by the FIR luminosity. The typical error bar is represented by the black cross. To compare the observations with the model contours \citet{stacey10} assume that 70\% of the \CII \ emission originate from PDRs. The grey shaded area represents the 1$\sigma$ spread of the \CIICO\ ratio in the SPT sample.\label{fig:ciifir_cofir}}
\end{figure*}

The structure of the gas in PDRs allows for different \CII \ and CO(1--0) excitation temperatures. 
The physical parameters for the ISM in low- and high-$z$ galaxies predicted by basic PDR models can be compared with observed data in a diagnostic diagram (Figure~\ref{fig:ciifir_cofir}) first presented by \cite{wolfire89} and updated by e.g. \cite{stacey91,stacey10} and \cite{hailey-dunsheath10}. The diagram has the advantage of plotting ratios where the lensing magnification factors for high-$z$ sources, beam-filling factors for low-$z$ sources and filling factors for the PDRs (which are unknown, but assumed the same for \CII \ and CO(1--0)) are divided out. It can be used to roughly estimate the strength of the FUV field and the gas density. One has to be cautious using this diagram, as the FIR luminosity can contain strong contributions from other sources not associated with PDRs (e.g. AGN tori). 

The SPT sample is ideal in this respect  as \CII \ has been measured for 17 out of 20, and low-$J$ CO for 11 (see Table~\ref{table:CII_data}) of these sources, so the CO luminosities for our sources (unlike the comparison samples)  do not depend on uncertain scaling factors. The SPT sample is the most complete high-$z$ sample included in this diagram. The comparison sample is also integrated over entire galaxies enabling a fair comparison to the SPT sample. 

The PDR model used in this diagram \citep{kaufman99} models a plane-parallel slab divided into a number of zones of different depths. 
The intensities are modeled for a parameter space of the FUV field strength ($G_0$) in units of the local Galactic interstellar radiation field (the `Habing Field', $1.6\times 10^{-3}$\,ergs \cmcube s$^{-1}$; \citealt{kaufman99}) in the range $10^{-0.5} \le G_0 \le 10^{6.5}$ and the gas densities in the range 10\,\cmcube$\le n \le 10^7$\,\cmcube. The emission from different species depends on the density, the field strength and the depth into the cloud.  
\cite{hailey-dunsheath10} assume that 70\% of the \CII \ emission originates from PDRs (see also \S~\ref{sec:nonPDR}), meaning that the points would move slightly down in Figure~\ref{fig:ciifir_cofir} if this was corrected for.

By comparing the SPT data points with the PDR model tracks in Figure~\ref{fig:ciifir_cofir}, we obtain a rough estimate of the radiation field strength and the gas density of $100<G_0<8000$ and $10^{2}\,$cm$^{-3}<n<10^5$\,cm$^{-3}$ (see Table~\ref{table:PDRsize}). These values are consistent with the ones found in previous samples of DSFGs \citep[e.g.][]{stacey10}. They imply PDR surface temperatures of $300 - 500$\,K \citep[Fig. 2 of][]{kaufman99}. As these surface temperatures are representative for regions up to $A_V\sim 2$, they cover most of the \CII\ emitting region. They are consistent with other derivations of the gas temperatures (see \S~\ref{sec:differentTex}). If the densitity exceeds the \CII\ critical density, these temperatures also represent \TexCII, which would imply optically thin \CII\ (Fig.~\ref{fig:Lratio_diff_Tex}). However, in half of our sources, the densities are below the critical densities, so a range of opacities up to unity remains possible.
\\\\
\textit{Sizes of PDRs}\\
Comparing the \LFIR/$\mu$ for the SPT sources\footnote{See \S\ref{sec:lensing} for a discussion on the lensing magnification factor $\mu$ in our sample.}, which lie in the range $(1.1-21.9)\mu^{-1}\times10^{12}$\,L$_{\odot}$, with the local starburst galaxy M82 (\LFIR$\sim(2.3-3.2)\times10^{10}$\,L$_{\odot}$, \citealt{rice88} and \citealt{colbert99}) show the significant difference in \LFIR \ at high and low redshifts. Using this comparison and estimates of $G_0$ for each SPT DSFG, we estimate approximate sizes of the PDRs populating the galaxies, following \cite{stacey10}. For this, we assume the molecular clouds are randomly mixed with young stellar clusters, acting as the radiation sources, within the galaxy \citep{wolfire90}. Assuming this structure, the relationship between the average $G_0$, the total size of the PDRs ($D$, diameter) and the total luminosity (\LFIR) of the source is given by $G_0\propto\lambda L_{\text{FIR}}/D^3$ for a short mean free path ($\lambda$) and $G_0\propto L_{\text{FIR}}/D^2$ for large mean free path of the FUV photons \citep[see][]{wolfire90}. 
To estimate the approximate sizes, we read off $G_0$ and \LFIR\ for the SPT DSFGs from Figure~\ref{fig:ciifir_cofir}, and scale these with the values obtained for M82 ($G_0\sim1000$ \citealt{lord96}, \LFIR$\sim2.8\times10^{10}$\,L$_{\odot}$, average of the values determined by \citealt{rice88} and \citealt{colbert99}), assuming the same mean free path for the SPT sources and M82. The exact size of the PDR region in M82 is rather uncertain with reported sizes ranging from 300\,pc \citep{joy87} to 600\,pc \citep{carlstrom91}. For consistency with \citet{stacey10}, we will assume $D\sim300$\,pc. 
We warn that differential lensing (if significant, see \S~\ref{sec:lensing}) could affect the positions of the sources in Figure~\ref{fig:ciifir_cofir} leading to different estimations of $G_0$ and therewith the sizes. The radii we estimate are listed in Table~\ref{table:PDRsize}, along with our estimated $G_0$ and $n$.
An estimate of the source radii from lens modelling is available for four sources in Table~\ref{table:PDRsize}. The estimated radii of the PDRs, while fairly uncertain, are comparable to the source sizes determined by the lens models.

\begin{table*}
 \begin{tabular}{@{}lcccccccc}
  \hline
  Source & $G_0$ & $n$ &  R(PDR) & R$_{1/2}$(lens model)  & R($L'_{\text{CO(1--0)}}$) & $\mu$ \\
  	      &  	    &  	  cm$^{-3}$           &	kpc  & kpc & kpc & \\
  \hline
SPT0113-46 &      $10^3$             &      $10^3$          &       $     0.6 -     1.1$ & --- 		 	      & $     0.85 -       1.16$ & ---$\dagger$ \\
SPT0345-47 &      $10^4$             &      $5\times10^4$        & $     0.4 -  0.7$ & --- 		 	      & $     0.39 -      0.52$ & ---$\dagger$ \\
SPT0346-52 &      $8\times10^3$ &      $10^5$      &       $     0.7 -       1.5$ & $0.59\pm0.03$       & $     0.18 -      0.24$ &    5.4$\pm$0.2 \\
SPT0243-49 &      $5\times10^3$ &      $10^3$          &       $     0.4 -      0.6$ & --- 		 	      & $     0.85 -       1.15$ & ---$\dagger$ \\
SPT0418-47 &      $10^3$             &      100            &       $     0.7 -       1.5$ & $1.07\pm0.17$      & $      2.26 -       3.07$ &   21.0$\pm$3.5 \\
SPT0441-46 &      $8\times10^3$ &      $10^4$        &       $     0.3 -      0.5$ & --- 		 	      & $     0.36 -      0.49$ & ---$\dagger$ \\
SPT2103-60 &      $10^3$             &      100            &       $     0.7 -       1.4$ & --- 		 	      & $      1.69 -       2.30$ & ---$\dagger$ \\
SPT2146-55 &      $5\times10^3$ &      $10^4$        &       $     0.5 -      0.8$ & ---                               & $     0.42 -      0.56$ & ---$\dagger$ \\
SPT2147-50 &      $5\times10^3$ &      $10^4$        &       $     0.4 -      0.6$ & ---                               & $     0.33 -      0.45$ & ---$\dagger$ \\
SPT0551-50 &      $10^2$             &      100            &      $     1.0 -       2.6$ & --- 		 	      & $      1.40 -       1.89$ & ---$\dagger$ \\
SPT0538-50 &      $10^2$             &      100            &      $     1.7 -       5.9$ &  $1.07\pm0.25$     & $      2.46 -       3.34$ & $20.9\pm4.2$ \\

\hline
 \end{tabular}
  \caption{The first column contains the names of the sources for which both \CII \ and low-$J$ CO lines have been detected. The second and third columns list $G_0$ and $n$ for the sources determined from the PDR models in Figure~\ref{fig:ciifir_cofir}. Note that especially $n$ can be very uncertain as the models are very degenerate in this part of the diagram. Column four gives the size range determined using the short and long mean free path assumed by Stacey et al. (2010). The fifth column lists the radii for the sources which have lens models (Hezaveh et al. 2013), and the sixth column gives the range in sizes of the molecular gas estimated from the molecular gas mass range given by $\alpha_{\text{CO}}=(0.8-2)\,$M$_{\odot}\,($K\,km/s\,pc$^{2})^{-1}$. The last column lists the lensing magnification factor from \citet{hezaveh13}; SPT0529-54 is not included in this table as low-$J$ CO lines have not been observed for this source. The sources marked with $\dagger$, we have assumed a mean of $\langle \mu \rangle=14.1$ (see \S\ref{sec:lensing}). \label{table:PDRsize}}
\end{table*}

A rough estimate of the size ranges occupied by the molecular gas is given in Table~\ref{table:PDRsize} as well. These sizes are estimated using the molecular gas mass estimated from $L'_{\text{CO(1--0)}}$. Using low and high CO-to-H$_2$ conversion factors often used in the literature of $\alpha_{\text{CO}}=0.8\,$M$_{\odot}\,($K\,km/s\,pc$^{2})^{-1}$ (see section~\ref{sec:deficit}, \cite{downes98}) and $\alpha_{\text{CO}}=2\,$M$_{\odot}\,($K\,km/s\,pc$^{-2})^{-1}$ (e.g. \citealt{swinbank11}), we estimate sizes of the molecular gas regions making the simplistic assumption that the gas is uniformly distributed in a sphere with radius $R$ and density $n$ as listed in Table~\ref{table:PDRsize}. 
The sizes estimated using this method are quite uncertain as we do not take into account non-uniform density profiles or non-unity volume filling factors of the gas. They are roughly consistent with both the sizes obtained from the PDR model, and the more accurate sizes derived from the lens models. We note that these kpc-scales are close to the typical sizes of host galaxies, and could be easily spatially resolved, especially given the lensing magnification, allowing detailed future studies of their spatial distributions.

\subsubsection{Other contributions to the \CII \ emission}\label{sec:nonPDR}
The \CII \ emission integrated over an entire galaxy will contain contributions from regions with different physical conditions such as XDRs, CRDRs, shock dominated regions, diffuse warm gas, \HII \ regions, and PDRs. 
Above, we have assumed that observed \CII \ emission in the SPT DSFGs is dominated by emission from PDRs on molecular cloud surfaces \citep[e.g.][]{stutzki88,stacey93}. We now consider the possible contributions from the alternative \CII\ emitting regions in increasing order of importance.

In the vicinity of an AGN, supplying X-ray radiation, we expect XDRs. X-ray photons penetrate deeper into the volume of the interstellar clouds than the FUV photons in PDRs as the absorption cross sections are smaller for X-ray energies. As argued in \S~\ref{subsec:AGN_content} and \ref{sec:obs_ratio}, the SPT sample does not show evidence for strong AGN activity, and XDR contributions are therefore expected to be negligible. 

In CRDRs, the gas heating and chemistry are controlled through interactions with high-energy particles. As the energy density in cosmic rays is low compared to photons, CRDRs are thought to trace the dense, innermost regions of  giant molecular clouds (GMCs), rather than the outer surfaces where \CII\ emission is assumed to be more prevalent \citep{viti13}. We therefore assume the CRDR contribution to be negligible.

Turbulence and shocks have been suggested by \cite{appleton13} to be an additional source of \CII \ emission. They suggest that this mechanism should be present in highly turbulent conditions such as colliding galaxies and the early stages of galaxy-disc build up. However, for low turbulent velocities, it becomes difficult to distinguish between this mechanism and low-density PDRs. The extreme \CIIFIR \ ratios \cite{appleton13} find for the intergalactic filament in Stephan's Quintet are $\sim 30 \times$ higher than those observed in the SPT sample. Smaller shock-ionised regions may still contribute significantly to the observed \CII\ emission. However, it would be difficult to explain the relatively narrow spread in the observed \CII \ to CO ratio if a range of such shock-ionised regions would be a frequent occurence in our SPT sources. We therefore do not expect this mechanism to be important in our samples.

The \CII \ emission could also originate from the diffuse warm low density medium in between the GMCs. Spatially resolved \CII\ and CO observations are required to differentiate between the PDR and diffuse components. Observations of another lensed DSFGs, HLSJ091828.6+514223 at $z=5.243$ \citep{rawle14}, do show that the \CII\ and CO(1--0) have a consistent structure and velocity profile, though the resolution may not go down to the scales of the GMCs (a few hundred pc). Observations of \CII\ and CO at spatial resolutions of $\sim$50\,mas are required to constrain this diffuse component in our DFSGs. 

\HII \ regions surround young O and B stars which emit Lyman continuum photons with energies exceeding the ionisation energy of hydrogren (13.6\,eV).
\cite{abel06} explore the contribution of \CII \ emission from \HII \ regions for a wide range in temperature, ionisation parameter ($U$) and electron density, and find that at least 10\%, and sometimes up to 50-60\% of the total \CII \ emission comes from within the \HII \ regions. Observations of other fine structure lines that only trace \HII\ regions are required to determine the exact contribution from \HII \ regions. \citet{oberst06,oberst11} used the observed \CII/[N{\scriptsize II}] 205\,$\mu$m ratio in the Carina nebula to constrain the contribution from \HII\ regions to 30\%. This technique has since also been applied to high redshift objects \citep{ferkinhoff11,decarli14}. 

\subsection{The \CII/FIR luminosity deficit}\label{sec:revisit}
Several studies have reported a `deficit' in the ratio of the \CII \ line strength to the FIR luminosity ratio (\CIIFIR) for LIRGs with \LFIR \ $\gtrsim10^{11}$\,L$_{\odot}$ (e.g. \citealt{malhotra97}, \citealt{luhman98}, \citealt{maiolino09, stacey10, graciacarpio11, sargsyan12}). 
Various physical explanations for this trend have been proposed, including an increased ionisation parameter \citep{malhotra01,abel09,graciacarpio11}, collisional de-excitation of \CII \ \citep{appleton13} and non-PDR contributions to the FIR luminosity \citep{luhman03}, possibly from AGN \citep{sargsyan12}.

\citet{stacey10} argued that the lower \CII/FIR ratio can be explained by the fact that the star-formation in local ULIRGs is confined and vigorous (leading to high $G_0$), while in the most distant objects, the star-formation is very large-scale, but of lower intensity (i.e. lower $G_0$). In PDR models, the \CII/FIR ratio is inversely proportional with $G_0$. Contrary to the FIR luminosity, which scales linearly with $G_0$, the \CII\ luminosity increases only slowly with $G_0$. This is because in the observed density regime, the C$^+$ column density scales only with dust extinction, while the emissivity is only weakly dependent on $G_0$ since the gas temperature is above the excitation potential of 92\,K. This leads to a ``saturation effect'' of the \CII\ emission at high luminosity in nearby ULIRGs, while the FIR remains unsaturated.
This is consistent with a recent study by \cite{diazsantos14} exploring the difference in the \CII/FIR luminosity deficit between the extended and compact (nuclei) regions in nearby LIRGs, revealing a larger deficit in the \CII/FIR luminosity ratio for the compact regions than for the extended regions. 
The `deficit' is mostly confined to the innermost compact regions, while \CII \ to FIR luminosity ratio for the extended regions is similar to that found in the extended disks of normal star-forming galaxies.

A similar saturation effect would also occur when the \CII\ line becomes (nearly) optically thick. The line then reaches its maximum brightness, in the sense that any additional incoming ionising photons will not increase the brightness of the line further. However the cooling of these additional ionising photons may still continue through optically thin processes, notably the dust continuum. An alternative or additional way of decreasing the \CII \ to FIR luminosity ratio is therefore to have optically thick \CII \ and optically thin dust continuum emission. The higher density in the more compact regions may then increase the probability of the \CII \ to become optically thick.

Finally we note that the \CII \ to FIR `deficit' at high luminosities is also reported for other fine structure lines such as [O{\scriptsize I}], [O{\scriptsize III}], [N{\scriptsize II}] and [N{\scriptsize III}], indicating that this `deficit' is a general aspect of all FIR fine structure lines, regardless of their origin in the ionised or neutral phase of the interstellar medium \citep[e.g.][]{graciacarpio11,farrah13}.


\section{Conclusions}\label{sec:conclusion}
We have presented the first uniformly selected \CII \ survey of lensed DSFGs covering the redshift range $z=2.1-5.7$. 
We have detected \CII \ for 17 out of 20 sources, 11 of which are also observed and detected in low-$J$ CO lines. 
This sample facilitates statistical studies of the ISM at high redshift. 
Our main results and conclusions are:
\\\\
\noindent
\textbf{1.} We fit single or double Gaussian functions to the CO and \CII \ velocity profiles, and find consistent velocity profiles in 13 out of 14 CO detections with ALMA. This suggest that differential lensing is not significant in these cases, and is consistent with the idea that the \CII \ and CO(1--0) emitting gas are spatially associated.
\\\\
\textbf{2.} The line luminosity ratio of the \CII \ and CO(1--0) detections for the SPT sources is $\sim5200\pm1800$, which agrees with the first reported ratio by \citet{crawford85}. The values derived from the SPT sample are consistent with both low-$z$ and high-$z$ comparison samples, but with significantly smaller dispersion. This is presumably due to the homogeneity of the SPT selection and followup observations and absence of any known AGN-dominated sources, which have lower \CII\ to CO(1--0) ratios.
\\\\
\textbf{3.} The SPT sample covers the same spread in the \CIIFIR \ ratio as the \LFIR \ $\gtrsim10^{11}\,$L$_{\odot}$ sources in both our low-$z$ and high-$z$ comparison samples.
AGN-dominated sources increase the scatter towards lower \CIIFIR \ ratios in the comparison sample. 
\\\\
\textbf{4.} We investigate the origin of the \CII \ emission using the observed \CIICO \ ratio, and conclude that the observed ratio is best described by a medium of \CII\ and CO(1--0) emitting gas with  \TexCII \ $>$ \TexCO, optically thick CO (\tauCO~$>$~1), and low to moderate \CII\ optical depth (\tauCII~$\lesssim$~1). 
The structure of PDRs allows for such different excitation temperatures of the \CII \ and CO(1--0) emitting gas.
Interestingly the PDR models converge to this \CII \ to CO(1--0) ratio for densities below $10^5$\,cm$^{-3}$. 
\\\\
\textbf{5.} We revisit the \CIIFIR \ `deficit' observed for sources with \LFIR \ $\gtrsim10^{11}$\,L$_{\odot}$, which has been explained as a ``saturation effect'' of the \CII\ emission in compact regions with higher $G_0$ factors. An alternative or additional explanation for this saturation effect is (nearly) optically thick \CII \ emission. In this case the \CII \ line becomes saturated and reaches the maximum \CII \ brightness, while cooling via the FIR continuum emission continues. The variation in the \LCII/\LFIR \ ratio is therefore dominated by the variation in \LFIR \ rather than \LCII. 
\\\\
\textbf{6.} We determine the FIR luminosity for both the SPT sample and the comparison low and high-$z$ sample in a consistent way, adding 11 SPT DSFGs to the  \LCII/\LFIR \ vs \LCO/\LFIR \ plot in Figure~\ref{fig:ciifir_cofir}. 
We compare the SPT sample with PDR models and estimate the radiation field strength and average gas density to be in the range $100<G_0<8000$ and $10^{2}\,$cm$^{-3}<n<10^5$\,cm$^{-3}$. 
\\\\
The reliability of \CII \ as a tracer of star formation rate (SFR) has been explored by e.g. \cite{delooze14}. They show that \CII \ is a good tracer for the SFR except for low metallicity sources. 
 Determining the metallicity using other fine structure lines such as [N{\scriptsize II}], [O{\scriptsize II}] and [O{\scriptsize III}], and the contribution to the \CII \ emission from \HII \ regions \citep{croxall12} are therefore key in finding the most reliable tracer of SFR in nearby and distant galaxies. 
Thus far, all high-$z$ fine structure line measurements have been unresolved, but thanks to lens shear of gravitationally lensed sources it will become possible to resolve structures down to 100\,pc scales \citep{swinbank10}. Hence, for PDRs of the sizes estimated here, future observations with ALMA will be able to spatially resolve \CII \ emission and other fine structure lines and provide new insight into how  \CII \ emission traces gas.

\section*{Acknowledgments}
We thank the referee for a very thorough reading of the paper and constructive comments which have significantly improved the paper. We are grateful to Javier Gracia-Carpio for providing us with a sample of low- and high-$z$ sources with both \CII \ and CO detections, ideal for comparison with the SPT sample of high-$z$ sources. We also thank Gordon Stacey and Padelis Papadopoulos for lively, educational, and profitable discussions.
We thank G\"oran Pilbratt, \textit{Herschel} Project scientist for the allocated \textit{Herschel} SPIRE Directors Discretionary Time and Richard George and Ivan Valtchanov for helpful comments on the \textit{Herschel} SPIRE FTS data reduction. 

This publication is based on data acquired with the Atacama Pathfinder Experiment (APEX). APEX is a collaboration between the Max-Planck-Institut fur Radioastronomie, the European Southern Observatory, and the Onsala Space Observatory.
This paper makes use of the following ALMA data: ADS/JAO.ALMA\#2011.0.00957.S, ADS/JAO.ALMA\#2011.0.00958.S and ADS/JAO.ALMA\#2012.1.00844.S. ALMA is a partnership of ESO (representing its member states), NSF (USA) and NINS (Japan), together with NRC (Canada) and NSC and ASIAA (Taiwan), in cooperation with Republic of Chile. The Joint ALMA Observatory is operated by ESO, AUI/NRAO and NAOJ.
The Australia Telescope Compact Array is part of the Australia Telescope National Facility which is funded by the Commonwealth of Australia for operation as a National Facility managed by CSIRO.
This research has made use of the NASA/IPAC Extragalactic Database (NED) which is operated by the Jet Propulsion Laboratory, California Institute of Technology, under contract with the National Aeronautics and Space Administration. This research has made use of NASA's Astrophysics Data System Bibliographic Services

This material is based on work supported by the U.S. National Science Foundation under grant No. AST-1312950. 
The South Pole Telescope is supported by the National Science Foundation through grant PLR-1248097.  Partial support is also provided by the NSF Physics Frontier Center grant PHY-1125897 to the Kavli Institute of Cosmological Physics at the University of Chicago, the Kavli Foundation and the Gordon and Betty Moore Foundation grant GBMF 947.

\bibliographystyle{mn2e}
\bibliography{spt_smg}

\begin{thebibliography}{126}
\expandafter\ifx\csname natexlab\endcsname\relax\def\natexlab#1{#1}\fi

\bibitem[{{Abel}(2006)}]{abel06}
{Abel} N.~P., 2006, \mnras, 368, 1949

\bibitem[{{Abel} {et~al}\mbox{.}(2009){Abel}, {Dudley}, {Fischer}, {Satyapal},
  \& {van Hoof}}]{abel09}
{Abel} N.~P., {Dudley} C., {Fischer} J., {Satyapal} S., {van Hoof} P.~A.~M.,
  2009, \apj, 701, 1147

\bibitem[{{Appleton} {et~al}\mbox{.}(2013){Appleton}, {Guillard}, {Boulanger},
  {Cluver}, {Ogle}, {Falgarone}, {Pineau des For{\^e}ts}, {O'Sullivan}, {Duc},
  {Gallagher}, {Gao}, {Jarrett}, {Konstantopoulos}, {Lisenfeld}, {Lord}, {Lu},
  {Peterson}, {Struck}, {Sturm}, {Tuffs}, {Valchanov}, {van der Werf}, \&
  {Xu}}]{appleton13}
{Appleton} P.~N. {et~al.}, 2013, \apj, 777, 66

\bibitem[{{Aravena} {et~al}\mbox{.}(2008){Aravena}, {Bertoldi}, {Schinnerer},
  {Weiss}, {Jahnke}, {Carilli}, {Frayer}, {Henkel}, {Brusa}, {Menten},
  {Salvato}, \& {Smolcic}}]{aravena08}
{Aravena} M. {et~al.}, 2008, \aap, 491, 173

\bibitem[{{Aravena} {et~al}\mbox{.}(2014){Aravena}, {Hodge}, {Wagg}, {Carilli},
  {Daddi}, {Dannerbauer}, {Lentati}, {Riechers}, {Sargent}, \&
  {Walter}}]{aravena14}
{Aravena} M. {et~al.}, 2014, \mnras, 442, 558

\bibitem[{{Aravena} {et~al}\mbox{.}(2013){Aravena}, {Murphy}, {Aguirre},
  {Ashby}, {Benson}, {Bothwell}, {Brodwin}, {Carlstrom}, {Chapman}, {Crawford},
  {de Breuck}, {Fassnacht}, {Gonzalez}, {Greve}, {Gullberg}, {Hezaveh},
  {Holder}, {Holzapfel}, {Keisler}, {Malkan}, {Marrone}, {McIntyre},
  {Reichardt}, {Sharon}, {Spilker}, {Stalder}, {Stark}, {Vieira}, \&
  {Wei{\ss}}}]{aravena13}
{Aravena} M. {et~al.}, 2013, \mnras, 433, 498

\bibitem[{{Barger} {et~al}\mbox{.}(1998){Barger}, {Cowie}, {Sanders}, {Fulton},
  {Taniguchi}, {Sato}, {Kawara}, \& {Okuda}}]{barger98}
{Barger} A.~J., {Cowie} L.~L., {Sanders} D.~B., {Fulton} E., {Taniguchi} Y.,
  {Sato} Y., {Kawara} K., {Okuda} H., 1998, \nat, 394, 248

\bibitem[{{Baugh} {et~al}\mbox{.}(2005){Baugh}, {Lacey}, {Frenk}, {Granato},
  {Silva}, {Bressan}, {Benson}, \& {Cole}}]{baugh05}
{Baugh} C.~M., {Lacey} C.~G., {Frenk} C.~S., {Granato} G.~L., {Silva} L.,
  {Bressan} A., {Benson} A.~J., {Cole} S., 2005, \mnras, 356, 1191

\bibitem[{{Benson}(2012)}]{benson12b}
{Benson} A.~J., 2012, \na, 17, 175

\bibitem[{{Boreiko} \& {Betz}(1996)}]{boreiko96}
{Boreiko} R.~T., {Betz} A.~L., 1996, \apjl, 467, L113

\bibitem[{{Bothwell} {et~al}\mbox{.}(2013{\natexlab{a}}){Bothwell}, {Aguirre},
  {Chapman}, {Marrone}, {Vieira}, {Ashby}, {Aravena}, {Benson}, {Bock},
  {Bradford}, {Brodwin}, {Carlstrom}, {Crawford}, {de Breuck}, {Downes},
  {Fassnacht}, {Gonzalez}, {Greve}, {Gullberg}, {Hezaveh}, {Holder},
  {Holzapfel}, {Ibar}, {Ivison}, {Kamenetzky}, {Keisler}, {Lupu}, {Ma},
  {Malkan}, {McIntyre}, {Murphy}, {Nguyen}, {Reichardt}, {Rosenman}, {Spilker},
  {Stalder}, {Stark}, {Strandet}, {Vernet}, {Wei{\ss}}, \&
  {Welikala}}]{bothwell13b}
{Bothwell} M.~S. {et~al.}, 2013{\natexlab{a}}, \apj, 779, 67

\bibitem[{{Bothwell} {et~al}\mbox{.}(2013{\natexlab{b}}){Bothwell}, {Smail},
  {Chapman}, {Genzel}, {Ivison}, {Tacconi}, {Alaghband-Zadeh}, {Bertoldi},
  {Blain}, {Casey}, {Cox}, {Greve}, {Lutz}, {Neri}, {Omont}, \&
  {Swinbank}}]{bothwell13}
{Bothwell} M.~S. {et~al.}, 2013{\natexlab{b}}, \mnras, 429, 3047

\bibitem[{{Braine} {et~al}\mbox{.}(2012){Braine}, {Gratier}, {Kramer},
  {Israel}, {van der Tak}, {Mookerjea}, {Boquien}, {Tabatabaei}, {van der
  Werf}, \& {Henkel}}]{braine12}
{Braine} J. {et~al.}, 2012, \aap, 544, A55

\bibitem[{{Brauher} {et~al}\mbox{.}(2008){Brauher}, {Dale}, \&
  {Helou}}]{brauher08}
{Brauher} J.~R., {Dale} D.~A., {Helou} G., 2008, \apjs, 178, 280

\bibitem[{{Brisbin} {et~al}\mbox{.}(2015){Brisbin}, {Ferkinhoff}, {Nikola},
  {Parshley}, {Stacey}, {Spoon}, {Hailey-Dunsheath}, \& {Verma}}]{brisbin14}
{Brisbin} D., {Ferkinhoff} C., {Nikola} T., {Parshley} S., {Stacey} G.~J.,
  {Spoon} H., {Hailey-Dunsheath} S., {Verma} A., 2015, \apj, 799, 13

\bibitem[{{Carilli} {et~al}\mbox{.}(2002{\natexlab{a}}){Carilli}, {Cox},
  {Bertoldi}, {Menten}, {Omont}, {Djorgovski}, {Petric}, {Beelen}, {Isaak}, \&
  {McMahon}}]{carilli02b}
{Carilli} C.~L. {et~al.}, 2002{\natexlab{a}}, \apj, 575, 145

\bibitem[{{Carilli} {et~al}\mbox{.}(2002{\natexlab{b}}){Carilli}, {Kohno},
  {Kawabe}, {Ohta}, {Henkel}, {Menten}, {Yun}, {Petric}, \&
  {Tutui}}]{carilli02a}
{Carilli} C.~L. {et~al.}, 2002{\natexlab{b}}, \aj, 123, 1838

\bibitem[{{Carilli} \& {Walter}(2013)}]{carilli13}
{Carilli} C.~L., {Walter} F., 2013, \araa, 51, 105

\bibitem[{{Carlstrom} {et~al}\mbox{.}(2011){Carlstrom}, {Ade}, {Aird},
  {Benson}, {Bleem}, {Busetti}, {Chang}, {Chauvin}, {Cho}, {Crawford},
  {Crites}, {Dobbs}, {Halverson}, {Heimsath}, {Holzapfel}, {Hrubes}, {Joy},
  {Keisler}, {Lanting}, {Lee}, {Leitch}, {Leong}, {Lu}, {Lueker}, {Luong-van},
  {McMahon}, {Mehl}, {Meyer}, {Mohr}, {Montroy}, {Padin}, {Plagge}, {Pryke},
  {Ruhl}, {Schaffer}, {Schwan}, {Shirokoff}, {Spieler}, {Staniszewski},
  {Stark}, {Tucker}, {Vanderlinde}, {Vieira}, \& {Williamson}}]{carlstrom11}
{Carlstrom} J.~E. {et~al.}, 2011, \pasp, 123, 568

\bibitem[{{Carlstrom} \& {Kronberg}(1991)}]{carlstrom91}
{Carlstrom} J.~E., {Kronberg} P.~P., 1991, \apj, 366, 422

\bibitem[{{Casey} {et~al}\mbox{.}(2014){Casey}, {Narayanan}, \&
  {Cooray}}]{casey14}
{Casey} C.~M., {Narayanan} D., {Cooray} A., 2014, \physrep, 541, 45

\bibitem[{{Chapman} {et~al}\mbox{.}(2005){Chapman}, {Blain}, {Smail}, \&
  {Ivison}}]{chapman05}
{Chapman} S.~C., {Blain} A.~W., {Smail} I., {Ivison} R.~J., 2005, \apj, 622,
  772

\bibitem[{{Colbert} {et~al}\mbox{.}(1999){Colbert}, {Malkan}, {Clegg}, {Cox},
  {Fischer}, {Lord}, {Luhman}, {Satyapal}, {Smith}, {Spinoglio}, {Stacey}, \&
  {Unger}}]{colbert99}
{Colbert} J.~W. {et~al.}, 1999, \apj, 511, 721

\bibitem[{{Coppin} {et~al}\mbox{.}(2010){Coppin}, {Chapman}, {Smail},
  {Swinbank}, {Walter}, {Wardlow}, {Weiss}, {Alexander}, {Brandt},
  {Dannerbauer}, {De Breuck}, {Dickinson}, {Dunlop}, {Edge}, {Emonts}, {Greve},
  {Huynh}, {Ivison}, {Knudsen}, {Menten}, {Schinnerer}, \& {van der
  Werf}}]{coppin10}
{Coppin} K.~E.~K. {et~al.}, 2010, \mnras, 407, L103

\bibitem[{{Cox} {et~al}\mbox{.}(2011){Cox}, {Krips}, {Neri}, {Omont},
  {G{\"u}sten}, {Menten}, {Wyrowski}, {Wei{\ss}}, {Beelen}, {Gurwell},
  {Dannerbauer}, {Ivison}, {Negrello}, {Aretxaga}, {Hughes}, {Auld}, {Baes},
  {Blundell}, {Buttiglione}, {Cava}, {Cooray}, {Dariush}, {Dunne}, {Dye},
  {Eales}, {Frayer}, {Fritz}, {Gavazzi}, {Hopwood}, {Ibar}, {Jarvis}, {Maddox},
  {Micha{\l}owski}, {Pascale}, {Pohlen}, {Rigby}, {Smith}, {Swinbank}, {Temi},
  {Valtchanov}, {van der Werf}, \& {de Zotti}}]{cox11}
{Cox} P. {et~al.}, 2011, \apj, 740, 63

\bibitem[{{Crawford} {et~al}\mbox{.}(1985){Crawford}, {Genzel}, {Townes}, \&
  {Watson}}]{crawford85}
{Crawford} M.~K., {Genzel} R., {Townes} C.~H., {Watson} D.~M., 1985, \apj, 291,
  755

\bibitem[{{Crawford} {et~al}\mbox{.}(1986){Crawford}, {Lugten}, {Fitelson},
  {Genzel}, \& {Melnick}}]{crawford86}
{Crawford} M.~K., {Lugten} J.~B., {Fitelson} W., {Genzel} R., {Melnick} G.,
  1986, \apjl, 303, L57

\bibitem[{{Croxall} {et~al}\mbox{.}(2012){Croxall}, {Smith}, {Wolfire},
  {Roussel}, {Sandstrom}, {Draine}, {Aniano}, {Dale}, {Armus}, {Beir{\~a}o},
  {Helou}, {Bolatto}, {Appleton}, {Brandl}, {Calzetti}, {Crocker}, {Galametz},
  {Groves}, {Hao}, {Hunt}, {Johnson}, {Kennicutt}, {Koda}, {Krause}, {Li},
  {Meidt}, {Murphy}, {Rahman}, {Rix}, {Sauvage}, {Schinnerer}, {Walter}, \&
  {Wilson}}]{croxall12}
{Croxall} K.~V. {et~al.}, 2012, \apj, 747, 81

\bibitem[{{Dalgarno} \& {McCray}(1972)}]{dalgarno72}
{Dalgarno} A., {McCray} R.~A., 1972, \araa, 10, 375

\bibitem[{{De Breuck} {et~al}\mbox{.}(2011){De Breuck}, {Maiolino}, {Caselli},
  {Coppin}, {Hailey-Dunsheath}, \& {Nagao}}]{debreuck11}
{De Breuck} C., {Maiolino} R., {Caselli} P., {Coppin} K., {Hailey-Dunsheath}
  S., {Nagao} T., 2011, \aap, 530, L8

\bibitem[{{De Breuck} {et~al}\mbox{.}(2014){De Breuck}, {Williams}, {Swinbank},
  {Caselli}, {Coppin}, {Davis}, {Maiolino}, {Nagao}, {Smail}, {Walter},
  {Wei{\ss}}, \& {Zwaan}}]{debreuck14}
{De Breuck} C. {et~al.}, 2014, \aap, 565, A59

\bibitem[{{De Looze} {et~al}\mbox{.}(2014){De Looze}, {Cormier},
  {Lebouteiller}, {Madden}, {Baes}, {Bendo}, {Boquien}, {Boselli}, {Clements},
  {Cortese}, {Cooray}, {Galametz}, {Galliano}, {Graci{\'a}-Carpio}, {Isaak},
  {Karczewski}, {Parkin}, {Pellegrini}, {R{\'e}my-Ruyer}, {Spinoglio}, {Smith},
  \& {Sturm}}]{delooze14}
{De Looze} I. {et~al.}, 2014, \aap, 568, A62

\bibitem[{{Decarli} {et~al}\mbox{.}(2014){Decarli}, {Walter}, {Carilli},
  {Bertoldi}, {Cox}, {Ferkinhoff}, {Groves}, {Maiolino}, {Neri}, {Riechers}, \&
  {Weiss}}]{decarli14}
{Decarli} R. {et~al.}, 2014, \apjl, 782, L17

\bibitem[{{D{\'{\i}}az-Santos} {et~al}\mbox{.}(2014){D{\'{\i}}az-Santos},
  {Armus}, {Charmandaris}, {Stacey}, {Murphy}, {Haan}, {Stierwalt}, {Malhotra},
  {Appleton}, {Inami}, {Magdis}, {Elbaz}, {Evans}, {Mazzarella}, {Surace}, {van
  der Werf}, {Xu}, {Lu}, {Meijerink}, {Howell}, {Petric}, {Veilleux}, \&
  {Sanders}}]{diazsantos14}
{D{\'{\i}}az-Santos} T. {et~al.}, 2014, \apjl, 788, L17

\bibitem[{{D{\'{\i}}az-Santos} {et~al}\mbox{.}(2013){D{\'{\i}}az-Santos},
  {Armus}, {Charmandaris}, {Stierwalt}, {Murphy}, {Haan}, {Inami}, {Malhotra},
  {Meijerink}, {Stacey}, {Petric}, {Evans}, {Veilleux}, {van der Werf}, {Lord},
  {Lu}, {Howell}, {Appleton}, {Mazzarella}, {Surace}, {Xu}, {Schulz},
  {Sanders}, {Bridge}, {Chan}, {Frayer}, {Iwasawa}, {Melbourne}, \&
  {Sturm}}]{diazsantos13}
{D{\'{\i}}az-Santos} T. {et~al.}, 2013, \apj, 774, 68

\bibitem[{{Downes} \& {Solomon}(1998)}]{downes98}
{Downes} D., {Solomon} P.~M., 1998, \apj, 507, 615

\bibitem[{{Ellis} \& {Werner}(1984)}]{ellis84}
{Ellis}, Jr. H.~B., {Werner} M.~W., 1984, in Bulletin of the American
  Astronomical Society, Vol.~16, Bulletin of the American Astronomical Society,
  p. 463

\bibitem[{{Evans} {et~al}\mbox{.}(1996){Evans}, {Sanders}, {Mazzarella},
  {Solomon}, {Downes}, {Kramer}, \& {Radford}}]{evans96}
{Evans} A.~S., {Sanders} D.~B., {Mazzarella} J.~M., {Solomon} P.~M., {Downes}
  D., {Kramer} C., {Radford} S.~J.~E., 1996, \apj, 457, 658

\bibitem[{{Farrah} {et~al}\mbox{.}(2013){Farrah}, {Lebouteiller}, {Spoon},
  {Bernard-Salas}, {Pearson}, {Rigopoulou}, {Smith}, {Gonz{\'a}lez-Alfonso},
  {Clements}, {Efstathiou}, {Cormier}, {Afonso}, {Petty}, {Harris}, {Hurley},
  {Borys}, {Verma}, {Cooray}, \& {Salvatelli}}]{farrah13}
{Farrah} D. {et~al.}, 2013, \apj, 776, 38

\bibitem[{{Ferkinhoff} {et~al}\mbox{.}(2011){Ferkinhoff}, {Brisbin}, {Nikola},
  {Parshley}, {Stacey}, {Phillips}, {Falgarone}, {Benford}, {Staguhn}, \&
  {Tucker}}]{ferkinhoff11}
{Ferkinhoff} C. {et~al.}, 2011, \apjl, 740, L29

\bibitem[{{Fixsen} {et~al}\mbox{.}(1999){Fixsen}, {Bennett}, \&
  {Mather}}]{fixsen99}
{Fixsen} D.~J., {Bennett} C.~L., {Mather} J.~C., 1999, \apj, 526, 207

\bibitem[{{Frayer} {et~al}\mbox{.}(2011){Frayer}, {Harris}, {Baker}, {Ivison},
  {Smail}, {Negrello}, {Maddalena}, {Aretxaga}, {Baes}, {Birkinshaw},
  {Bonfield}, {Burgarella}, {Buttiglione}, {Cava}, {Clements}, {Cooray},
  {Dannerbauer}, {Dariush}, {De Zotti}, {Dunlop}, {Dunne}, {Dye}, {Eales},
  {Fritz}, {Gonzalez-Nuevo}, {Herranz}, {Hopwood}, {Hughes}, {Ibar}, {Jarvis},
  {Lagache}, {Leeuw}, {Lopez-Caniego}, {Maddox}, {Micha{\l}owski}, {Omont},
  {Pohlen}, {Rigby}, {Rodighiero}, {Scott}, {Serjeant}, {Smith}, {Swinbank},
  {Temi}, {Thompson}, {Valtchanov}, {van der Werf}, \& {Verma}}]{frayer11}
{Frayer} D.~T. {et~al.}, 2011, \apjl, 726, L22

\bibitem[{{Frayer} {et~al}\mbox{.}(2008){Frayer}, {Koda}, {Pope}, {Huynh},
  {Chary}, {Scott}, {Dickinson}, {Bock}, {Carpenter}, {Hawkins}, {Hodges},
  {Lamb}, {Plambeck}, {Pound}, {Scott}, {Scoville}, \& {Woody}}]{frayer08}
{Frayer} D.~T. {et~al.}, 2008, \apjl, 680, L21

\bibitem[{{George} {et~al}\mbox{.}(2013){George}, {Ivison}, {Hopwood},
  {Riechers}, {Bussmann}, {Cox}, {Dye}, {Krips}, {Negrello}, {Neri},
  {Serjeant}, {Valtchanov}, {Baes}, {Bourne}, {Clements}, {De Zotti}, {Dunne},
  {Eales}, {Ibar}, {Maddox}, {Smith}, {Valiante}, \& {van der Werf}}]{george13}
{George} R.~D. {et~al.}, 2013, \mnras, 436, L99

\bibitem[{{Goldreich} \& {Kwan}(1974)}]{goldreich74}
{Goldreich} P., {Kwan} J., 1974, \apj, 189, 441

\bibitem[{{Goldsmith} {et~al}\mbox{.}(2012){Goldsmith}, {Langer}, {Pineda}, \&
  {Velusamy}}]{goldsmith12}
{Goldsmith} P.~F., {Langer} W.~D., {Pineda} J.~L., {Velusamy} T., 2012, \apjs,
  203, 13

\bibitem[{{Graci{\'a}-Carpio} {et~al}\mbox{.}(2011){Graci{\'a}-Carpio},
  {Sturm}, {Hailey-Dunsheath}, {Fischer}, {Contursi}, {Poglitsch}, {Genzel},
  {Gonz{\'a}lez-Alfonso}, {Sternberg}, {Verma}, {Christopher}, {Davies},
  {Feuchtgruber}, {de Jong}, {Lutz}, \& {Tacconi}}]{graciacarpio11}
{Graci{\'a}-Carpio} J. {et~al.}, 2011, \apjl, 728, L7

\bibitem[{{Graf} {et~al}\mbox{.}(2012){Graf}, {Simon}, {Stutzki}, {Colgan},
  {Guan}, {G{\"u}sten}, {Hartogh}, {Honingh}, \& {H{\"u}bers}}]{graf12}
{Graf} U.~U. {et~al.}, 2012, \aap, 542, L16

\bibitem[{{Greve} {et~al}\mbox{.}(2005){Greve}, {Bertoldi}, {Smail}, {Neri},
  {Chapman}, {Blain}, {Ivison}, {Genzel}, {Omont}, {Cox}, {Tacconi}, \&
  {Kneib}}]{greve05}
{Greve} T.~R. {et~al.}, 2005, \mnras, 359, 1165

\bibitem[{{Greve} {et~al}\mbox{.}(2012){Greve}, {Vieira}, {Wei{\ss}},
  {Aguirre}, {Aird}, {Ashby}, {Benson}, {Bleem}, {Bradford}, {Brodwin},
  {Carlstrom}, {Chang}, {Chapman}, {Crawford}, {de Breuck}, {de Haan}, {Dobbs},
  {Downes}, {Fassnacht}, {Fazio}, {George}, {Gladders}, {Gonzalez},
  {Halverson}, {Hezaveh}, {High}, {Holder}, {Holzapfel}, {Hoover}, {Hrubes},
  {Johnson}, {Keisler}, {Knox}, {Lee}, {Leitch}, {Lueker}, {Luong-Van},
  {Malkan}, {Marrone}, {McIntyre}, {McMahon}, {Mehl}, {Menten}, {Meyer},
  {Montroy}, {Murphy}, {Natoli}, {Padin}, {Plagge}, {Pryke}, {Reichardt},
  {Rest}, {Rosenman}, {Ruel}, {Ruhl}, {Schaffer}, {Sharon}, {Shaw},
  {Shirokoff}, {Stalder}, {Stanford}, {Staniszewski}, {Stark}, {Story},
  {Vanderlinde}, {Walsh}, {Welikala}, \& {Williamson}}]{greve12}
{Greve} T.~R. {et~al.}, 2012, \apj, 756, 101

\bibitem[{{Griffin} {et~al}\mbox{.}(2010){Griffin}, {Abergel}, {Abreu}, {Ade},
  {Andr{\'e}}, {Augueres}, {Babbedge}, {Bae}, {Baillie}, {Baluteau}, {Barlow},
  {Bendo}, {Benielli}, {Bock}, {Bonhomme}, {Brisbin}, {Brockley-Blatt},
  {Caldwell}, {Cara}, {Castro-Rodriguez}, {Cerulli}, {Chanial}, {Chen},
  {Clark}, {Clements}, {Clerc}, {Coker}, {Communal}, {Conversi}, {Cox},
  {Crumb}, {Cunningham}, {Daly}, {Davis}, {de Antoni}, {Delderfield}, {Devin},
  {di Giorgio}, {Didschuns}, {Dohlen}, {Donati}, {Dowell}, {Dowell}, {Duband},
  {Dumaye}, {Emery}, {Ferlet}, {Ferrand}, {Fontignie}, {Fox}, {Franceschini},
  {Frerking}, {Fulton}, {Garcia}, {Gastaud}, {Gear}, {Glenn}, {Goizel},
  {Griffin}, {Grundy}, {Guest}, {Guillemet}, {Hargrave}, {Harwit}, {Hastings},
  {Hatziminaoglou}, {Herman}, {Hinde}, {Hristov}, {Huang}, {Imhof}, {Isaak},
  {Israelsson}, {Ivison}, {Jennings}, {Kiernan}, {King}, {Lange}, {Latter},
  {Laurent}, {Laurent}, {Leeks}, {Lellouch}, {Levenson}, {Li}, {Li},
  {Lilienthal}, {Lim}, {Liu}, {Lu}, {Madden}, {Mainetti}, {Marliani}, {McKay},
  {Mercier}, {Molinari}, {Morris}, {Moseley}, {Mulder}, {Mur}, {Naylor},
  {Nguyen}, {O'Halloran}, {Oliver}, {Olofsson}, {Olofsson}, {Orfei}, {Page},
  {Pain}, {Panuzzo}, {Papageorgiou}, {Parks}, {Parr-Burman}, {Pearce},
  {Pearson}, {P{\'e}rez-Fournon}, {Pinsard}, {Pisano}, {Podosek}, {Pohlen},
  {Polehampton}, {Pouliquen}, {Rigopoulou}, {Rizzo}, {Roseboom}, {Roussel},
  {Rowan-Robinson}, {Rownd}, {Saraceno}, {Sauvage}, {Savage}, {Savini},
  {Sawyer}, {Scharmberg}, {Schmitt}, {Schneider}, {Schulz}, {Schwartz},
  {Shafer}, {Shupe}, {Sibthorpe}, {Sidher}, {Smith}, {Smith}, {Smith},
  {Spencer}, {Stobie}, {Sudiwala}, {Sukhatme}, {Surace}, {Stevens}, {Swinyard},
  {Trichas}, {Tourette}, {Triou}, {Tseng}, {Tucker}, {Turner}, {Vaccari},
  {Valtchanov}, {Vigroux}, {Virique}, {Voellmer}, {Walker}, {Ward}, {Waskett},
  {Weilert}, {Wesson}, {White}, {Whitehouse}, {Wilson}, {Winter}, {Woodcraft},
  {Wright}, {Xu}, {Zavagno}, {Zemcov}, {Zhang}, \& {Zonca}}]{griffin10}
{Griffin} M.~J. {et~al.}, 2010, \aap, 518, L3

\bibitem[{{Guilloteau} {et~al}\mbox{.}(1999){Guilloteau}, {Omont}, {Cox},
  {McMahon}, \& {Petitjean}}]{guilloteau99}
{Guilloteau} S., {Omont} A., {Cox} P., {McMahon} R.~G., {Petitjean} P., 1999,
  \aap, 349, 363

\bibitem[{{Hailey-Dunsheath} {et~al}\mbox{.}(2010){Hailey-Dunsheath}, {Nikola},
  {Stacey}, {Oberst}, {Parshley}, {Benford}, {Staguhn}, \&
  {Tucker}}]{hailey-dunsheath10}
{Hailey-Dunsheath} S., {Nikola} T., {Stacey} G.~J., {Oberst} T.~E., {Parshley}
  S.~C., {Benford} D.~J., {Staguhn} J.~G., {Tucker} C.~E., 2010, \apjl, 714,
  L162

\bibitem[{{Hainline} {et~al}\mbox{.}(2011){Hainline}, {Blain}, {Smail},
  {Alexander}, {Armus}, {Chapman}, \& {Ivison}}]{hainline11}
{Hainline} L.~J., {Blain} A.~W., {Smail} I., {Alexander} D.~M., {Armus} L.,
  {Chapman} S.~C., {Ivison} R.~J., 2011, \apj, 740, 96

\bibitem[{{Hayward} {et~al}\mbox{.}(2013){Hayward}, {Narayanan}, {Kere{\v s}},
  {Jonsson}, {Hopkins}, {Cox}, \& {Hernquist}}]{hayward13}
{Hayward} C.~C., {Narayanan} D., {Kere{\v s}} D., {Jonsson} P., {Hopkins}
  P.~F., {Cox} T.~J., {Hernquist} L., 2013, \mnras, 428, 2529

\bibitem[{{Helou} {et~al}\mbox{.}(1988){Helou}, {Khan}, {Malek}, \&
  {Boehmer}}]{helou88}
{Helou} G., {Khan} I.~R., {Malek} L., {Boehmer} L., 1988, \apjs, 68, 151

\bibitem[{{Heyminck} {et~al}\mbox{.}(2006){Heyminck}, {Kasemann}, {G{\"u}sten},
  {de Lange}, \& {Graf}}]{heyminck06}
{Heyminck} S., {Kasemann} C., {G{\"u}sten} R., {de Lange} G., {Graf} U.~U.,
  2006, \aap, 454, L21

\bibitem[{{Hezaveh} \& {Holder}(2011)}]{hezaveh11}
{Hezaveh} Y.~D., {Holder} G.~P., 2011, \apj, 734, 52

\bibitem[{{Hezaveh} {et~al}\mbox{.}(2013){Hezaveh}, {Marrone}, {Fassnacht},
  {Spilker}, {Vieira}, {Aguirre}, {Aird}, {Aravena}, {Ashby}, {Bayliss},
  {Benson}, {Bleem}, {Bothwell}, {Brodwin}, {Carlstrom}, {Chang}, {Chapman},
  {Crawford}, {Crites}, {De Breuck}, {de Haan}, {Dobbs}, {Fomalont}, {George},
  {Gladders}, {Gonzalez}, {Greve}, {Halverson}, {High}, {Holder}, {Holzapfel},
  {Hoover}, {Hrubes}, {Husband}, {Hunter}, {Keisler}, {Lee}, {Leitch},
  {Lueker}, {Luong-Van}, {Malkan}, {McIntyre}, {McMahon}, {Mehl}, {Menten},
  {Meyer}, {Mocanu}, {Murphy}, {Natoli}, {Padin}, {Plagge}, {Reichardt},
  {Rest}, {Ruel}, {Ruhl}, {Sharon}, {Schaffer}, {Shaw}, {Shirokoff}, {Stalder},
  {Staniszewski}, {Stark}, {Story}, {Vanderlinde}, {Wei{\ss}}, {Welikala}, \&
  {Williamson}}]{hezaveh13}
{Hezaveh} Y.~D. {et~al.}, 2013, \apj, 767, 132

\bibitem[{{Hezaveh} {et~al}\mbox{.}(2012){Hezaveh}, {Marrone}, \&
  {Holder}}]{hezaveh12a}
{Hezaveh} Y.~D., {Marrone} D.~P., {Holder} G.~P., 2012, \apj, 761, 20

\bibitem[{{Hollenbach} \& {Tielens}(1999)}]{hollenbach99}
{Hollenbach} D.~J., {Tielens} A.~G.~G.~M., 1999, Reviews of Modern Physics, 71,
  173

\bibitem[{{Hughes} {et~al}\mbox{.}(1998){Hughes}, {Serjeant}, {Dunlop},
  {Rowan-Robinson}, {Blain}, {Mann}, {Ivison}, {Peacock}, {Efstathiou}, {Gear},
  {Oliver}, {Lawrence}, {Longair}, {Goldschmidt}, \& {Jenness}}]{hughes98}
{Hughes} D.~H. {et~al.}, 1998, \nat, 394, 241

\bibitem[{{Huynh} {et~al}\mbox{.}(2014){Huynh}, {Kimball}, {Norris}, {Smail},
  {Chow}, {Coppin}, {Emonts}, {Ivison}, {Smol{\v c}i{\'c}}, \&
  {Swinbank}}]{huynh14}
{Huynh} M.~T. {et~al.}, 2014, \mnras, 443, L54

\bibitem[{{Iono} {et~al}\mbox{.}(2006){Iono}, {Yun}, {Elvis}, {Peck}, {Ho},
  {Wilner}, {Hunter}, {Matsushita}, \& {Muller}}]{iono06}
{Iono} D. {et~al.}, 2006, \apjl, 645, L97

\bibitem[{{Ivison} {et~al}\mbox{.}(2010){Ivison}, {Swinbank}, {Swinyard},
  {Smail}, {Pearson}, {Rigopoulou}, {Polehampton}, {Baluteau}, {Barlow},
  {Blain}, {Bock}, {Clements}, {Coppin}, {Cooray}, {Danielson}, {Dwek}, {Edge},
  {Franceschini}, {Fulton}, {Glenn}, {Griffin}, {Isaak}, {Leeks}, {Lim},
  {Naylor}, {Oliver}, {Page}, {P{\'e}rez Fournon}, {Rowan-Robinson}, {Savini},
  {Scott}, {Spencer}, {Valtchanov}, {Vigroux}, \& {Wright}}]{ivison10}
{Ivison} R.~J. {et~al.}, 2010, \aap, 518, L35+

\bibitem[{{Joy} {et~al}\mbox{.}(1987){Joy}, {Lester}, \& {Harvey}}]{joy87}
{Joy} M., {Lester} D.~F., {Harvey} P.~M., 1987, \apj, 319, 314

\bibitem[{{Kaufman} {et~al}\mbox{.}(1999){Kaufman}, {Wolfire}, {Hollenbach}, \&
  {¤}}]{kaufman99}
{Kaufman} M.~J., {Wolfire} M.~G., {Hollenbach} D.~J., {¤} M.~L., 1999, \apj,
  527, 795

\bibitem[{{Klein} {et~al}\mbox{.}(2006){Klein}, {Philipp}, {Kr{\"a}mer},
  {Kasemann}, {G{\"u}sten}, \& {Menten}}]{klein06}
{Klein} B., {Philipp} S.~D., {Kr{\"a}mer} I., {Kasemann} C., {G{\"u}sten} R.,
  {Menten} K.~M., 2006, \aap, 454, L29

\bibitem[{{Komatsu} {et~al}\mbox{.}(2011){Komatsu}, {Smith}, {Dunkley},
  {Bennett}, {Gold}, {Hinshaw}, {Jarosik}, {Larson}, {Nolta}, {Page},
  {Spergel}, {Halpern}, {Hill}, {Kogut}, {Limon}, {Meyer}, {Odegard}, {Tucker},
  {Weiland}, {Wollack}, \& {Wright}}]{komatsu11}
{Komatsu} E. {et~al.}, 2011, \apjs, 192, 18

\bibitem[{{Lacey} {et~al}\mbox{.}(2010){Lacey}, {Baugh}, {Frenk}, {Benson},
  {Orsi}, {Silva}, {Granato}, \& {Bressan}}]{lacey10}
{Lacey} C.~G., {Baugh} C.~M., {Frenk} C.~S., {Benson} A.~J., {Orsi} A., {Silva}
  L., {Granato} G.~L., {Bressan} A., 2010, \mnras, 405, 2

\bibitem[{{Lavalley} {et~al}\mbox{.}(1992){Lavalley}, {Isobe}, \&
  {Feigelson}}]{lavalley92}
{Lavalley} M.~P., {Isobe} T., {Feigelson} E.~D., 1992, in Bulletin of the
  American Astronomical Society, Vol.~24, Bulletin of the American Astronomical
  Society, pp. 839--840

\bibitem[{{Lord} {et~al}\mbox{.}(1996){Lord}, {Hollenbach}, {Haas}, {Rubin},
  {Colgan}, \& {Erickson}}]{lord96}
{Lord} S.~D., {Hollenbach} D.~J., {Haas} M.~R., {Rubin} R.~H., {Colgan}
  S.~W.~J., {Erickson} E.~F., 1996, \apj, 465, 703

\bibitem[{{Luhman} {et~al}\mbox{.}(1998){Luhman}, {Satyapal}, {Fischer},
  {Wolfire}, {Cox}, {Lord}, {Smith}, {Stacey}, \& {Unger}}]{luhman98}
{Luhman} M.~L. {et~al.}, 1998, \apjl, 504, L11

\bibitem[{{Luhman} {et~al}\mbox{.}(2003){Luhman}, {Satyapal}, {Fischer},
  {Wolfire}\, {Sturm}, {Dudley}, {Lutz}, \& {Genzel}}]{luhman03}
{Luhman} M.~L., {Satyapal} S., {Fischer} J., {Wolfire}\ M.~G., {Sturm} E.,
  {Dudley} C.~C., {Lutz} D., {Genzel} R., 2003, \apj, 594, 758

\bibitem[{{Madden} {et~al}\mbox{.}(1997){Madden}, {Geis}, {Genzel}, {Nikola},
  {Poglitsch}, {Stacey}, \& {Townes}}]{madden97}
{Madden} S., {Geis} N., {Genzel} R., {Nikola} T., {Poglitsch} A., {Stacey}
  G.~J., {Townes} C., 1997, in ESA Special Publication, Vol. 401, The Far
  Infrared and Submillimetre Universe., {Wilson} A., ed., p. 111

\bibitem[{{Maiolino} {et~al}\mbox{.}(2009){Maiolino}, {Caselli}, {Nagao},
  {Walmsley}, {De Breuck}, \& {Meneghetti}}]{maiolino09}
{Maiolino} R., {Caselli} P., {Nagao} T., {Walmsley} M., {De Breuck} C.,
  {Meneghetti} M., 2009, \aap, 500, L1

\bibitem[{{Maiolino} {et~al}\mbox{.}(2005){Maiolino}, {Cox}, {Caselli},
  {Beelen}, {Bertoldi}, {Carilli}, {Kaufman}, {Menten}, {Nagao}, {Omont},
  {Wei{\ss}}, {Walmsley}, \& {Walter}}]{maiolino05}
{Maiolino} R. {et~al.}, 2005, \aap, 440, L51

\bibitem[{{Malhotra} {et~al}\mbox{.}(1997){Malhotra}, {Helou}, {Stacey},
  {Hollenbach}, {Lord}, {Beichman}, {Dinerstein}, {Hunter}, {Lo}, {Lu},
  {Rubin}, {Silbermann}, {Thronson}, \& {Werner}}]{malhotra97}
{Malhotra} S. {et~al.}, 1997, \apjl, 491, L27

\bibitem[{{Malhotra} {et~al}\mbox{.}(2001){Malhotra}, {Kaufman}, {Hollenbach},
  {Helou}, {Rubin}, {Brauher}, {Dale}, {Lu}, {Lord}, {Stacey}, {Contursi},
  {Hunter}, \& {Dinerstein}}]{malhotra01}
{Malhotra} S. {et~al.}, 2001, \apj, 561, 766

\bibitem[{{Maloney} \& {Black}(1988)}]{maloney88}
{Maloney} P., {Black} J.~H., 1988, \apj, 325, 389

\bibitem[{{Mashian} {et~al}\mbox{.}(2013){Mashian}, {Sternberg}, \&
  {Loeb}}]{mashian13}
{Mashian} N., {Sternberg} A., {Loeb} A., 2013, \mnras, 435, 2407

\bibitem[{{Meijerink} {et~al}\mbox{.}(2007){Meijerink}, {Spaans}, \&
  {Israel}}]{meijerink07}
{Meijerink} R., {Spaans} M., {Israel} F.~P., 2007, \aap, 461, 793

\bibitem[{{Micha{\l}owski} {et~al}\mbox{.}(2012){Micha{\l}owski}, {Dunlop},
  {Cirasuolo}, {Hjorth}, {Hayward}, \& {Watson}}]{michalowski12}
{Micha{\l}owski} M.~J., {Dunlop} J.~S., {Cirasuolo} M., {Hjorth} J., {Hayward}
  C.~C., {Watson} D., 2012, \aap, 541, A85

\bibitem[{{Mittal} {et~al}\mbox{.}(2011){Mittal}, {O'Dea}, {Ferland}, {Oonk},
  {Edge}, {Canning}, {Russell}, {Baum}, {B{\"o}hringer}, {Combes}, {Donahue},
  {Fabian}, {Hatch}, {Hoffer}, {Johnstone}, {McNamara}, {Salom{\'e}}, \&
  {Tremblay}}]{mittal11}
{Mittal} R. {et~al.}, 2011, \mnras, 418, 2386

\bibitem[{{Mocanu} {et~al}\mbox{.}(2013){Mocanu}, {Crawford}, {Vieira}, {Aird},
  {Aravena}, {Austermann}, {Benson}, {B{\'e}thermin}, {Bleem}, {Bothwell},
  {Carlstrom}, {Chang}, {Chapman}, {Cho}, {Crites}, {de Haan}, {Dobbs},
  {Everett}, {George}, {Halverson}, {Harrington}, {Hezaveh}, {Holder},
  {Holzapfel}, {Hoover}, {Hrubes}, {Keisler}, {Knox}, {Lee}, {Leitch},
  {Lueker}, {Luong-Van}, {Marrone}, {McMahon}, {Mehl}, {Meyer}, {Mohr},
  {Montroy}, {Natoli}, {Padin}, {Plagge}, {Pryke}, {Rest}, {Reichardt}, {Ruhl},
  {Sayre}, {Schaffer}, {Shirokoff}, {Spieler}, {Spilker}, {Stalder},
  {Staniszewski}, {Stark}, {Story}, {Switzer}, {Vanderlinde}, \&
  {Williamson}}]{mocanu13}
{Mocanu} L.~M. {et~al.}, 2013, \apj, 779, 61

\bibitem[{{Negishi} {et~al}\mbox{.}(2001){Negishi}, {Onaka}, {Chan}, \&
  {Roellig}}]{negishi01}
{Negishi} T., {Onaka} T., {Chan} K.-W., {Roellig} T.~L., 2001, \aap, 375, 566

\bibitem[{{Neri} {et~al}\mbox{.}(2014){Neri}, {Downes}, {Cox}, \&
  {Walter}}]{neri14}
{Neri} R., {Downes} D., {Cox} P., {Walter} F., 2014, \aap, 562, A35

\bibitem[{{Oberst} {et~al}\mbox{.}(2011){Oberst}, {Parshley}, {Nikola},
  {Stacey}, {L{\"o}hr}, {Lane}, {Stark}, \& {Kamenetzky}}]{oberst11}
{Oberst} T.~E., {Parshley} S.~C., {Nikola} T., {Stacey} G.~J., {L{\"o}hr} A.,
  {Lane} A.~P., {Stark} A.~A., {Kamenetzky} J., 2011, \apj, 739, 100

\bibitem[{{Oberst} {et~al}\mbox{.}(2006){Oberst}, {Parshley}, {Stacey},
  {Nikola}, {L{\"o}hr}, {Harnett}, {Tothill}, {Lane}, {Stark}, \&
  {Tucker}}]{oberst06}
{Oberst} T.~E. {et~al.}, 2006, \apjl, 652, L125

\bibitem[{{Orr} {et~al}\mbox{.}(2014){Orr}, {Pineda}, \& {Goldsmith}}]{orr14}
{Orr} M.~E., {Pineda} J.~L., {Goldsmith} P.~F., 2014, \apj, 795, 26

\bibitem[{{Ossenkopf} {et~al}\mbox{.}(2013){Ossenkopf}, {R{\"o}llig},
  {Neufeld}, {Pilleri}, {Lis}, {Fuente}, {van der Tak}, \&
  {Bergin}}]{ossenkopf13}
{Ossenkopf} V., {R{\"o}llig} M., {Neufeld} D.~A., {Pilleri} P., {Lis} D.~C.,
  {Fuente} A., {van der Tak} F.~F.~S., {Bergin} E., 2013, \aap, 550, A57

\bibitem[{{Penzias} {et~al}\mbox{.}(1972){Penzias}, {Solomon}, {Jefferts}, \&
  {Wilson}}]{penzias72}
{Penzias} A.~A., {Solomon} P.~M., {Jefferts} K.~B., {Wilson} R.~W., 1972,
  \apjl, 174, L43

\bibitem[{{Pilbratt} {et~al}\mbox{.}(2010){Pilbratt}, {Riedinger}, {Passvogel},
  {Crone}, {Doyle}, {Gageur}, {Heras}, {Jewell}, {Metcalfe}, {Ott}, \&
  {Schmidt}}]{pilbratt10}
{Pilbratt} G.~L. {et~al.}, 2010, \aap, 518, L1

\bibitem[{{Rawle} {et~al}\mbox{.}(2014){Rawle}, {Egami}, {Bussmann}, {Gurwell},
  {Ivison}, {Boone}, {Combes}, {Danielson}, {Rex}, {Richard}, {Smail},
  {Swinbank}, {Altieri}, {Blain}, {Clement}, {Dessauges-Zavadsky}, {Edge},
  {Fazio}, {Jones}, {Kneib}, {Omont}, {P{\'e}rez-Gonz{\'a}lez}, {Schaerer},
  {Valtchanov}, {van der Werf}, {Walth}, {Zamojski}, \& {Zemcov}}]{rawle14}
{Rawle} T.~D. {et~al.}, 2014, \apj, 783, 59

\bibitem[{{Rice} {et~al}\mbox{.}(1988){Rice}, {Lonsdale}, {Soifer},
  {Neugebauer}, {Kopan}, {Lloyd}, {de Jong}, \& {Habing}}]{rice88}
{Rice} W., {Lonsdale} C.~J., {Soifer} B.~T., {Neugebauer} G., {Kopan} E.~L.,
  {Lloyd} L.~A., {de Jong} T., {Habing} H.~J., 1988, \apjs, 68, 91

\bibitem[{{Riechers} {et~al}\mbox{.}(2013){Riechers}, {Bradford}, {Clements},
  {Dowell}, {P{\'e}rez-Fournon}, {Ivison}, {Bridge}, {Conley}, {Fu}, {Vieira},
  {Wardlow}, {Calanog}, {Cooray}, {Hurley}, {Neri}, {Kamenetzky}, {Aguirre},
  {Altieri}, {Arumugam}, {Benford}, {B{\'e}thermin}, {Bock}, {Burgarella},
  {Cabrera-Lavers}, {Chapman}, {Cox}, {Dunlop}, {Earle}, {Farrah}, {Ferrero},
  {Franceschini}, {Gavazzi}, {Glenn}, {Solares}, {Gurwell}, {Halpern},
  {Hatziminaoglou}, {Hyde}, {Ibar}, {Kov{\'a}cs}, {Krips}, {Lupu}, {Maloney},
  {Martinez-Navajas}, {Matsuhara}, {Murphy}, {Naylor}, {Nguyen}, {Oliver},
  {Omont}, {Page}, {Petitpas}, {Rangwala}, {Roseboom}, {Scott}, {Smith},
  {Staguhn}, {Streblyanska}, {Thomson}, {Valtchanov}, {Viero}, {Wang},
  {Zemcov}, \& {Zmuidzinas}}]{riechers13}
{Riechers} D.~A. {et~al.}, 2013, \nat, 496, 329

\bibitem[{{Sargsyan} {et~al}\mbox{.}(2012){Sargsyan}, {Lebouteiller},
  {Weedman}, {Spoon}, {Bernard-Salas}, {Engels}, {Stacey}, {Houck}, {Barry},
  {Miles}, \& {Samsonyan}}]{sargsyan12}
{Sargsyan} L. {et~al.}, 2012, \apj, 755, 171

\bibitem[{{Sargsyan} {et~al}\mbox{.}(2014){Sargsyan}, {Samsonyan},
  {Lebouteiller}, {Weedman}, {Barry}, {Bernard-Salas}, {Houck}, \&
  {Spoon}}]{sargsyan14}
{Sargsyan} L., {Samsonyan} A., {Lebouteiller} V., {Weedman} D., {Barry} D.,
  {Bernard-Salas} J., {Houck} J., {Spoon} H., 2014, \apj, 790, 15

\bibitem[{{Smail} {et~al}\mbox{.}(1997){Smail}, {Ivison}, \& {Blain}}]{smail97}
{Smail} I., {Ivison} R.~J., {Blain} A.~W., 1997, \apjl, 490, L5+

\bibitem[{{Solomon} \& {Vanden Bout}(2005)}]{solomon05}
{Solomon} P.~M., {Vanden Bout} P.~A., 2005, \araa, 43, 677

\bibitem[{{Spilker} {et~al}\mbox{.}(2014){Spilker}, {Marrone}, {Aguirre},
  {Aravena}, {Ashby}, {B{\'e}thermin}, {Bradford}, {Bothwell}, {Brodwin},
  {Carlstrom}, {Chapman}, {Crawford}, {de Breuck}, {Fassnacht}, {Gonzalez},
  {Greve}, {Gullberg}, {Hezaveh}, {Holzapfel}, {Husband}, {Ma}, {Malkan},
  {Murphy}, {Reichardt}, {Rotermund}, {Stalder}, {Stark}, {Strandet}, {Vieira},
  {Wei{\ss}}, \& {Welikala}}]{spilker14}
{Spilker} J.~S. {et~al.}, 2014, \apj, 785, 149

\bibitem[{{Stacey} {et~al}\mbox{.}(1991{\natexlab{a}}){Stacey}, {Geis},
  {Genzel}, {Lugten}, {Poglitsch}, {Sternberg}, \& {Townes}}]{stacey91}
{Stacey} G.~J., {Geis} N., {Genzel} R., {Lugten} J.~B., {Poglitsch} A.,
  {Sternberg} A., {Townes} C.~H., 1991{\natexlab{a}}, \apj, 373, 423

\bibitem[{{Stacey} {et~al}\mbox{.}(2010){Stacey}, {Hailey-Dunsheath},
  {Ferkinhoff}, {Parshley}, {Benford}, {Staguhn}, \& {Fiolet}}]{stacey10}
{Stacey} G.~J., {Hailey-Dunsheath} S., {Ferkinhoff}, C. an\ d~{Nikola} T.,
  {Parshley} S.~C., {Benford} D.~J., {Staguhn} J.~G., {Fiolet} N., 2010, \apj,
  724, 957

\bibitem[{{Stacey} {et~al}\mbox{.}(1993){Stacey}, {Jaffe}, {Geis}, {Grenzel},
  {Harris}, {Poglitsch}, {Stutzki}, \& {Townes}}]{stacey93}
{Stacey} G.~J., {Jaffe} D.~T., {Geis} N., {Grenzel} R., {Harris} A.~I.,
  {Poglitsch} A., {Stutzki} J., {Townes} C.~H., 1993, \apj, 404, 219

\bibitem[{{Stacey} {et~al}\mbox{.}(1983){Stacey}, {Smyers}, {Kurtz}, \&
  {Harwit}}]{stacey83}
{Stacey} G.~J., {Smyers} S.~D., {Kurtz} N.~T., {Harwit} M., 1983, \apjl, 265,
  L7

\bibitem[{{Stacey} {et~al}\mbox{.}(1991{\natexlab{b}}){Stacey}, {Townes},
  {Geis}, {Madden}, {Herrmann}, {Genzel}, {Poglitsch}, \&
  {Jackson}}]{stacey91b}
{Stacey} G.~J., {Townes} C.~H., {Geis} N., {Madden} S.~C., {Herrmann} F.,
  {Genzel} R., {Poglitsch} A., {Jackson} J.~M., 1991{\natexlab{b}}, \apjl, 382,
  L37

\bibitem[{{Stutzki} {et~al}\mbox{.}(1988){Stutzki}, {Stacey}, {Genzel},
  {Harris}, {Jaffe}, \& {Lugten}}]{stutzki88}
{Stutzki} J., {Stacey} G.~J., {Genzel} R., {Harris} A.~I., {Jaffe} D.~T.,
  {Lugten} J.~B., 1988, \apj, 332, 379

\bibitem[{{Swinbank} {et~al}\mbox{.}(2011){Swinbank}, {Papadopoulos}, {Cox},
  {Krips}, {Ivison}, {Smail}, {Thomson}, {Neri}, {Richard}, \&
  {Ebeling}}]{swinbank11}
{Swinbank} A.~M. {et~al.}, 2011, \apj, 742, 11

\bibitem[{{Swinbank} {et~al}\mbox{.}(2010){Swinbank}, {Smail}, {Longmore},
  {Harris}, {Baker}, {De Breuck}, {Richard}, {Edge}, {Ivison}, {Blundell},
  {Coppin}, {Cox}, {Gurwell}, {Hainline}, {Krips}, {Lundgren}, {Neri}, {Siana},
  {Siringo}, {Stark}, {Wilner}, \& {Younger}}]{swinbank10}
{Swinbank} A.~M. {et~al.}, 2010, \nat, 464, 733

\bibitem[{{Swinbank} {et~al}\mbox{.}(2012){Swinbank}, {Smail}, {Sobral},
  {Theuns}, {Best}, \& {Geach}}]{swinbank12}
{Swinbank} A.~M., {Smail} I., {Sobral} D., {Theuns} T., {Best} P.~N., {Geach}
  J.~E., 2012, \apj, 760, 130

\bibitem[{{Tacconi} {et~al}\mbox{.}(2010){Tacconi}, {Genzel}, {Neri}, {Cox},
  {Cooper}, {Shapiro}, {Bolatto}, {Bouch{\'e}}, {Bournaud}, {Burkert},
  {Combes}, {Comerford}, {Davis}, {Schreiber}, {Garcia-Burillo},
  {Gracia-Carpio}, {Lutz}, {Naab}, {Omont}, {Shapley}, {Sternberg}, \&
  {Weiner}}]{tacconi10}
{Tacconi} L.~J. {et~al.}, 2010, \nat, 463, 781

\bibitem[{{Valtchanov} {et~al}\mbox{.}(2011){Valtchanov}, {Virdee}, {Ivison},
  {Swinyard}, {van der Werf}, {Rigopoulou}, {da Cunha}, {Lupu}, {Benford},
  {Riechers}, {Smail}, {Jarvis}, {Pearson}, {Gomez}, {Hopwood}, {Altieri},
  {Birkinshaw}, {Coia}, {Conversi}, {Cooray}, {de Zotti}, {Dunne}, {Frayer},
  {Leeuw}, {Marston}, {Negrello}, {Portal}, {Scott}, {Thompson}, {Vaccari},
  {Baes}, {Clements}, {Micha{\l}owski}, {Dannerbauer}, {Serjeant}, {Auld},
  {Buttiglione}, {Cava}, {Dariush}, {Dye}, {Eales}, {Fritz}, {Ibar}, {Maddox},
  {Pascale}, {Pohlen}, {Rigby}, {Rodighiero}, {Smith}, {Temi}, {Carpenter},
  {Bolatto}, {Gurwell}, \& {Vieira}}]{valtchanov11}
{Valtchanov} I. {et~al.}, 2011, \mnras, 415, 3473

\bibitem[{{Venemans} {et~al}\mbox{.}(2012){Venemans}, {McMahon}, {Walter},
  {Decarli}, {Cox}, {Neri}, {Hewett}, {Mortlock}, {Simpson}, \&
  {Warren}}]{venemans12}
{Venemans} B.~P. {et~al.}, 2012, \apjl, 751, L25

\bibitem[{{Vieira} {et~al}\mbox{.}(2010){Vieira}, {Crawford}, {Switzer}, {Ade},
  {Aird}, {Ashby}, {Benson}, {Bleem}, {Brodwin}, {Carlstrom}, {Chang}, {Cho},
  {Crites}, {de Haan}, {Dobbs}, {Everett}, {George}, {Gladders}, {Hall},
  {Halverson}, {High}, {Holder}, {Holzapfel}, {Hrubes}, {Joy}, {Keisler},
  {Knox}, {Lee}, {Leitch}, {Lueker}, {Marrone}, {McIntyre}, {McMahon}, {Mehl},
  {Meyer}, {Mohr}, {Montroy}, {Padin}, {Plagge}, {Pryke}, {Reichardt}, {Ruhl},
  {Schaffer}, {Shaw}, {Shirokoff}, {Spieler}, {Stalder}, {Staniszewski},
  {Stark}, {Vanderlinde}, {Walsh}, {Williamson}, {Yang}, {Zahn}, \&
  {Zenteno}}]{vieira10}
{Vieira} J.~D. {et~al.}, 2010, \apj, 719, 763

\bibitem[{{Vieira} {et~al}\mbox{.}(2013){Vieira}, {Marrone}, {Chapman}, {De
  Breuck}, {Hezaveh}, {Wei{$\beta$}}, {Aguirre}, {Aird}, {Aravena}, {Ashby},
  {Bayliss}, {Benson}, {Biggs}, {Bleem}, {Bock}, {Bothwell}, {Bradford},
  {Brodwin}, {Carlstrom}, {Chang}, {Crawford}, {Crites}, {de Haan}, {Dobbs},
  {Fomalont}, {Fassnacht}, {George}, {Gladders}, {Gonzalez}, {Greve},
  {Gullberg}, {Halverson}, {High}, {Holder}, {Holzapfel}, {Hoover}, {Hrubes},
  {Hunter}, {Keisler}, {Lee}, {Leitch}, {Lueker}, {Luong-van}, {Malkan},
  {McIntyre}, {McMahon}, {Mehl}, {Menten}, {Meyer}, {Mocanu}, {Murphy},
  {Natoli}, {Padin}, {Plagge}, {Reichardt}, {Rest}, {Ruel}, {Ruhl}, {Sharon},
  {Schaffer}, {Shaw}, {Shirokoff}, {Spilker}, {Stalder}, {Staniszewski},
  {Stark}, {Story}, {Vanderlinde}, {Welikala}, \& {Williamson}}]{vieira13}
{Vieira} J.~D. {et~al.}, 2013, \nat, 495, 344

\bibitem[{{Viti} {et~al}\mbox{.}(2013){Viti}, {Bayet}, {Hartquist}, {Bell},
  {Williams}, \& {Banerji}}]{viti13}
{Viti} S., {Bayet} E., {Hartquist} T.~W., {Bell} T.~A., {Williams} D.~A.,
  {Banerji} M., 2013, in Advances in Solid State Physics, Vol.~34, Cosmic Rays
  in Star-Forming Environments, {Torres} D.~F., {Reimer} O., eds., p.~7

\bibitem[{{Wagg} {et~al}\mbox{.}(2010){Wagg}, {Carilli}, {Wilner}, {Cox}, {De
  Breuck}, {Menten}, {Riechers}, \& {Walter}}]{wagg10}
{Wagg} J., {Carilli} C.~L., {Wilner} D.~J., {Cox} P., {De Breuck} C., {Menten}
  K., {Riechers} D.~A., {Walter} F., 2010, \aap, 519, L1

\bibitem[{{Wagg} {et~al}\mbox{.}(2012){Wagg}, {Wiklind}, {Carilli}, {Espada},
  {Peck}, {Riechers}, {Walter}, {Wootten}, {Aravena}, {Barkats}, {Cortes},
  {Hills}, {Hodge}, {Impellizzeri}, {Iono}, {Leroy}, {Mart{\'{\i}}n},
  {Rawlings}, {Maiolino}, {McMahon}, {Scott}, {Villard}, \&
  {Vlahakis}}]{wagg12}
{Wagg} J. {et~al.}, 2012, \apjl, 752, L30

\bibitem[{{Walter} {et~al}\mbox{.}(2012){Walter}, {Decarli}, {Carilli},
  {Bertoldi}, {Cox}, {da Cunha}, {Daddi}, {Dickinson}, {Downes}, {Elbaz},
  {Ellis}, {Hodge}, {Neri}, {Riechers}, {Weiss}, {Bell}, {Dannerbauer},
  {Krips}, {Krumholz}, {Lentati}, {Maiolino}, {Menten}, {Rix}, {Robertson},
  {Spinrad}, {Stark}, \& {Stern}}]{walter12}
{Walter} F. {et~al.}, 2012, \nat, 486, 233

\bibitem[{{Wang} {et~al}\mbox{.}(2013){Wang}, {Wagg}, {Carilli}, {Walter},
  {Lentati}, {Fan}, {Riechers}, {Bertoldi}, {Narayanan}, {Strauss}, {Cox},
  {Omont}, {Menten}, {Knudsen}, {Neri}, \& {Jiang}}]{wang13}
{Wang} R. {et~al.}, 2013, \apj, 773, 44

\bibitem[{{Wei{\ss}} {et~al}\mbox{.}(2013){Wei{\ss}}, {De Breuck}, {Marrone},
  {Vieira}, {Aguirre}, {Aird}, {Aravena}, {Ashby}, {Bayliss}, {Benson},
  {B{\'e}thermin}, {Biggs}, {Bleem}, {Bock}, {Bothwell}, {Bradford}, {Brodwin},
  {Carlstrom}, {Chang}, {Chapman}, {Crawford}, {Crites}, {de Haan}, {Dobbs},
  {Downes}, {Fassnacht}, {George}, {Gladders}, {Gonzalez}, {Greve},
  {Halverson}, {Hezaveh}, {High}, {Holder}, {Holzapfel}, {Hoover}, {Hrubes},
  {Husband}, {Keisler}, {Lee}, {Leitch}, {Lueker}, {Luong-Van}, {Malkan},
  {McIntyre}, {McMahon}, {Mehl}, {Menten}, {Meyer}, {Murphy}, {Padin},
  {Plagge}, {Reichardt}, {Rest}, {Rosenman}, {Ruel}, {Ruhl}, {Schaffer},
  {Shirokoff}, {Spilker}, {Stalder}, {Staniszewski}, {Stark}, {Story},
  {Vanderlinde}, {Welikala}, \& {Williamson}}]{weiss13}
{Wei{\ss}} A. {et~al.}, 2013, \apj, 767, 88

\bibitem[{{Willott} {et~al}\mbox{.}(2013){Willott}, {Omont}, \&
  {Bergeron}}]{willott13}
{Willott} C.~J., {Omont} A., {Bergeron} J., 2013, \apj, 770, 13

\bibitem[{{Wolfire} {et~al}\mbox{.}(1989){Wolfire}, {Hollenbach}, \&
  {Tielens}}]{wolfire89}
{Wolfire} M.~G., {Hollenbach} D., {Tielens} A.~G.~G.~M., 1989, \apj, 344, 770

\bibitem[{{Wolfire} {et~al}\mbox{.}(1993){Wolfire}, {Hollenbach}, \&
  {Tielens}}]{wolfire93}
{Wolfire} M.~G., {Hollenbach} D., {Tielens} A.~G.~G.~M., 1993, \apj, 402, 195

\bibitem[{{Wolfire} {et~al}\mbox{.}(1990){Wolfire}, {Tielens}, \&
  {Hollenbach}}]{wolfire90}
{Wolfire} M.~G., {Tielens} A.~G.~G.~M., {Hollenbach} D., 1990, \apj, 358, 116

\bibitem[{{Young} {et~al}\mbox{.}(1995){Young}, {Xie}, {Tacconi}, {Knezek},
  {Viscuso}, {Tacconi-Garman}, {Scoville}, {Schneider}, {Schloerb}, {Lord},
  {Lesser}, {Kenney}, {Huang}, {Devereux}, {Claussen}, {Case}, {Carpenter},
  {Berry}, \& {Allen}}]{young95}
{Young} J.~S. {et~al.}, 1995, \apjs, 98, 219

\end{thebibliography}

\appendix 
\section{New redshifts}\label{app:new_sources}
All but two sources observed in \CII\ were published by \citet{weiss13}, where the full source names are listed. The two new sources were selected from the list of \citet{mocanu13}, and observed as part of our ongoing ALMA Cycle~1 project to determine redshifts of additional SPT DSFGs. Table~\ref{table:newSPTz} lists their full names and ALMA band~3 continuum positions. The redshift of SPT2319-55 is based on detections of CO(5-4) and CO(6-5), while the redshift of SPT2311-54 is based on CO(5-4) confirmed by our APEX \CII\ detection (Figure~\ref{fig:compare_spec}).
\begin{table}
 \begin{tabular}[!ht]{@{}|lccc}
 \hline
Short name & Source & R.A. & Dec.\\
                    &            &   \multicolumn{2}{c}{J2000} \\
\hline
SPT2311-54  & SPT-S J231124-5450.5 & 23:11:23.94 & -54:50:30.0 \\
SPT2319-55  & SPT-S J231922-5557.9 & 23:19:21.67 & -55:57:57.8 \\            
\hline
\end{tabular}
 \caption{Source names are based on positions measured with the SPT (Mocanu et al. 2013). Source positions are
based on the ALMA 3\,mm continuum data.}
 \label{table:newSPTz}
\end{table}

\section{High-$z$ comparison galaxies}\label{app:high_z_sample}
\begin{table*}
 \label{table:Highzcompare}
 \begin{tabular}[!hb]{@{}lcccccclll}
  \hline
  Source & $z$ & L$_{\text{\CII}}$ 		&  L$_{\text{CO(1--0)}}$ & \LFIR \ & \Td & lensing & \CII \ and CO reference \\
		&      &   [$10^9$\,L$_{\odot}$]	&  [$10^6$\,L$_{\odot}$]  & 	 [$10^{12}$\,L$_{\odot}$] & [K] & magnification & \\
  \hline
SMMJ2135-0102 (Eyelash)		&       2.33 &   $5.8 \pm 1.3$   & $0.8\pm 0.04$  & $40.9\pm 9.1$ &$ 36.0\pm 2.3$ & $32.5\pm4.5$ & {\scriptsize \citealt{swinbank10,ivison10}}\\ 
SDP.130 			                         &       2.63 &   $<65$               & $10.1\pm 2.9$  & $17.5\pm 3.6$ & $38.6\pm2.0$ & $6\pm1$ &{\scriptsize \citealt{valtchanov11,frayer11}}\\ 
SDP.81 				                 &       3.04 &   $275\pm 39$    & $22.0\pm 5.0$ & $33.1\pm 6.3$ & $40.0\pm 1.8$ & $25\pm7$ &{\scriptsize \citealt{valtchanov11,frayer11}}\\ 
PSSJ2322+1944$^\dagger$		&       4.12 &   $<8.0$              & $6.1\pm 2.6$   & $ 28.6\pm 10.2$ & $ 55.8\pm6.1$ & $5.4\pm0.3$ & {\scriptsize  Benford et al. in prep., \citealt{carilli02b}}\\ 
SDP.141 			                         &       4.24 &   $60\pm 9$      & $38.7\pm 4.6$ & $66.2\pm 17.7$& $46.1\pm 3.2$ & $10-30$ &{\scriptsize \citealt{cox11}}\\ 
BRI1335-0417$^\dagger$		        &       4.41 &   $15.7\pm 2.5$  & $4.4\pm 0.8$  & $17.7\pm  4.9$ & $51.3\pm 4.4$ & --- &{\scriptsize \citealt{wagg10, carilli02a}}\\
ALESS65.1                                        &       4.45  &  $3.2\pm 0.4$   & $<1.11$             & $3.12\pm1.40$ & $44.3\pm5.1$ & --- & {\scriptsize \citealt{swinbank12, huynh14}}\\
BR1202-0725$^{\bigstar}$$^\dagger$ &     4.69 &    $15.8\pm 1.8$  & $5.4\pm 0.7$ & $34.1\pm 8.4$ & $55.1\pm 4.3$ & --- &{\scriptsize \citealt{wagg12, carilli02a}}\\ 
HDF850.1                                          &       5.19 &     $7.5\pm 0.8$   & $2.2\pm 0.5$ & $1.7 \pm 1.6$    & $29.9\pm7.7$ & $1.5-1.7$ & {\scriptsize \citealt{neri14, walter12}}\\ 
HLSJ091828.6+514223                    &       5.24 &    $82.4\pm 2.3$ & $43.0\pm 3.3$& $97.5\pm 24.4$ & $48.7\pm 3.0$ & $8.9\pm1.9$ & {\scriptsize \citealt{rawle14}}\\ 
SDSSJ1044-0125$^\dagger$           &       5.78 &    $1.6\pm0.4$    & $1.6\pm 0.3$ & $5.6\pm5.1$ & $59.8\pm 25.8$ & --- &{\scriptsize \citealt{wang13}}\\ 
SDSSJ2310+1855$^\dagger$         & 	 6.00 &   $8.7\pm1.4$    & $12.1\pm1.0$ & $20.0\pm 14.5$ & $61.8\pm20.7$ & --- &{\scriptsize \citealt{wang13}}\\ 
HerMESFLS3                                  & 	 6.34 &   $14.9\pm 3.1$  & $4.6\pm1.5$  & $29.9\pm9.6$ & $52.8\pm 6.2$& $2.2\pm0.3$ &  {\scriptsize \citealt{riechers13}}\\ 
SDSSJ1148+5251$^\dagger$ 	&       6.42 &    $4.6\pm0.5$ 	 & $6.4\pm0.7$  & $9.8\pm 6.7$ & $60.9\pm15.7$ & --- &{\scriptsize \citealt{maiolino05}}\\ 
 \hline 
IRASF10026+4949 		&       1.12 &    $25.8 \pm 3.2$ & $<0.8$	    & --- & --- &  & {\scriptsize \citealt{stacey10}}\\ 
3C368 				&       1.13 &    $9.3 \pm 1.5$	 & $<1.9$          & --- & --- &  & {\scriptsize \citealt{stacey10, evans96}}\\
SMMJ123634.51+621241.0 		&       1.22 &    $14.4 \pm 2.0$ & $3.76\pm 0.12$ & --- & --- &  & {\scriptsize \citealt{stacey10, frayer08}}\\
MIPSJ142824.0+352619 		&       1.32 &    $10.5 \pm 3.1$ & $1.10\pm 0.03$ & --- & --- &  & {\scriptsize \citealt{hailey-dunsheath10, iono06}}\\ 
SDSSJ100038.01+020822.4$^\dagger$&       1.83 &    $10.6 \pm 2.4$ & $3.10\pm 0.17$ & --- & --- &  & {\scriptsize \citealt{stacey10, aravena08}}\\
SWIREJ104704.97+592332.3 	&       1.95 &   $20.0 \pm 2.6$  & $8.9$	  & --- & --- & & {\scriptsize \citealt{stacey10}}\\ 
SWIREJ104738.32+591010.0 	&       1.96 &   $12.1 \pm 3.2$  & $3.5$	  & --- & --- &   & {\scriptsize \citealt{stacey10}}\\ 
BRI0952-0115 $^\dagger$		&       4.44 &   $4.5 \pm 2.6$	 & $0.43\pm 0.11$ & --- & --- &  & {\scriptsize \citealt{maiolino09, guilloteau99}}\\
LESSJ033229.4-275619$^\dagger$	&       4.76 &   $9.9 \pm 1.5$	 & $1.05\pm 0.25$ & --- & --- &  & {\scriptsize \citealt{debreuck11,coppin10}}\\
SDSSJ0129-0035$^\dagger$		&       5.78 &    $1.8\pm0.3$    & $2.8\pm0.5$    & --- & --- &  & {\scriptsize \citealt{wang13}}\\ 
SDSSJ2054-0005$^\dagger$         & 	 6.04 &   $3.3\pm0.5$ 	 & $2.7\pm0.6$    & --- & --- &  & {\scriptsize \citealt{wang13}}\\ 
ULASJ1319+0950$^\dagger$          & 	 6.13 &   $4.4\pm0.9$ 	 & $3.5\pm0.7$    & --- & --- & & {\scriptsize \citealt{wang13}}\\ 
CFHQSJ0210-0456$^\dagger$        & 	 6.43 &   $0.30\pm0.04$  & $<0.88$ 	  & --- & --- &  & {\scriptsize \citealt{willott13}}\\ 
\hline 
ALESS61.1                                         &      4.42  &        $1.48\pm 0.23$       &       ---                       & $3.0\pm 1.4$ & $43.7\pm5.1$  & --- & {\scriptsize \citealt{swinbank12}}\\
 \end{tabular}
  \caption{The high-$z$ sources in the comparison sample. All but one sources (ALESS61.1) have published \CII \ and CO detections. The observed CO luminosities have been scaled to CO(1--0) luminosities using the ratios from Stacey et al. 2010. The first 14 sources have a sufficient amount of photometric data published for the determination of \LFIR \ and \Td, while the remaining are unconstrained. ALESS61.1 has good photometry, but no published low-$J$ CO observations.\newline
$^{\bigstar}$ A sum of the [CII] and CO emission from the north and south source.\newline
$^\dagger$ AGN dominated source.}
\end{table*}

\bsp

\label{lastpage}

\end{document}